\documentclass[12pt,a4paper abstracton]{scrartcl} 
\usepackage{amssymb,amsmath,amsfonts,mathrsfs}
\usepackage[utf8]{inputenc}
\usepackage[T1]{fontenc}
\usepackage{wasysym}
\usepackage[ amsmath, thmmarks, hyperref]{ntheorem}
\usepackage{mathrsfs}
\usepackage{fullpage}
\usepackage{pstricks}
\usepackage{microtype}
\usepackage[ colorlinks=true, citecolor=blue]{hyperref}
\usepackage{caption}
\usepackage{graphicx,color}
\usepackage{array}
\usepackage{tikz}
\usetikzlibrary{shapes.geometric}
\usetikzlibrary{decorations.pathmorphing}
\usetikzlibrary{arrows,shapes,positioning}
\usetikzlibrary{decorations.markings}
\tikzstyle arrowstyle=[scale=1]
\tikzstyle directed=[postaction={decorate,decoration={markings,mark=at position .65 with {\arrow[arrowstyle]{stealth}}}}]
\tikzstyle reverse directed=[postaction={decorate,decoration={markings,mark=at position .65 with {\arrowreversed[arrowstyle]{stealth};}}}]

\usepackage{mathptmx}
\DeclareRobustCommand\hbar{\mathchar'26\mkern-9mu h}
\def\ben{\begin{equation}}
\def\een{\end{equation}}
\def\bena{\begin{eqnarray}}
\def\eena{\end{eqnarray}}

\newcommand{\non}{\nonumber}

\theoremstyle{plain}
\theoremseparator{:}
\theoremheaderfont{\normalfont\sffamily\bfseries}
 \theorembodyfont{\itshape}

\newtheorem{thm}{Theorem}
\newtheorem{lemma}{Lemma}[section]

\newtheorem{definition}{Definition}[section]

\newcommand{\lc}{-\frac{1}{4}}
\newcommand{\eH}{\mathscr{H}}
\newcommand{\eB}{\mathscr{B}}
\newcommand{\eC}{\mathscr{C}}
\newcommand{\M}{\mathscr{M}}
\newcommand{\D}{\mbox{d}}
\newcommand{\E}{\mathcal{E}}
\newcommand{\cO}{\mathcal{O}}

\newcommand{\cD}{\text{\dh}}
\newcommand{\cP}{\text{\th}}
\newcommand{\cL}{\mathcal{L}}
\newcommand{\cA}{\mathcal{A}}
\newcommand{\bbE}{\mathbb{E}}
\newcommand{\bbF}{\mathbb{F}}
\newcommand{\cS}{\mathcal{S}}
\newcommand{\cR}{\mathcal{R}}
\newcommand{\cT}{\mathcal{T}}
\newcommand{\cC}{\mathcal{C}}
\newcommand{\bC}{{\bf C}}
\newcommand{\NN}{\mathbb{N}}
\newcommand{\RR}{\mathbb{R}}
\newcommand{\CC}{\mathbb{C}}

\newcommand{\ZZ}{\mathbb{Z}}
\newcommand{\eI}{\mathscr{I}}
\newcommand{\eps}{\varepsilon}
\newcommand{\half}{\frac{1}{2}}
\newcommand{\thalf}{\tfrac{1}{2}}

\newcommand{\bD}{\mathcal{D}}

\newcommand{\uk}{\underline{k}}
\newcommand{\um}{\underline{m}}
\renewcommand{\hbar}{h}

\newcommand{\bX}{{\bf X}}

\begin{document}

\title{Instabilities of extremal rotating black holes in higher dimensions}

\author{Stefan Hollands, Akihiro Ishibashi}

\author{Stefan Hollands$^{1}$\thanks{\tt stefan.hollands@uni-leipzig.de}
and Akihiro Ishibashi$^{2}$\thanks{\tt akihiro@phys.kindai.ac.jp}
\\ \\
{\it ${}^{1}$Institut f\" ur Theoretische Physik,}\\ 
{\it Universit\" at Leipzig, }\\
{\it Br\" uderstrasse 16, D-04103 Leipzig, Germany} \\
{\it ${}^{2}$Department of Physics, Kinki University, } \\
{\it Higashi-Osaka, 577-8502, Japan} \\
}

\maketitle

\begin{abstract}
Recently, Durkee and Reall have conjectured a criterion for linear instability of rotating, extremal,
asymptotically Minkowskian black holes in $d\ge 4$ dimensions, such as the Myers-Perry black holes.
They considered a certain elliptic operator, $\cA$, acting on symmetric trace-free tensors intrinsic
to the horizon. Based in part on numerical evidence, they suggested that if the lowest eigenvalue of this operator is less than
the critical value $-1/4$ ( called ``effective BF-bound''), then  the black hole is linearly unstable.
In this paper, we prove an extended version of their conjecture. Our proof uses a combination of methods such as (i)
the ``canonical energy method'' of Hollands-Wald,
(ii) algebraically special properties of the near horizon geometries associated with the black hole, (iii)
the Corvino-Schoen technique, and (iv) semiclassical analysis. 
Our method of proof is also applicable to rotating, extremal
asymptotically Anti-deSitter black holes. In that case, we find additional instabilities for ultra-spinning
black holes. Although we explicitly discuss in this paper only extremal black holes, we argue that our results
can be generalized to {\em near} extremal black holes.  
\end{abstract}

\section{Introduction}

Whether one believes that extra dimensions ought to play a role in fundamental theories of Nature, or
whether one merely employs them as a tool in holographic approaches to
strongly correlated real-life systems~\cite{Herzog,Hartnoll}, one needs to understand the nature of black holes
in higher dimensional spacetimes. Apart from the obvious interest in finding new, in particular stationary, black
hole solutions, it is also very important to understand the stability properties of known solutions,
see~\cite{Emparan-Reall} for a review. Stable black holes are of obvious
relevance. But also unstable ones are interesting, because instabilities can evolve to new, as yet
unknown, black holes, or they can correspond to new stationary black holes branching off a known solution.

To analyze the (in)stability of a background, the first step is to study linear perturbations, i.e. 
solutions to the linearized Einstein equations (in this paper we consider
the vacuum Einstein equations with $\Lambda$). If these settle down in a sufficiently strong sense, then one can hope that small non-linear perturbations will do the same. 
On the other hand, if there are linear perturbations which do not settle down, then the background is clearly unstable, 
although a linear analysis cannot be used to predict what might be the endpoint of the non-linear evolution. In this paper, we 
want to identify criteria for linear instabilities of (rotating) black hole backgrounds in $d \ge 4$ dimensions. 

Unfortunately, understanding the long-time behavior of solutions to the linearized Einstein equations on black hole backgrounds is a highly non-trivial problem. 
It has been solved in generality only
for the -- already far from trivial -- case of Schwarzschild spacetime~\cite{Regge-Wheeler, Zerilli,Moncrief}, and its higher-dimensional
cousins~\cite{Ishibashi-Kodama}, where no unstable modes\footnote{It is very important to note that
mode stability does not imply uniform boundedness in time of generic perturbations with bounded initial data.
This problem has been studied e.g. in~\cite{Dafermos-Rodnianski1,Dafermos-Rodnianski2,Blue}.
} were found. For the Kerr-metric,
one can cast the perturbation equations in Teukolsky form~\cite{Teukolsky}, and thereby analyze stability. Again,
no unstable modes were found~\cite{Whiting}. This success suggests to search for an analogous `Teukolsky' form for the perturbation
equations of rotating black holes also in higher dimensions, e.g. for the Myers-Perry solutions~\cite{Myers-Perry,Lu}, which can be viewed as generalizations of Kerr/Kerr-AdS, or the black rings~\cite{Emparan,Senkov}.
Since the existence of the Teukolsky form appears to be related to the profound algebraically special properties of the
Kerr metric, one is naturally led to generalize such notions to higher dimensions, as was in fact done in a series of papers by~\cite{Pravda1,Pravda2,Pravda3,Durkee,Durkee2}. Unfortunately, the bottom line of these investigations is that the known rotating black holes are
not of a sufficiently algebraically special nature to cast the perturbation equations in Teukolsky form. It appears that,
mainly for this reason, there has been limited success in the analytical understanding of the (even linear)
stability of generic rotating black holes in higher dimensions, although there are by now several very interesting partial numerical results~\cite{Santos,Santos2}.

In~\cite{Durkee} Durkee and Reall observed that, while the perturbation equations on the known asymptotically flat rotating backgrounds in $d>4$
cannot be put in Teukolsky form, this is possible for their {\em near horizon (NH) limits}~\cite{Lucietti1,Lucietti2,Lucietti3,Lucietti4}. 
In fact,~\cite{Durkee} showed that the Teukolsky equations on the NH geometry separate into an ``$(R,T)$''-part obeying a
charged Klein-Gordon equation
in an auxiliary\footnote{Note that the original black hole background has vanishing cosmological constant, and is asymptotically Minkowskian,
rather than asymptotically anti-deSitter.} $AdS_2$-space, and an ``angular part''. The modes of the angular part are eigenfunctions of an
elliptic operator, $\cA$, acting on symmetric trace-free tensors intrinsic to the $(d-2)$-dimensional horizon cross section, $\eB$. Its eigenvalues
effectively become a mass term in the $AdS_2$-Klein-Gordon equation for the $(R,T)$-part. By  looking at the
properties of that equation,~\cite{Durkee} made a conjecture about the stability properties of the corresponding extremal 
black hole (assumed to be asymptotically flat, $\Lambda = 0$), which we rephrase as follows: \\

{\bf Conjecture~1}:{\em Assuming generic}\footnote{
\label{footnote1}
We call the angular velocities $\underline{\Omega} = (\Omega^1,\dots,\Omega^n)$ {\em generic} if the components are linearly independent over
$\mathbb{Q}$. This is the same as saying that there is no non-trivial vector of integers $\um$ such that $\um \cdot \underline{\Omega} = 0$. The generic values form a dense set of 
full Lebesgue measure. For non-generic values, a variant of the conjecture can be formulated, see part (ii) of thms.~\ref{thm2}, \ref{thm3}.} 
{\em values of the angular velocities of the black hole, if the lowest eigenvalue  $\lambda$ of the operator
$\cA$ defined in~\eqref{Adef} (acting on axisymmetric tensors) is below the critical value of $\lc$ (called the ``effective BF-bound''), then the original extremal black hole is
unstable.} \\

To support
their conjecture,~\cite{Durkee} worked out explicitly the spectrum of $\cA$ in the case of the cohomogneity-1 Myers-Perry black holes, and
compared the implications of their conjecture to the numerical results of~\cite{Santos,Santos2}. The conjecture was thereby found to
hold up to dimension $d=15$. 
(For further support for their conjecture, see, e.g., \cite{Murata1103.5635,Tanahashi1208.0981}.)
In this paper, we prove conjecture~1.  The precise
statements  are given below in thm.~\ref{thm2}, which also includes an extension concerning ``nongeneric'' values
of the spin parameter, relevant for the stability of ultraspinning black holes. 

To show conjecture~1, the first idea might be to look at the explicit form of the $(R,T)$-part of
the perturbations in the NH geometry corresponding to a mass below the effective BF-bound. Unfortunately, while
these modes can be given in closed form (see e.g.~\cite{Horowitz-Marolf,Dias:2009ex} and also appendix~\ref{appB}), it is hard to see what one learns
from them directly about the behavior of perturbations on the original black hole background. 
The point is that the  modes 
fail to be $L^2$-normalizable at the ``infinity'', $R \to \infty$, of the NH geometry. But the NH geometry is
supposed to be a reasonable description (``blow up'') of the black hole only for finite $R$, so it is rather unclear how one could use those 
modes directly to prove or disprove the above conjecture. It is also unclear how to implement the dynamical evolution of compactly 
supported initial data for the $AdS_2$-Klein-Gordon equation, because the case $\lambda < \lc$ corresponds precisely to the situation
where it is essentially impossible to construct a well-defined $AdS_2$-dynamics~\cite{Ishibashi-Wald}.

For these reasons, we will use a different approach which is based on a method introduced in~\cite{HW}\footnote{In~\cite{HW} this approach was introduced in the context of
general non-extremal black holes with $\Lambda=0$. This method suitably generalizes to the extremal case, and it also generalizes 
to $\Lambda<0$.  Theories with various additional types of 
matter fields were considered in~\cite{Keir,Greene}.}.
The method is a sort of variational principle associated with the so-called ``canonical energy'', $\E$, of the perturbation, $\gamma_{ab}$.
$\E$ is a quadratic expression depending on up to two derivatives of the perturbation,
and depending on a Cauchy surface $\Sigma$ outside the black hole. Its concrete form
and key properties are recalled below in sec.~\ref{sec1}. These are: 1) $\E$ is gauge invariant,
2) $\E$ is monotonically decreasing for any axi-symmetric perturbation, in the sense that $\E(\Sigma_2) \le \E(\Sigma_1)$ as long as  $\Sigma_2$
is later than $\Sigma_1$ [see fig.~\ref{DOC1}]. 3) $\E$ vanishes if and only if $\gamma_{ab}$ represents a perturbation towards another black hole in the family up to a gauge transformation. Properties 1),2),  and 3) together
imply that if we can find a perturbation with $\E < 0$, then such a perturbation
cannot settle down to a perturbation to another stationary black hole in the family.
Hence, such a perturbation must correspond to a linear instability.

Thus, to establish an instability, we must find a perturbation $\gamma_{ab}$
for which $\E<0$. Since $\E$ can be expressed in terms of the initial data
for the perturbation, we basically have a variational problem involving initial data. 
However, a major complication
arises from the fact that the initial data must satisfy the linearized constraint equations.
Since these have a rather complicated structure, this may at first sight appear to render our
method rather impractical. Fortunately, it turns out that, in order 
to construct the desired perturbation with $\E<0$, we can proceed by a roundabout
route which effectively avoids having to solve the constraints explicitly.  
There are basically three steps:

\begin{enumerate}
\item We pass to the NH limit of the black hole. In the NH
limit,  linearized perturbations can be constructed via a
higher dimensional generalization~\cite{Madi} of the ``Hertz-potential'' ansatz~\cite{Hertz1,Hertz2}. Using the
Hertz-potential ansatz, and the separability property of the linearized Einstein
equations on the NH geometry background established by~\cite{Durkee}, we reduce the canonical energy
$\E$ in the NH geometry to an ``energy-like''\footnote{
The $AdS_2$ Klein-Gordon field involves a {\em complex} charge parameter. This implies, 
among other things, that the energy-like expression has an unusual form, containing up to 3 derivatives.
} expression involving only the complex scalar Klein-Gordon-type field on the
auxiliary $AdS_2$-space. It
involves the lowest eigenvalue, $\lambda$ of the elliptic, second order, hermitian operator
$\cA$ [see eq.~\eqref{Adef}] on the $(d-2)$-dimensional horizon cross section. It is shown that the $AdS_2$-energy can become negative
for compactly supported data outside the horizon
if $\lambda$ is below the critical value $\lc$. From these data, we get a gravitational perturbation
with compactly supported initial data in the NH geometry, having $\E<0$.

\item The perturbation of the NH geometry obtained in step 1) is next scaled, using
the isometries of the NH geometry, to a perturbation having support in a neighborhood
of ``size'' $\eps$ near the  horizon. In such a neighborhood, the NH geometry is by construction approximately equal
to the original black hole geometry. It is therefore plausible -- and will be shown --
that the initial data of the scaled perturbation satisfy, to within a small error of order $\epsilon$, the linearized 
constraint equations of the original geometry.

\item We show by the powerful methods of Corvino-Schoen~\cite{Corvino} and also~\cite{Chrusciel} that for sufficiently small $\eps$, 
the initial data of the scaled perturbation on the NH geometry can be modified
to give a perturbation on the original black hole geometry still having $\E<0$.
\end{enumerate}

The above mentioned properties of $\E$ then imply that conjecture~1 is true for any of the known extremal, asymptotically flat black holes, i.e. 
the Myers-Perry black holes and the black rings. Finding the lowest eigenvalue of $\cA$ in those concrete 
geometries is a {\em much} simpler problem than that of analyzing the perturbed Einstein equations \eqref{einstein}, although
even this problem probably has to be solved on a computer for generic values of the spin-parameters. 

The techniques of this paper also apply to 
the case of rotating, extremal, asymptotically $AdS$ black holes ($\Lambda<0$) of the MP-type. In this case, 
the black hole is found to be not only unstable for an eigenvalue below the effective BF-bound, but also 
under more general conditions including the case of ultra-spinning black holes. The precise statement is given in thm.~\ref{thm3}. The dS-case is
briefly discussed in the conclusions section, where we also discuss the extension of 
conjecture~1 to near extremal black holes. The same methods as for the gravitational perturbations also work 
for a test Maxwell field, and analogues of all of the above results are shown to be true for that case, too.

\medskip
\noindent
{\bf Conventions:} Our conventions for the signature and definition of the Riemann tensor are identical with those used in Wald's text~\cite{waldbook}. Letters $a,b, \dots$ from 
the beginning of the Roman alphabet refer to tensor structures on spacetime $\M$, whereas indices $i,j,\dots$ from the middle of the Roman alphabet
to tensor structures on a Cauchy surface $\Sigma$. Capital letters $A,B, \dots$ from the beginning of the Roman alphabet 
refer to tensors on the horizon cross section $\eB$, whereas letters $I,J, \dots$ from the middle of the Roman alphabet 
run between $1$ and $n$ and label the rotational isometries. 

\section{Stationary black holes and canonical energy} \label{sec0}

\subsection{Stationary black holes and their perturbations}

In this paper, we consider $d$-dimensional stationary black hole spacetimes $(\M, g)$ with Killing horizons satisfying the Einstein equations $G_{ab} = R_{ab} - \half R g_{ab} = -\Lambda g_{ab}$ with cosmological constant $\Lambda$.
A stationary spacetime with Killing horizon by definition has a Killing vector field (KVF) $K$ that is tangent to the generators of the horizon $\eH = \eH^+ \cup \eH^-$, where $\pm$ means the future/past horizon. This implies that, on $\eH$, we have
\ben
K^b \nabla_b K^a = \kappa K^a \ .
\een
The quantity $\kappa \ge 0$ is the surface gravity and is shown to be constant on $\eH$~\cite{waldbook}. A black
hole is called ``extremal'' if $\kappa=0$, in which case the flow of $K$ coincides with the geodesic flow of affinely
parameterized null-geodesics of $\eH$. {\em  For the remainder of this paper until sec.~\ref{conclusions},
we restrict attention to extremal black holes (BH's).} If the black hole is
rotating, i.e. if $K$ does not coincide with the asymptotically time-like Killing
vector field, then one can show in a very general setting~\cite{HIW} that there must exist
rotational Killing vector fields, written $\partial/\partial \phi^I, I=1, \dots ,n>0$ in suitable coordinates, 
such that
\ben
K = \frac{\partial}{\partial t} + \Omega^I \frac{\partial}{\partial \phi^I} \ .
\een
The constants $\Omega^I \in \RR, I=1, \dots, n>0$ are called the ``angular velocities''
of the horizon, and ``rotational'' means that $\partial/\partial \phi^I$ should generate an isometric
action of ${\rm U}(1)^n$ on the spacetime corresponding to shifts in the angular coordinates $\phi^I$.
Concrete examples of such black holes are the Myers-Perry solutions~\cite{Myers-Perry,Lu} (briefly reviewed in sec.~\ref{sec0}), or the black rings~\cite{Emparan,Senkov}. 
In these examples, $n = \lfloor (d-1)/2 \rfloor$. 

When $\Lambda = 0$, one is dealing with asymptotically flat spacetimes, see~\cite{Hollands-Ishibashi,Tanabe}
and~\cite{Thorne} for a precise definition of this concept in higher dimensions. For even
$d$, this notion can be formulated within the formalism of conformal infinity, used throughout this paper. In this framework, 
one considers a conformal compactification $(\tilde \M, \tilde g = f^2 g)$ of $(\M, g)$. Future/past infinity correspond to the 
conformal boundary $\eI = \eI^+ \cup \eI^-$, which is a (conformal) null surface defined by $f=0$. The definition of the canonical 
energy given in the next section uses the framework of conformal infinity. Since our general arguments rely on the properties of 
the canonical energy (see sec.~\ref{sec2}) which are derived using that framework, the results of this paper apply, strictly speaking, only to even $d$, 
which is from now on assumed in the asymptotically flat context. It is highly likely that this technical assumption can be removed 
by replacing the framework of conformal infinity by that of~\cite{Tanabe}, but we shall not attempt to do this here.

When $\Lambda < 0$, the spacetime is asymptotically $AdS$.
In this case, the conformal boundary $\eI$ is timelike, see e.g.~\cite{Marolf} for further explanation.

A metric perturbation is a  solution to the linearized Einstein equations around a background satisfying $G_{ab} + \Lambda g_{ab} = 0$. Denoting the 
linearization of $G_{ab} + \Lambda g_{ab}$ by the linear operator $\gamma_{ab} \mapsto (\cL \gamma)_{ab}$, the linearized Einstein 
equations can be written as:
\ben\label{einstein}
\begin{split}
0 = (\cL \gamma)_{ab}  \equiv&
- \half \nabla_a \nabla_b \gamma - \half \nabla^c \nabla_c \gamma_{ab} +
\nabla^c \nabla_{(a} \gamma_{b)c} \\
&- \half g_{ab}
   \left(
         \nabla^c\nabla^d \gamma_{cd}
         -\nabla^c \nabla_c \gamma
         - \frac{2\Lambda}{d-2} \gamma
   \right)
 - \frac{2\Lambda}{d-2} \gamma_{ab} \ ,
\end{split}
\een
where $\gamma = \gamma_a{}^a$ in this equation, and where indices are raised and lowered with $g_{ab}$. 
This equation has a gauge-invariance in the sense that $\gamma_{ab} = \pounds_X g_{ab}$ is a solution to $(\cL \gamma)_{ab}=0$
for any smooth vector field $X^a$.
In this paper, we will consider only perturbations having initial data of compact support on some Cauchy surface of the exterior region, 
$\Sigma$, i.e. the support is bounded away from the black hole $\eB = \eH \cap \Sigma$
and infinity $\eC = \eI \cap \Sigma$. See fig.~\ref{DOC1} for an illustration of this situation with $\Sigma = \Sigma_1$.

\begin{figure}
\begin{center}
\begin{tikzpicture}[scale=.8, transform shape]
\draw (-2,6) decorate[decoration=snake] {--(-2,-6)};
\draw (-2,6) ..controls (-0.2,4) and (-0.2,4) .. (-2,2);
\filldraw[fill=gray,opacity=.5,draw=black] (-2,6) -- (0,4) -- (-2,2) ..controls (-0.2,4) and (-0.2,4) .. (-2,6);
\filldraw[fill=gray,opacity=.5,draw=black] (-2,2) -- (0,0) -- (-2,-2) ..controls (-0.2,0) and (-0.2,0) .. (-2,2);
\filldraw[fill=gray,opacity=.5,draw=black] (-2,-6) -- (0,-4) -- (-2,-2) ..controls (-0.2,-4) and (-0.2,-4) .. (-2,-6);
\filldraw[fill=gray,opacity=.5,draw=black] (0,4) -- (-2,2) -- (0,0) ..controls (-1.8,2) and (-1.8,2) .. (0,4);
\filldraw[fill=gray,opacity=.5,draw=black] (0,-4) -- (-2,-2) -- (0,0) ..controls (-1.8,-2) and (-1.8,-2) .. (0,-4);
\filldraw[fill=gray,opacity=.5,draw=black] (-2,6) -- (-1.4,6) ..controls (-1.4,5.8) and (-1.4,5.8) .. (0,4);
\filldraw[fill=gray,opacity=.5,draw=black] (-2,-6) -- (-1.4,-6) ..controls (-1.4,-5.8) and (-1.4,-5.8) .. (0,-4);
\draw (-2,6) -- (0,4) -- (-2,2) -- (0,0) -- (-2,-2) -- (0,-4) -- (-2,-6);
\draw (-2,6) node[draw,shape=circle,scale=0.3,fill=black]{};
\draw (-2,2) node[draw,shape=circle,scale=0.3,fill=black]{};
\draw (-2,-2) node[draw,shape=circle,scale=0.3,fill=black]{};
\draw (-2,-6) node[draw,shape=circle,scale=0.3,fill=black]{};
\node at (-1.4,4.6) {$\mathscr{H}^-$};
\node at (-1.4,3.4) {$\mathscr{H}^+$};
\node at (-1.4,0.6) {$\mathscr{H}^-$};
\node at (-1.4,-3.4) {$\mathscr{H}^+$};
\node at (-1.4,-0.6) {$\mathscr{H}^-$};
\node at (-1.4,-4.6) {$\mathscr{H}^+$};
\node at (-2.8,2) {$\mathscr{B}$};
\draw[red,thick] (-2,2) -- (2,2);
\draw (0,0) ..controls (-1.4,2) and (-1.4,2) .. (0,4);
\draw (0,0) ..controls (-1.0,2) and (-1.0,2) .. (0,4);
\draw (0,0) ..controls (-0.5,2) and (-.5,2) .. (0,4);
\draw (0,0) ..controls (0,2) and (0,2) .. (0,4);
\draw (0,0) ..controls (0.5,2) and (.5,2) .. (0,4);
\draw (0,0) ..controls (1.0,2) and (1,2) .. (0,4);
\draw (0,0) ..controls (1.4,2) and (1.4,2) .. (0,4);
\draw (0,0) ..controls (1.8,2) and (1.8,2) .. (0,4);
\draw (-2,2) ..controls (0,2.5) and (0,2.5) .. (2,2);
\draw (-2,2) ..controls (0,1.5) and (0,1.5) .. (2,2);
\draw (-2,2) ..controls (0,3) and (0,3) .. (2,2);
\draw (-2,2) ..controls (0,1) and (0,1) .. (2,2);
\draw (-2,2) ..controls (0,3.4) and (0,3.4) .. (2,2);
\draw (-2,2) ..controls (0,.6) and (0,.6) .. (2,2);
\draw (-2,2) ..controls (0,3.8) and (0,3.8) .. (2,2);
\draw (-2,2) ..controls (0,.2) and (0,.2) .. (2,2);
\node at (-.8,1.6) {\textcolor{red}{$\Sigma$}};
\filldraw[fill=white,draw=white] (0,6) -- (2,6) -- (2,-6) -- (0,-6) -- (0,6);
\draw[double] (0,6) -- (0,-6);
\draw (0,4) node[draw,shape=circle,scale=0.4,fill=black]{};
\draw (0,0) node[draw,shape=circle,scale=0.4,fill=black]{};
\draw (0,-4) node[draw,shape=circle,scale=0.4,fill=black]{};
\node at (.5,2) {$\mathscr{I}$};
\node at (.5,-2) {$\mathscr{I}$};
\end{tikzpicture}
\end{center}
\caption{
\label{MP0}
Conformal diagram of the extremal $\Lambda<0$ MP spacetime. 
}
\end{figure}
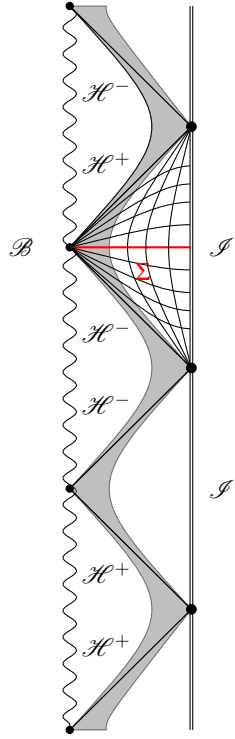

The linearized Einstein equation is not hyperbolic in nature due to its gauge invariance.
But, as is well known, if one fixes the gauge (e.g. the transverse-trace-free gauge), then
the system becomes hyperbolic, and possesses a well-posed initial value formulation. This means that, if we prescribe compactly supported initial data on $\Sigma$ (satisfying the linearized
constraints, see below), then the solution $\gamma_{ab}$ exists, is smooth, and is unique inside the domain of dependence $D(\Sigma)$. In the asymptotically flat ($\Lambda = 0$)
spacetimes considered in this paper such as the extremal Myers-Perry black holes, if we take $\Sigma$ to be a slice as shown in fig.~\ref{MP1}, then $D(\Sigma)$ comprises an
entire exterior region. 
In the asymptotically $AdS$ black hole spacetimes $(\Lambda<0)$ considered in this paper, if we take $\Sigma$ to be a slice as shown in fig.~\ref{MP0}, 
then $D(\Sigma)$ is only a subset of an exterior region. 
This corresponds to the well-known fact that these regions are not globally hyperbolic. In order to get a solution in an entire exterior region, we must specify what happens 
at the $AdS$-conformal boundary ${\mathscr I}$. For this, one has to specify (conformal) boundary conditions on $\gamma_{ab}$, which for an exact $AdS$-background
were motivated thoroughly in~\cite{Ishibashi-Wald}, and correspond to keeping the conformal metric fixed to first order. The boundary-initial value problem for asymptotically $AdS$-spacetimes in the fully non-linear regime has been analyzed by~\cite{friedrich}. His results  imply that the initial-boundary value problem for the linearized problem 
has a globally regular solution in any exterior region, with the standard asymptotic expansions near $\mathscr I$ as given e.g. in~\cite{Marolf}.

\subsection{Canonical energy of gravitational perturbations}\label{sec1}

\begin{figure}
\begin{center}
\begin{tikzpicture}[scale=1, transform shape]
\filldraw[fill=gray,opacity=.2,draw=black] (-1,0) -- (-1.5,0.5) -- (0,2) -- (1.5,.5) -- (1,0) -- (1.5,-.5)--(0,-2)--(-1.5,-.5) --(-1,0);
\filldraw[fill=gray,opacity=.6,draw=black] (-2,0) -- (2,0) -- (1,1) -- (-1,1) -- (-2,0);
\draw[double] (0,2) -- (2,0) -- (0,-2);
\draw (0,2) -- (-2,0) -- (0,-2);
\draw[red,thick] (-2,0) -- (2,0);
\node at (0,-.4) {$\Sigma_1$};
\draw[red,thick] (-1,1) -- (1,1);
\node at (0,.6) {$\Sigma_2$};
\draw (-2,0) node[draw,shape=circle,scale=0.3,fill=black]{};
\draw (2,0) node[draw,shape=circle,scale=0.3,fill=black]{};
\draw (0,2) node[draw,shape=circle,scale=0.3,fill=black]{};
\draw (0,-2) node[draw,shape=circle,scale=0.3,fill=black]{};
\node at (-2.1,.5) {$\mathscr{H}_{12}$};
\node at (2.11,.5) {$\mathscr{I}_{12}$};
\node at (-1.65,1) {$\mathscr{B}_2$};
\node at (1.5,1) {$\mathscr{C}_2$};
\node at (-2.45,0) {$\mathscr{B}_1$};
\node at (2.47,0) {$\mathscr{C}_1$};
\draw (-1,1) node[draw,shape=circle,scale=0.3,fill=red]{};
\draw (1,1) node[draw,shape=circle,scale=0.3,fill=red]{};
\end{tikzpicture}
\end{center}
\caption{
\label{DOC1}
Conformal diagram of the exterior of the black hole.  To obtain the balance equation, we integrate $\nabla^a w_a = 0$ over 
 rectangle shaded in dark grey. The region shaded in light grey indicates the support of a perturbation 
$\gamma_{ab}$ whose initial data are compactly supported. 
}
\end{figure}
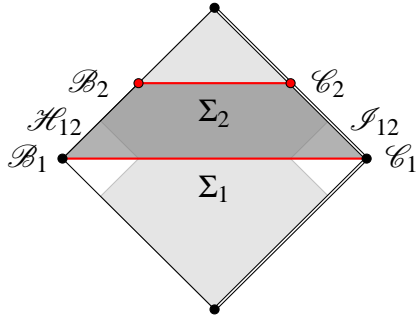

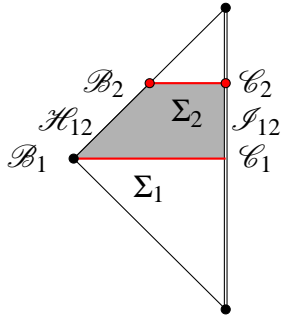
\begin{figure}
\begin{center}
\begin{tikzpicture}[scale=1, transform shape]
\filldraw[fill=gray,opacity=.6,draw=black] (-2,0) -- (0,0) -- (0,1) -- (-1,1) -- (-2,0);
\draw[double] (0,2) --  (0,-2);
\draw (0,2) -- (-2,0) -- (0,-2);
\draw[red,thick] (-2,0) -- (0,0);
\node at (-1,-.4) {$\Sigma_1$};
\draw[red,thick] (-1,1) -- (0,1);
\node at (-.5,.6) {$\Sigma_2$};
\draw (-2,0) node[draw,shape=circle,scale=0.3,fill=black]{};
\draw (0,2) node[draw,shape=circle,scale=0.3,fill=black]{};
\draw (0,-2) node[draw,shape=circle,scale=0.3,fill=black]{};
\node at (-2.1,.5) {$\mathscr{H}_{12}$};
\node at (.4,.5) {$\mathscr{I}_{12}$};
\node at (-1.6,1) {$\mathscr{B}_2$};
\node at (-2.6,0) {$\mathscr{B}_1$};
\node at (.4,1) {$\mathscr{C}_2$};
\node at (.4,0) {$\mathscr{C}_1$};
\draw (-1,1) node[draw,shape=circle,scale=0.3,fill=red]{};
\draw (0,1) node[draw,shape=circle,scale=0.3,fill=red]{};
\end{tikzpicture}
\end{center}
\caption{
\label{DOC2}
Conformal diagram of the exterior of the $AdS$ black hole.  To obtain the balance equation, we integrate $\nabla^a w_a = 0$ over 
the shaded rectangle. In this case, there is no flux across $\mathscr{I}_{12}$ due to the $AdS$-boundary conditions.
}
\end{figure}

We next recall the definition of the canonical energy of a perturbation of a
stationary asymptotically {\em AdS} ($\Lambda<0$) or flat ($\Lambda=0$) black hole with
Killing horizon, and its key
properties, referring to~\cite{HW} for details. The main ingredient is the ``{\em symplectic
current}'' of two solutions to the linearized Einstein equations, given by
\begin{equation}
  \label{omegadef}
w^a =\frac{1}{16\pi}  g^{abcdef} (\gamma_{2 \, bc} \nabla_d \gamma_{1 \, ef} - \gamma_{1 \, bc} \nabla_d \gamma_{2 \, ef}) \ , 
\end{equation}
where
\begin{equation}
  g^{abcdef}
  = g^{ae} g^{fb} g^{cd} - \frac{1}{2} g^{ad} g^{be} g^{fc} - \frac{1}{2} g^{ab} g^{cd} g^{ef}
  - \frac{1}{2} g^{bc} g^{ae} g^{fd} + \frac{1}{2} g^{bc} g^{ad} g^{ef}.
\end{equation}
This current is shown to be conserved, $\nabla^a w_a=0$.
The {\em symplectic form} $W(\Sigma; \gamma_1, \gamma_2)$ is defined
by integrating the dual $\star w$ over a $(d-1)$-dimensional submanifold $\Sigma$,
\ben
W(\Sigma; \gamma_1, \gamma_2) \equiv \int_\Sigma \star w(g; \gamma_1, \gamma_2) \, .
\label{symform}
\een
We
typically take $\Sigma$ to run between a cut $\eC$ of infinity $\eI$,  and a section $\eB$ of the future horizon $\eH^+$, or a slice ``running down the throat'', 
see $\Sigma=\Sigma_1$ or $=\Sigma_2$ in figs.~\ref{DOC1},~\ref{DOC2} for examples of such slices. 

In order to define the canonical energy associated with such a slice, we need to introduce
 two boundary terms, and we need to fix the gauge at $\eH$ and $\eI$. The gauge conditions and 
boundary terms are needed, as in~\cite{HW}, so that (i) $\E$ has appropriate gauge invariance properties, and 
such that (ii) $\E$ has suitable monotonicity properties. 
We begin by stating our gauge conditions. In the asymptotically flat case, 
we impose, near $\eI^\pm$, that the perturbation is in transverse-trace-free gauge. The decay near the null-infinities ${\mathscr I}^\pm$ of solutions with compactly supported data on a Cauchy surface $\Sigma=\Sigma_1$ (see fig.~\ref{DOC1}) in this gauge has been analyzed in sec.~2 of~\cite{Hollands-Ishibashi}. The analysis shows 
in particular that the integral~\eqref{symform} converges also for a Cauchy surface of the type $\Sigma=\Sigma_2$, see fig.~\ref{DOC1}. In the asymptotically $AdS$-case, we impose on $\gamma_{ab}$ the linearized 
version of the Graham-Fefferman type gauge, implying again convergence of~\eqref{symform} (see e.g.~\cite{Marolf}). Near $\eH^\pm$, we can first impose the linearized
``Gaussian normal null form'' gauge conditions described in~\cite{HW}.  As in that reference, we would additionally like to 
impose as a gauge condition that the perturbed expansion\footnote{
Here we use the standard convention that $\delta X$ denotes the first order perturbation of a quantity $X$. More precisely, if $g_{ab}(\lambda)$ is 
a differentiable 1-parameter family of metrics with $\gamma_{ab} = dg_{ab}(\lambda)/d\lambda |_{\lambda = 0}$, and if $X$ depends on $g_{ab}$ 
in a differentiable manner, then $\delta X = d X(g(\lambda))/d\lambda |_{\lambda = 0}$.}, 
$\delta \vartheta$, of  $\gamma_{ab}$, vanishes on $\eH^\pm$. In~\cite{HW} a proof was given that 
such a gauge always exists,  but this proof does not appear to generalize to extremal black holes. We circumvent this problem in 
the present paper by only considering perturbations $\gamma_{ab}$ having compact support on a slice ``going down the throat'', as shown 
by $\Sigma=\Sigma_1$ in fig.~\ref{DOC1}. In this situation $\delta \vartheta = 0$ on $\eH^\pm$  can be established via the linearized 
Raychaudhuri equation, 
\ben\label{ray}
\frac{\D }{\D u} \delta \vartheta = - \frac{2}{d-2} \vartheta \delta \vartheta - 2\sigma_{ab} \delta \sigma^{ab} - \delta R_{ab} K^a K^b  = 0 \ ,
\een
where $\sigma_{ab}$ and $\vartheta$ are the (vanishing)  shear and expansion of the background 
and $\delta \sigma_{ab}$ and $\delta \vartheta$ their first order variation under $\gamma_{ab}$.
The point is that, for example in the Lorentz gauge, $\gamma_{ab}$ must be supported in the region shaded in red in fig.~\ref{DOC1}
by the usual rules for the propagation of disturbances for hyperbolic PDE's.  Thus, 
$\delta \vartheta$ must clearly vanish for sufficiently negative values of the affine parameter $u$ on $\eH^+$, and therefore, 
by~\eqref{ray}, for all $u$ (and similarly for $\eH^-$). It then also follows that the perturbed area, $\delta A$, 
of a horizon cross section, must vanish on $\eH^\pm$, so 
\ben\label{horgauge}
\delta A |_{\eB} = 0 = \delta \vartheta|_\eB \ , 
\een
for any cross section $\eB \subset \eH$. 
It is not hard to see that the vector fields $X^a$ preserving this gauge under $\gamma_{ab} \to \gamma_{ab} + \pounds_X g_{ab}$ must be 
tangent to $\eH$. 
We next define the boundary terms. 
The first boundary term, $B(\eB, \gamma)$ is associated
with the section $\eB$ of the future horizon, and is defined as 
\ben\label{Bgammadef}
B(\eB, \gamma) = \frac{1}{32\pi} \int_\eB \gamma^{ab} \delta \sigma_{ab}  \ .
\een
The volume element $vol_{\eB}$ understood under the integral is defined by 
contracting $K^a$ into the first entry of $vol_{\mathscr H}$,  where $vol_{\mathscr H}$ is defined in turn implicitly by
$vol_{\mathscr H} \wedge K= vol_{g}$. The definition of the
boundary term from infinity, $B(\eC, \gamma)$, depends on the asymptotic structure.
In the asymptotically $AdS$-case, it is simply zero. In the asymptotically flat case,
the boundary term at infinity is given by replacing, roughly speaking, $\delta \sigma_{ab}$
with the perturbed news tensor\footnote{
It has been shown in~\cite{Hollands-Ishibashi} that the decay of $\gamma_{ab}$ in the transverse-trace-free gauge 
is sufficiently strong that the (linearized) Bondi news tensor at ${\mathscr I}^\pm$ is finite. 
}, 
\ben
\delta N_{ab} = \tilde q_a^c \tilde q_b^d f^{-\frac{d-4}{2}} \delta \left(\frac{2}{d-2} \tilde R_{cd} - \frac{1}{(d-1)(d-2)} \tilde g_{cd} \tilde R  \right) - 
\frac{1}{d-2}\tilde q_{ab}({\rm trace}) , 
\een
where $\tilde q^a_b$ is the projector onto a cross section $\eC$ of ${\mathscr I}$ defined using the 
conformal metric $\tilde g_{ab} = f^2 g_{ab}$, and where ``trace'' denotes the trace with respect to $\tilde q^{ab}$ of the first term. 
It is understood in the formula that the conformal factor $f$ has been chosen 
such that $\tilde n^a = \tilde \nabla^a f = (\partial/\partial t)^a$ is an affinely parameterized null field tangent to $\mathscr I$. 
$\tilde R_{ab}$ is the Ricci tensor of this conformal metric and $\tilde R$ 
the Ricci scalar. Letting $\tilde \gamma_{ab} = f^{-(d-6)/2} \gamma_{ab}$ -- which is shown to be smooth at $\mathscr I$ -- we set
\ben\label{Cgammadef}
C(\eC, \gamma) = -\frac{1}{32\pi} \int_{\eC} \tilde \gamma^{ab} \delta \tilde N_{ab} \ . 
\een
The volume element $vol_{\eC}$ understood under the integral is defined by 
contracting $\tilde n^a$ into the first entry of $vol_{\mathscr I}$,  where $vol_{\mathscr I}$ is defined in turn implicitly by
$vol_{\mathscr I} \wedge \D f = vol_{\tilde g}$. Indices in the formula have been raised 
with $\tilde g^{ab}$. For details see~\cite{HW}. With these notions in hand, we can make the following

\begin{definition}
The canonical energy of a
perturbation is defined as the quadratic form
\ben\label{Edef}
\E(\Sigma, \gamma) \equiv W(\Sigma; \gamma, \pounds_K \gamma) - B(\eB, \gamma) - C(\eC, \gamma) \ .
\een
\end{definition}

The boundary terms are added in the definition of $\E$ in order for $\E$ to have a
very important {\em monotonicity property} under `time evolution'. This property comes about as follows. Since the symplectic current is conserved, it follows that $\D (\star w) = 0$.
We can integrate this equation over a `quadrangle-shaped' domain of $\M$ as shown in fig.~\ref{DOC1}.
By Stokes' theorem, the result is a contribution from the boundaries. The contributions
from $\Sigma_1$ respectively $\Sigma_2$ give $W(\Sigma_1, \gamma, \pounds_K \gamma)$ respectively
$-W(\Sigma_2, \gamma, \pounds_K \gamma)$, whereas the contributions from $\eH_{12}$
respectively $\eI_{12}$ represent `fluxes'. One can compute these fluxes using the
consequences of the linearized Raychaudhuri equation on $\eH$, and the asymptotic
expansion of the metric and perturbation near $\eI$.
Combining these with the boundary terms in the definition of $\E$,
one reaches the following important conclusion:

\begin{lemma}\label{fluxlemma}
Let $\gamma$ be a perturbation having smooth compactly supported initial data on
$\Sigma_1$ (i.e. with support intersecting neither $\eH$ nor $\eI$). Let $\Sigma_2
\subset J^+(\Sigma_1)$, as in figs.~\ref{DOC1},~\ref{DOC2}.
\begin{enumerate}
 \item In the asymptotically flat case, assume that the perturbation is axi-symmetric in the sense
that
\ben\label{axisymmetry}
\pounds_\psi \gamma_{ab} = 0 \ , \quad
\psi = \Omega^I \frac{\partial}{\partial \phi^I} \ .
\een
Then it follows that\footnote{Here natural integration elements on $\eH$ and $\eI$ are understood, see the remarks below 
\eqref{Bgammadef},\eqref{Cgammadef}.} 
\ben
\E(\Sigma_1) - \E(\Sigma_2) = \frac{1}{4\pi} \int_{\eH_{12}} \delta \sigma_{ab} \delta \sigma^{ab}  + \frac{1}{16\pi} \int_{\eI_{12}} \delta \tilde N_{ab} \delta \tilde N^{ab}  \ge 0 \ , 
\een
meaning that
$\E(\Sigma_2) \le \E(\Sigma_1)$.

\item In the asymptotically $AdS$-case, we have $\E(\Sigma_2) \le \E(\Sigma_1)$
also for non-axi-symmetric perturbations.
\end{enumerate}
\end{lemma}

A proof  is given for the case of asymptotically flat non-extremal black holes in thm.~1 of~\cite{HW}, and 
the proof for extremal black holes generalizes straightforwardly (modulo the change regarding how to show $\delta \vartheta = 0$ mentioned earlier). 
The axi-symmetry restriction in the asymptotically flat case is imposed, as in~\cite{HW}, to eliminate any
indefinite flux terms at infinity, corresponding physically to the radiation of angular momentum. The same proof also
works for the asymptotically $AdS$-case, illustrated in fig.~\ref{DOC2}. The key difference in the $AdS$-case is due to the fact that there simply is
no flux at infinity, due to the ``reflecting nature'' of the $AdS$ boundary conditions,~\cite{Marolf}. Therefore, no ``axi-symmetry'' restriction needs to be imposed
in the asymptotically $AdS$-case.

A key property of $\E$ is its gauge invariance under $\gamma_{ab} \to \gamma_{ab} + \pounds_X g_{ab}$, see lemma~2 of~\cite{HW}. Although that 
lemma was formulated for stationary, non-extremal black holes, inspection of the proof shows that the lemma 
also applies to a slice $\Sigma$ ``going down the throat'' in an extremal black 
hole such as $\Sigma = \Sigma_1$ in figs.~\ref{DOC1},~\ref{DOC2}, if the perturbation $\gamma_{ab}$ has compact support on $\Sigma$, as will be the case in 
our applications. Gauge invariance also holds for a slice $\Sigma = \Sigma_2$ as drawn in figs.~\ref{DOC1},~\ref{DOC2} intersecting the 
future horizon if $X^a$ becomes tangent to the generators on $\eH$ and approaches a BMS-transformation at ${\mathscr I}^\pm$. This follows 
again by inspecting the proof of lemma~2 of~\cite{HW}, noting that the perturbed area and expansion of $\eH$ must vanish in our case. By arguing as in 
prop.~4 of~\cite{HW}, it then follows also that $\E(\gamma, \Sigma)$ is a perturbation towards another stationary black hole (where $\Sigma=\Sigma_1$ or 
$=\Sigma_2$). 

With these properties of $\E$ at hand, we may now explain how one can use $\E$ to obtain a sufficient condition for 
the linearized instability of a black hole spacetime, for details see \cite{HW}.  
Suppose that, on a slice $\Sigma = \Sigma_1$ as in figs.~\ref{DOC1},~\ref{DOC2} (for asymptotically flat respectively $AdS$ black holes), we can find a compactly supported 
perturbation -- axi-symmetric in the asymptotically flat case -- for which $\E(\Sigma_1,\gamma) <0$. By lemma~\eqref{fluxlemma}, $\E(\Sigma_2,\gamma)$ must 
be less than or equal to $\E(\Sigma_1, \gamma)$ for any later slice as drawn in figs.~\ref{DOC1} respectively~\ref{DOC2}. On the other hand, if $\gamma_{ab}$ is to approach 
a pure gauge transformation (compatible with our gauge conditions on $\eH$ and $\eI$), then $\E(\Sigma_2, \gamma)$ must go to zero on a 
sufficiently late slice. This cannot be the case, and so $\gamma_{ab}$ cannot go 
to a pure gauge transformation at late times. Likewise, $\gamma_{ab}$ cannot converge to a perturbation to another stationary black hole. Thus, in this sense, the black holes is 
linearly unstable.  

Below, it is useful to work also with a formulation of $\E$ in terms of the initial data
of the background and perturbations. Let $\Sigma$ be a spatial slice, with unit
normal $\nu^a$. We denote the induced metric by $h_{ij}$ and an extrinsic curvature by $\chi_{ij}$. Recall that the
canonical momentum $p^{ij}$ is defined in terms of the extrinsic curvature of $\Sigma$ as
\ben
p^{ij} = h^\half (\chi^{ij} - \chi^k{}_k h^{ij}) \ .
\een
Here and in the following we introduce a fixed (e.g. coordinate-) $(d-1)$-form field
$\D^{d-1} x$ on $\M$ related to the volume form on $\Sigma$ by $\D vol_h=
h^\half \D^{d-1} x$. With these definitions,
$h_{ij}$ and $p^{ij}$ are canonically conjugate pairs. The lapse and shift of $K^a$
are denoted by $N$ respectively $N^j$.
The initial data of a perturbation $\gamma_{ab}$ are written $(\delta h_{ij}, \delta p^{ij})$ and are, 
throughout this article, assumed to be of compact support on $\Sigma$. 
In terms of these, we have $\E = (1/16\pi)  \int_\Sigma \rho \D vol_h$, where
$\rho$ is given by:
\bena
\rho&=& N \ [ \thalf  Ric(h)_{ij} \delta h_k{}^k \delta h^{ij}
- 2  Ric(h)_{ik} \delta h^{ij} \delta h_j{}^k + \thalf  (D_i \delta h^{ik})  D_k
\delta h_l{}^l - \non\\
&& \hspace{1.2cm} \thalf \ (D^j \delta h^{ik}) D_j \delta h_{ik}
 - (D^j \delta h^{ik}) D_k \delta h_{jk} ]  +\non\\
&& N \ [ 2  \delta p_{ij} \delta p^{ij} +
\thalf  p_{ij} p^{ij} (\delta h_k{}^k)^2 -  p_{ij} \delta p^{ij} \delta h_k{}^k -
3  p^i{}_j p^{jk} \delta h_l{}^l \delta h_{ik} -
\non\\
&& \hspace{1.2cm} \tfrac{2}{d-2}  (\delta p_i{}^i)^2 +  \tfrac{3}{d-2}   p_i{}^i \delta p_j{}^j \delta h_k{}^k + \tfrac{3}{d-2}   p_k{}^k p^{ij} \delta h \delta h_{ij} +
8 \ p^k{}_j \delta h_{ik} \delta p^{ij} + \non\\
&& \hspace{1.2cm} p_{kl} p^{kl} \delta h_{ij} \delta h^{ij} +
2  p^{ij} p^{kl} \delta h_{ik} \delta h_{jl} -\tfrac{1}{d-2}   (p_k{}^k)^2 \delta h_{ij} \delta h^{ij}
- \tfrac{1}{2(d-2)}  (p_i{}^i)^2 (\delta h_j{}^j)^2 -\non\\
&& \hspace{1.2cm} \tfrac{4}{d-2}  \ p_j{}^j \delta p^{ik} \delta h_{ik} -
\tfrac{2}{d-2}  (p^{ij} \delta h_{ij})^2 - \tfrac{4}{d-2}  p_{ij} \delta p_k{}^k \delta h^{ij} ] h^{-1}  - \non\\
&& N^i [ -2  \delta p^{jk} D_i \delta h_{jk} + 4  \delta p^{jk} D_j \delta h_{ik} +2  \delta h_{ik} D_j \delta p^{jk} - \non\\
&&\hspace{1.2cm} 2  p^{jk} \delta h_{il}D_j \delta h_k{}^l + p^{jk} \delta h_{il} D^l \delta h_{jk}  ) ]  h^{-\half} \  \ ,
\label{Eexpr}
\eena
see~\cite{HW} for a derivation\footnote{Note that the boundary terms in $\E$ in~\cite{HW} can be omitted for perturbations 
having compact support on $\Sigma$.}.

\subsection{Canonical energy of electromagnetic perturbations}

One may also study a test electromagnetic field, $A_a$, propagating on the background black hole
spacetime $(\M, g_{ab})$. We will call these ``electromagnetic perturbations''. The field equation is the Maxwell
equation $0 = \nabla^a \nabla_{[a} A_{b]}$, and the field strength is as usual $F_{ab} = 2\nabla_{[a} A_{b]}$. The symplectic $(d-1)$-form for two perturbations $A_1, A_2$
can be derived from the Lagrangian formulation as described in~\cite{GaoWald}, with the result
\begin{equation}
  \label{omegadef1}
w_a = \frac{1}{2\pi}( A_1^b \nabla_{[a} A_{2 \, b]} - A_2^b \nabla_{[a} A_{1 \, b]} )\ .
\end{equation}
As always, $\nabla^a w_a = 0$.
The symplectic form $W(\Sigma; A_1, A_2)$ is obtained, just as in the gravitational case,
by integrating $\star w$ over a $(d-1)$-dimensional submanifold $\Sigma$,
\ben
W(\Sigma; A_1, A_2) \equiv \int_\Sigma \star w(g; A_1, A_2) \, .
\label{symform1}
\een
As in the gravitational case, we impose gauge conditions on $A_a$ near infinity and near the horizon. 
Near the horizon, our gauge condition on $A_a$ analogous to~\eqref{horgauge} is that perturbed electrostatic potiential vanishes, 
$-K^a A_a |_{\eH} = 0$. Similar to the case of gravitational perturbations, if this condition 
is satisfied on one cross section $\eB$ of $\eH^+$, then it is automatically satisfied everywhere on $\eH^+$~\cite{GaoWald}.  
Near infinity, we impose the Lorentz gauge condition $\nabla^a A_a = 0$. 
It is shown in appendix~\ref{app:E} that this condition implies the following behavior of $A_a$ near infinity 
in the asymptotically flat case ($\Lambda = 0$): In terms of the unphysical metric $\tilde g_{ab} = f^2 g_{ab}$, 
we have that $\tilde A_a = f^{(d-4)/2} A_a$ and $f^{-1} \tilde n^a \tilde A_a$ are finite and smooth 
at $\mathscr I$, where $\tilde n^b = \tilde g^{ab} \tilde \nabla_a f$. 

In order to define the canonical energy for an electromagnetic perturbation
associated with such a slice, we must, as in the gravitational case, introduce
certain boundary terms. The boundary term on the horizon is
\ben\label{Bdef1}
B(\eB, A) = \frac{1}{2\pi} \int_{\eB} A^a \pounds_K A_a  \ , 
\een
whereas the boundary term at infinity is
\ben\label{Cdef1}
C(\eC, A) = \frac{1}{2\pi} \int_{\eC} \tilde A^a \pounds_{\tilde n} \tilde A_a  \ , 
\een
where, as in the gravitational case, natural integration elements are understood and indices on tilded quantities are raised with $\tilde g_{ab}$. 
For asymptotically $AdS$-spacetimes ($\Lambda<0$), 
the boundary term from infinity is again simply set to zero. 
The canonical energy in the electromagnetic case is then defined in precise analogy to the 
gravitational case:
\begin{definition}
The canonical energy of an electromagnetic
perturbation is defined as the quadratic form
\ben\label{Edef1}
\E(\Sigma, A) \equiv W(\Sigma; A, \pounds_K A) - B(\eB, A) - C(\eC, A) \ .
\een
\end{definition}
Proceeding as in the case of gravitational perturbation, one can derive a monotonicity property analogous to that  described in
lemma~\eqref{fluxlemma}. In the case $\Lambda = 0$, we assume, as in the case of gravitational perturbations, that $A_a$ is axi-symmetric, 
$\pounds_\psi A_a = 0$, compare eq.~\eqref{axisymmetry}. One obtains the balance equation ($n^a = K^a$ on $\eH$)
\ben\label{flux1}
\E(\Sigma_1) - \E(\Sigma_2) = \frac{1}{2\pi} \int_{\eH_{12}} (\pounds_n A^a) \pounds_n A_a  + \frac{1}{2\pi} \int_{\eI_{12}} (\pounds_{\tilde n} \tilde A^a) 
\pounds_{\tilde n} \tilde A_a \ge 0 \ , 
\een
meaning that
$\E(\Sigma_2) \le \E(\Sigma_1)$ as in the gravitational case. For $\Lambda<0$, one obtains the same balance equation without the second term on the right side
even for non-axi-symmetric perturbations. More details on how to derive~\eqref{flux1} are given in~\cite{Keir} and appendix~\ref{app:E}. 

Again, one can also write $\E$ in terms of the initial data
of the perturbations. In the case of electromagnetic perturbations these are given by $(A_i, E^i)$, where 
$E^i$ is the densitized\footnote{We choose $E^i$ to be a density so that it is canonically conjugate to $A_i$, but 
of course we could also work with the undensitized electric field.} electric field. 
In terms of these, we have $\E = (1/4\pi) \int_\Sigma \rho \D vol_h$, where
$\rho$ is given by:
\ben
\rho = N( \tfrac{1}{2} h^{-1} E_i E^i + \tfrac{1}{4} F_{ij} F^{ij} ) + N^i E^j F_{ij} h^{-\half} \ . 
\een

\section{Extremal black holes and their near horizon limit} \label{sec2}

A key role is played in conjecture~1 by the notion of near horizon (NH) limit of an extremal black hole. We now 
recall this construction, and establish some notation used in the subsequent sections. A good review of 
NH geometries is~\cite{Lucietti4}. In an open neighborhood of the horizon $\eH^+$ we may
introduce Gaussian normal coordinates as follows. We pick a section $\eB \subset \eH^+$ and choose arbitrary local coordinates $x^A, A=1, \dots, d-2$
on $\eB$. We then complement these by $\rho,u$, where the coordinate $u$ by definition parameterizes affine null geodesics ruling $\eH^+$,
whereas $\rho$ parameterizes null geodesics transversal to $\eH^+$ and orthogonal to $\eB$. It can be shown~\cite{HIW} that the metric takes the form
\ben\label{gnc}
\D s^2 = 2 \D u( \D \rho - \tfrac{1}{2} \rho^2 \alpha \, \D u - \rho \beta_A \, \D x^A) + \mu_{AB} \, \D x^A \D x^B \ 
\een
in these coordinates. The tensor fields $\alpha, \beta_A \D x^A , \mu_{AB} \D x^A \D x^B$ are
defined independently of our arbitrary choice of the coordinates $x^A$.
The Gaussian null coordinates may further be chosen, by adjusting $\eB$ if necessary, so that
the Killing field is $K = \partial/\partial u$.
A key role is played  in this paper by the 1-parameter group of diffoemorphisms defined in
a neighborhood of $\eH^+$ by
\ben\label{phieps}
\phi_\eps: (u,\rho,x^A) \mapsto \left( \eps u, \frac{1}{\eps} \rho, x^A \right) \ .
\een
The form for the metric~\eqref{gnc} and the fact that $K=\partial/\partial u$ is a Killing field imply that the limit
\ben\label{nhgdef}
g^{\rm NH} = \lim_{\eps \to 0} \phi_{1/\eps}^* \, g
\een
defines a new metric solving the Einstein equations. It is called ``NH limit''.
In the following, we will omit the superscript ``NH'' to avoid clutter.
The near horizon metric can again
be represented in the form~\eqref{gnc}. In these coordinates, $\beta_A, \mu_{AB}$
are independent of the coordinates $u,\rho$ and are obtained from their
counterparts in the original BH metric simply by setting $\rho=0$. The diffeomorphisms $\phi_\eps$ by construction form
a 1-parameter group of isometries of the NH geometry, which together with the group generated by
$K$ forms generates an action of the 2-dimensional group $\RR \ltimes \RR_+$.

These general constructions can be applied, in particular, to the known extremal, vacuum stationary black holes, 
i.e. the Myers-Perry (MP) black holes and black rings~\cite{Lucietti2}. The former are known in any dimension $d \ge 5$, 
whereas the latter only in $d=5$.  For definiteness, we will focus on the MP black holes. We recall 
the results in the case $\Lambda = 0$ following Ref.~\cite{Lucietti2}, and refer to~\cite{Lu2} for the case $\Lambda \neq 0$. 
First we describe the MP black holes themselves. These solutions are parameterized by a mass parameter
$M >0$ and rotation parameters  $a_I \in \RR$ where the index $I$ runs between $1,\dots,n$. Their properties differ somewhat in
even and odd dimension, so for definiteness and simplicity we focus on the odd dimensional case, where $d=2n+1$.
The horizon topology is $\eB \cong S^{2n-1}$, and the topology of a Cauchy surface for the exterior region is
$\RR^{d-1}$ minus a ball. The exterior region is parameterized by coordinates $t, r>r_+$, $n$ azimuthal
coordinates $\phi^I \in [0,2\pi]$ and $n$ latitudinal coordinates $\mu_I \in [0,1]$ subject to $\sum \mu_I^2 = 1$. In
terms of these, the MP metric is
\ben\label{MP}
\begin{split}
g =& -\D t^2 + \frac{M r}{\Pi F} \left( \D t + \sum_{I=1}^n a_I \mu_I^2 \D \phi_I \right)^2
+ \frac{\Pi F}{\Pi - M r^2} \D r^2 \\
&+ \sum_{I=1}^n (r^2 + a_I^2) (\D \mu_I^2 + \mu_I^2 \D \phi_I^2) \ .
\end{split}
\een
Here,
\ben
\Pi = \prod_{I=1}^n (r^2 + a_I^2) \ , \quad F = 1 - \sum_{I=1}^n \frac{a_I^2 \mu_I^2}{r^2 + a_I^2} \ .
\een
The location of the event horizon $\eB$  is at the value $r=r_+>0$ defined by $\Pi(r_+) = M r_+^2$, and the angular velocities of the horizon are given by
\ben\label{Omega}
\Omega^I  = -\frac{a_I}{r_+^2 + a_I^2} \ .
\een
The isometry group of the MP black holes is $\RR \times {\rm U}(1)^n$, corresponding to
shifts in $t$ respectively in $\phi^I$.

 As in the case of the Kerr metric,
there are extremal limits. In odd $d =2n+1 \ge 5$ these are characterized by the condition\footnote{Note that this requires that all $a_I \neq 0$.}
\ben
1 = \sum_{I=1}^n \frac{r_+^2}{a_I^2 + r_+^2} \ .
\een
A Penrose diagram for the extreme MP spacetime is given in figure~\ref{MP1}.

\begin{figure}
\begin{center}
\begin{tikzpicture}[scale=.8, transform shape]
\draw (-2,6) decorate[decoration=snake] {--(-2,-6)};
\draw (-2,6) ..controls (-0.2,4) and (-0.2,4) .. (-2,2);
\filldraw[fill=gray,opacity=.5,draw=black] (-2,6) -- (0,4) -- (-2,2) ..controls (-0.2,4) and (-0.2,4) .. (-2,6);
\filldraw[fill=gray,opacity=.5,draw=black] (-2,2) -- (0,0) -- (-2,-2) ..controls (-0.2,0) and (-0.2,0) .. (-2,2);
\filldraw[fill=gray,opacity=.5,draw=black] (-2,-6) -- (0,-4) -- (-2,-2) ..controls (-0.2,-4) and (-0.2,-4) .. (-2,-6);
\filldraw[fill=gray,opacity=.5,draw=black] (0,4) -- (-2,2) -- (0,0) ..controls (-1.8,2) and (-1.8,2) .. (0,4);
\filldraw[fill=gray,opacity=.5,draw=black] (0,-4) -- (-2,-2) -- (0,0) ..controls (-1.8,-2) and (-1.8,-2) .. (0,-4);
\filldraw[fill=gray,opacity=.5,draw=black] (-2,6) -- (-1.4,6) ..controls (-1.4,5.8) and (-1.4,5.8) .. (0,4);
\filldraw[fill=gray,opacity=.5,draw=black] (-2,-6) -- (-1.4,-6) ..controls (-1.4,-5.8) and (-1.4,-5.8) .. (0,-4);
\draw (-2,6) -- (0,4) -- (-2,2) -- (0,0) -- (-2,-2) -- (0,-4) -- (-2,-6);
\draw[double] (2,6) -- (0,4) -- (2,2) -- (0,0) -- (2,-2) -- (0,-4) -- (2,-6);
\draw (-2,6) node[draw,shape=circle,scale=0.3,fill=black]{};
\draw (2,6) node[draw,shape=circle,scale=0.3,fill=black]{};
\draw (-2,2) node[draw,shape=circle,scale=0.3,fill=black]{};
\draw (2,2) node[draw,shape=circle,scale=0.3,fill=black]{};
\draw (-2,-2) node[draw,shape=circle,scale=0.3,fill=black]{};
\draw (2,-2) node[draw,shape=circle,scale=0.3,fill=black]{};
\draw (-2,-6) node[draw,shape=circle,scale=0.3,fill=black]{};
\draw (2,-6) node[draw,shape=circle,scale=0.3,fill=black]{};
\draw (0,4) node[draw,shape=circle,scale=0.3,fill=black]{};
\draw (0,0) node[draw,shape=circle,scale=0.3,fill=black]{};
\draw (0,-4) node[draw,shape=circle,scale=0.3,fill=black]{};
\node at (-1.4,4.6) {$\mathscr{H}^-$};
\node at (-1.4,3.4) {$\mathscr{H}^+$};
\node at (-1.4,0.6) {$\mathscr{H}^-$};
\node at (-1.4,-3.4) {$\mathscr{H}^+$};
\node at (-1.4,-0.6) {$\mathscr{H}^-$};
\node at (-1.4,-4.6) {$\mathscr{H}^+$};
\node at (1.4,4.6) {$\mathscr{I}^-$};
\node at (1.4,3.4) {$\mathscr{I}^+$};
\node at (1.4,0.6) {$\mathscr{I}^-$};
\node at (1.4,-3.4) {$\mathscr{I}^+$};
\node at (1.4,-0.6) {$\mathscr{I}^-$};
\node at (1.4,-4.6) {$\mathscr{I}^+$};
\node at (-2.8,2) {$\mathscr{B}$};
\draw[red,thick] (-2,2) -- (2,2);
\draw (0,0) ..controls (-1.4,2) and (-1.4,2) .. (0,4);
\draw (0,0) ..controls (-1.0,2) and (-1.0,2) .. (0,4);
\draw (0,0) ..controls (-0.5,2) and (-.5,2) .. (0,4);
\draw (0,0) ..controls (0,2) and (0,2) .. (0,4);
\draw (0,0) ..controls (0.5,2) and (.5,2) .. (0,4);
\draw (0,0) ..controls (1.0,2) and (1,2) .. (0,4);
\draw (0,0) ..controls (1.4,2) and (1.4,2) .. (0,4);
\draw (0,0) ..controls (1.8,2) and (1.8,2) .. (0,4);
\draw (-2,2) ..controls (0,2.5) and (0,2.5) .. (2,2);
\draw (-2,2) ..controls (0,1.5) and (0,1.5) .. (2,2);
\draw (-2,2) ..controls (0,3) and (0,3) .. (2,2);
\draw (-2,2) ..controls (0,1) and (0,1) .. (2,2);
\draw (-2,2) ..controls (0,3.4) and (0,3.4) .. (2,2);
\draw (-2,2) ..controls (0,.6) and (0,.6) .. (2,2);
\draw (-2,2) ..controls (0,3.8) and (0,3.8) .. (2,2);
\draw (-2,2) ..controls (0,.2) and (0,.2) .. (2,2);
\node at (0,1.6) {\textcolor{red}{$\Sigma$}};
\end{tikzpicture}
\end{center}
\caption{
\label{MP1}
Conformal diagram of the extremal $\Lambda=0$ MP spacetime~\cite{Myers-Perry}. The Cauchy surface for the exterior is a complete manifold, i.e. the 
proper distance of $\eB$ from any point on $\Sigma$ is infinite. The near horizon region is shaded, and the upward curvy lines are 
the orbits of the Killing field $K$.  
}
\end{figure}
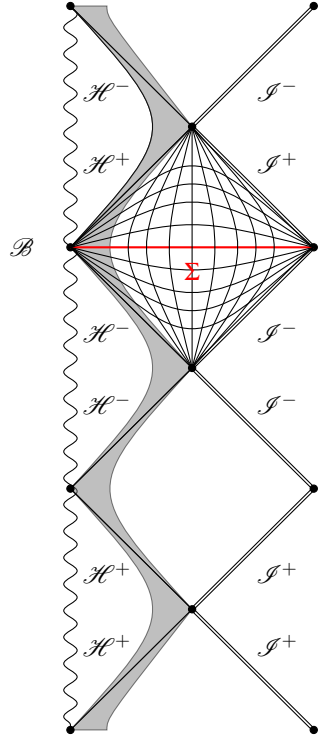

\begin{figure}
\begin{center}
\begin{tikzpicture}[scale=.8, transform shape]
\draw (-1,6) -- (-1,-6);
\draw (-1.05,6) -- (-1.05,-6);
\draw (1,6) -- (1,-6);
\draw (1.05,6) -- (1.05,-6);
\fill[fill=gray,opacity=.3] (-1,6) -- (-1,-6) -- (1,-6) -- (1,6);
\fill[fill=gray,opacity=.6] (-1,2) -- (1,0) -- (1,4) -- (-1,2);
\draw (-1,6) -- (1,4) -- (-1,2) -- (1,0) -- (-1,-2) -- (1,-4) -- (-1,-6);
\draw (-1,6) .. controls (0.5,4) and (0.5,4) .. (-1,2);
\draw (-1,6) .. controls (0,4) and (0,4) .. (-1,2);
\draw (-1,6) .. controls (-0.5,4) and (-0.5,4) .. (-1,2);
\draw (-1,6) .. controls (1,4) and (1,4) .. (-1,2);
\draw (1,4) .. controls (-1,2) and (-1,2) .. (1,0);
\draw (1,4) .. controls (0.5,2) and (0.5,2) .. (1,0);
\draw (1,4) .. controls (0,2) and (0,2) .. (1,0);
\draw (1,4) .. controls (-0.5,2) and (-0.5,2) .. (1,0);
\draw (-1,2) .. controls (1,0) and (1,0) .. (-1,-2);
\draw (-1,2) .. controls (0.5,0) and (0.5,0) .. (-1,-2);
\draw (-1,2) .. controls (0,0) and (0,0) .. (-1,-2);
\draw (-1,2) .. controls (-0.5,0) and (-0.5,0) .. (-1,-2);
\draw (1,0) .. controls (-1,-2) and (-1,-2) .. (1,-4);
\draw (1,0) .. controls (-0.5,-2) and (-0.5,-2) .. (1,-4);
\draw (1,0) .. controls (0,-2) and (0,-2) .. (1,-4);
\draw (1,0) .. controls (0.5,-2) and (0.5,-2) .. (1,-4);
\draw (-1,-2) .. controls (1,-4) and (1,-4) .. (-1,-6);
\draw (-1,-2) .. controls (0.5,-4) and (0.5,-4) .. (-1,-6);
\draw (-1,-2) .. controls (0,-4) and (0,-4) .. (-1,-6);
\draw (-1,-2) .. controls (-0.5,-4) and (-0.5,-4) .. (-1,-6);
\draw (1,-4) .. controls (-0.65,-5.8) and (-0.65,-5.8) .. (-0.65,-6);
\draw (1,-4) .. controls (-0.25,-5.8) and (-0.25,-5.8) .. (-0.25,-6);
\draw (1,-4) .. controls (0.2,-5.8) and (0.2,-5.8) .. (0.2,-6);
\draw (1,-4) .. controls (0.6,-5.8) and (0.6,-5.8) .. (0.6,-6);
\draw (1,4) .. controls (-0.65,5.8) and (-0.65,5.8) .. (-0.65,6);
\draw (1,4) .. controls (-0.25,5.8) and (-0.25,5.8) .. (-0.25,6);
\draw (1,4) .. controls (0.2,5.8) and (0.2,5.8) .. (0.2,6);
\draw (1,4) .. controls (0.6,5.8) and (0.6,5.8) .. (0.6,6);
\draw (1,4) .. controls (-0.3,5.0) and (-0.3,5.0) .. (-1,5.3);
\draw (1,4) .. controls (-0.3,3.0) and (-0.3,3.0) .. (-1,2.7);
\draw (1,4) .. controls (-0.3,4.6) and (-0.3,4.6) .. (-1,4.7);
\draw (1,4) .. controls (-0.3,3.4) and (-0.3,3.4) .. (-1,3.3);
\draw (1,4) .. controls (-0.3,4.2) and (-0.3,4.2) .. (-1,4.2);
\draw (1,4) .. controls (-0.3,3.8) and (-0.3,3.8) .. (-1,3.8);
\draw (1,-4) .. controls (-0.3,-5.0) and (-0.3,-5.0) .. (-1,-5.3);
\draw (1,-4) .. controls (-0.3,-3.0) and (-0.3,-3.0) .. (-1,-2.7);
\draw (1,-4) .. controls (-0.3,-4.6) and (-0.3,-4.6) .. (-1,-4.7);
\draw (1,-4) .. controls (-0.3,-3.4) and (-0.3,-3.4) .. (-1,-3.3);
\draw (1,-4) .. controls (-0.3,-4.2) and (-0.3,-4.2) .. (-1,-4.2);
\draw (1,-4) .. controls (-0.3,-3.8) and (-0.3,-3.8) .. (-1,-3.8);
\draw (1,0) .. controls (-0.3,1.0) and (-0.3,1.0) .. (-1,1.3);
\draw (1,0) .. controls (-0.3,-1.0) and (-0.3,-1.0) .. (-1,-1.3);
\draw (1,0) .. controls (-0.3,0.6) and (-0.3,0.6) .. (-1,0.7);
\draw (1,0) .. controls (-0.3,-0.6) and (-0.3,-0.6) .. (-1,-0.7);
\draw (1,0) .. controls (-0.3,0.2) and (-0.3,0.2) .. (-1,0.2);
\draw (1,0) .. controls (-0.3,-0.2) and (-0.3,-0.2) .. (-1,-0.2);
\draw (-1,2) .. controls (0.3,3.0) and (0.3,3.0) .. (1,3.3);
\draw (-1,2) .. controls (0.3,1.0) and (0.3,1.0) .. (1,0.7);
\draw (-1,2) .. controls (0.3,2.6) and (0.3,2.6) .. (1,2.7);
\draw (-1,2) .. controls (0.3,1.4) and (0.3,1.4) .. (1,1.3);
\draw (-1,2) .. controls (0.3,2.2) and (0.3,2.2) .. (1,2.2);
\draw (-1,2) .. controls (0.3,1.8) and (0.3,1.8) .. (1,1.8);
\draw (-1,-2) .. controls (0.3,-3.0) and (0.3,-3.0) .. (1,-3.3);
\draw (-1,-2) .. controls (0.3,-1.0) and (0.3,-1.0) .. (1,-0.7);
\draw (-1,-2) .. controls (0.3,-2.6) and (0.3,-2.6) .. (1,-2.7);
\draw (-1,-2) .. controls (0.3,-1.4) and (0.3,-1.4) .. (1,-1.3);
\draw (-1,-2) .. controls (0.3,-2.2) and (0.3,-2.2) .. (1,-2.2);
\draw (-1,-2) .. controls (0.3,-1.8) and (0.3,-1.8) .. (1,-1.8);
\draw (-1,-6) .. controls (0.3,-5.0) and (0.3,-5.0) .. (1,-4.7);
\draw (-1,-6) .. controls (0.3,-5.4) and (0.3,-5.4) .. (1,-5.3);
\draw (-1,-6) .. controls (0.3,-5.8) and (0.3,-5.8) .. (1,-5.8);
\draw (-1,6) .. controls (0.3,5.0) and (0.3,5.0) .. (1,4.7);
\draw (-1,6) .. controls (0.3,5.4) and (0.3,5.4) .. (1,5.3);
\draw (-1,6) .. controls (0.3,5.8) and (0.3,5.8) .. (1,5.8);
\draw (-1,6) node[draw,shape=circle,scale=0.5,fill=black]{};
\draw (-1,2) node[draw,shape=circle,scale=0.5,fill=black]{};
\draw (-1,-2) node[draw,shape=circle,scale=0.5,fill=black]{};
\draw (-1,-6) node[draw,shape=circle,scale=0.5,fill=black]{};
\draw (1,4) node[draw,shape=circle,scale=0.5,fill=black]{};
\draw (1,0) node[draw,shape=circle,scale=0.5,fill=black]{};
\draw (1,-4) node[draw,shape=circle,scale=0.5,fill=black]{};
\draw[red,thick] (-1,2) -- (1,2);
\node at (1.5,2) {\textcolor{red}{$\Sigma$}};
\node at (-1.5,2) {$\mathscr{B}$};
\end{tikzpicture}
\end{center}
\caption{
\label{MP2}
Conformal diagram of the NH limit of the extremal MP spacetime~\cite{Myers-Perry}, i.e. $AdS_2$. This should be thought of as corresponding to the shaded region in the 
diagrams~\ref{MP1} or~\ref{MP0} of the extreme MP black hole, to be taken ``infinitely thin''. The Cauchy surface $\Sigma$ in that conformal diagram corresponds to the 
surface $\Sigma$ drawn here. The curvy upward lines show the orbits of $K=\partial/\partial T$, whereas the curvy horizontal lines the surfaces of constant $T$.
}
\end{figure}
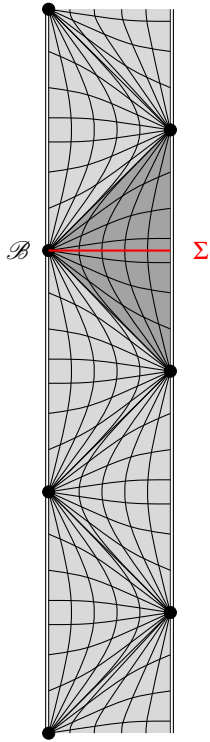

The NH-limit for the extremal MP solutions has been computed in~\cite{Lucietti2}. It has the general form
\ben\label{nhg}
\D s^2 = L^2 \ \D \hat s^2 + g_{IJ} (\D \phi^I + k^I \ \hat A)(\D \phi^J + k^J \ \hat A)
+ \D \sigma^2_{d-n-2} \ .
\een
The geometry can be thought of as a
fibration $\eB \to \M \to \hat \M$ with leaves $\eB \cong S^{2n-1}$, total space $\M$, and orbit space $\hat \M$. The quantities $L>0, \D \sigma^2_{d-n-2}$
are intrinsic to $\eB$, whereas hatted quantities refer to the base space $\hat \M$. The base space is geometrically $\hat \M = AdS_2$ with a uniform
electric field $\D \hat A = vol_{\hat g}$:
\ben\label{ads2}
\D \hat s^2 = -R^2 \D T^2 + \frac{\D R^2}{R^2} \ , \quad \hat A = -R \D T \ .
\een
The coordinates $R,T$ on $AdS_2$ are sometimes called ``Poincare-coordinates'', and cover  
the region shaded in dark grey with the slice $\Sigma = \{T=0\}$ in figure~\ref{MP2}. 
The other quantities appearing in the NH-metric of the extremal MP solutions are explicitly:
\bena\label{comparison}
L^2 &=& F(r_+)/C^2 \ , \\
k^I  &=& \frac{2r_+ a_I}{C^2(r_+^2 + a_I^2)^2} \ , \\
\D \sigma_{d-n-2}^2 &=& \sum_{I=1}^n (r_+^2 + a_I^2) \D \mu_I^2  \ , \\
g_{IJ} &=& (r_+^2 + a_I^2) \mu_I^2 \delta_{IJ} + \frac{a_I a_J \mu_I^2 \mu_J^2}{L^2}  \ ,
\eena
where $C^2 = \Pi''(r_+)/2\Pi(r_+) > 0$.
If the coordinates on $\eB = S^{2n-1}$ are denoted collectively by
$(x^A) = (\phi^I, \mu_J)$, the relationship to the
Gaussian null form~\eqref{gnc} is:
\bena\label{coordinates1}
\rho &=& L^2 \cdot R \ , \quad u = T -1/R\\
\beta_{A} \D x^A  &=& \frac{1}{L^2} \bigg( g_{IJ} k^I \D \phi^J + \D L^2 \bigg) \\
\mu_{AB} \D x^A \D x^B &=& g_{IJ} \D \phi^I \D \phi^J  + \D \sigma^2_{d-n-2} \\
\alpha &=& \frac{1}{L^2} \bigg(1 - \frac{1}{L^2} g_{IJ} k^I k^J \bigg)
\eena
implying in particular that the horizon Killing field in the coordinates~\eqref{nhg} is
$K = \partial/\partial T$. The relationship between the coordinates $(\rho,u,x^A)$
and those used to represent the MP metric~\eqref{MP} can be found in~\cite{Lucietti3}. A Penrose diagram of the NH geometry 
illustrating the relation to the extremal MP spacetime is shown in figure~\ref{MP2}.

The presence of the $AdS_2$ factor is crucial for the considerations of this paper.
It implies for example that the NH geometry has a larger isometry group than what can be inferred 
straightforwardly from the general construction leading to eq.~\eqref{nhgdef}~\cite{Lucietti1}.
This enhanced isometry group is ${\rm SL}(2,\RR) \times {\rm U}(1)^n$, with the ${\rm SL}(2, \RR)$ factor corresponding to $AdS_2$.
The metric $\D \sigma^2_{d-2-n}$ may be thought as that inherited on the orbit space $\M/[{\rm SL}(2,\RR) \times {\rm U}(1)^n]$.

For later purposes, it is useful to know the (asymptotic) forms of the induced metric and extrinsic curvature on the slice
$\Sigma = \{T=0\}$ ``going down the throat'' of the extreme MP metric respectively the NH geometry. These can
straightforwardly be calculated noting that, by~\eqref{coordinates1}, $R$ behaves like $\rho$ for small $R$, and recalling
that the components $\alpha, \beta_A, \mu_{AB}$ of the BH and NH metrics (see \eqref{gnc})
differ by terms of order $\rho$ from the BH metric. With $(x^i) = (y = \log R, x^A)$, and $h=h_{ij} \D x^i \D x^j, \chi = \chi_{ij} \D x^i \D x^j$, one finds
\bena\label{background}
h &=& L^2 \, \D y \otimes \D y + g_{IJ} \D \phi^I \otimes \D \phi^J +
\D \sigma_{d-n-2}^2
 + O(e^{y}) \ , \\
\chi &=& \frac{g_{IJ} k^I}{\sqrt{L^2 + g_{MN} k^M k^N}} \,  (\D \phi^J \otimes \D y + \D y \otimes \D \phi^J)
+ O(e^y) \ .
\eena
To derive the formula for $\chi_{ij}$, one can make use e.g. of the well-known formula (see e.g. appendix~E of~\cite{waldbook})
$\chi_{ij} = (2N)^{-1} [\partial_T h_{ij} - 2 D_{(i} N_{j)}]$ in terms of the lapse $N$ and
shift $N^j$ of $K=\partial/\partial T$, together with $\partial_T h_{ij} = 0$.
The expression $O(e^y)$ represents terms whose coordinate components with respect to $(y,x^A)$, including their $y$-derivatives, decay
as $e^y$ for $y \to -\infty$, i.e. in the throat. The form of $h_{ij}$ shows explicitly that the slice $\Sigma$ is a complete manifold, 
which is a characteristic feature of extremal black holes. 

\section{Hertz potentials}

\subsection{Gravitational perturbations}

The NH limits arising from the known black hole solutions in various dimensions (and
in particular of the Myers-Perry family) have further special properties that allow
one, to a certain extent, to decouple and separate the linearized Einstein equations~\eqref{einstein}.
These properties have to do with the presence of certain null vector fields with special optical properties,
and with the fact that the Weyl-tensor of the NH geometries is algebraically special in a sufficiently strong sense.

The properties are formulated in terms of a distinguished  pair of null vector fields $l^a, n^a$. They 
are normalized so that
\ben
n^a l_a = 1 \ , \quad
n^a n_a = l^a l_a = 0 \ , \quad
g_{ab} = 2 n_{(a} l_{b)} + q_{ab} \ ,
\een
so  $q_{a}{}^b$ projects onto the subspace of $T\M$ orthogonal to $n^a, l^a$. The `algebraically special property' of the
Weyl tensor, $C_{abcd}$ which we referred to is:
\ben\label{alg}
C_{abcd} q^a{}_e q^c{}_f l^b l^d = 0 = C_{abcd} q^a{}_e q^b{}_f q^c{}_h l^d \ , \quad
\text{and same for $n^a \leftrightarrow l^a$.}
\een
The `special optical properties' for $n^a, l^a$ are that they should be geodesic, shear free, expansion free, and twist free; in formulas:
\ben\label{optical}
q_{cb} l^a \nabla_a l^b = 0 \ , \quad
q_{ac} q_{bd} \nabla^c l^d=0 \ , \quad
\text{and same for $n^a \leftrightarrow l^a$.}
\een
In the terminology
of~\cite{Durkee}, spacetimes satisfying~\eqref{alg} and~\eqref{optical} are ``doubly Kundt''. The NH geometries~\eqref{nhg}
studied in this paper all fall into this class, with $l^a, n^a$ concretely given by~\eqref{nldef}. For the considerations of 
this section, the explicit forms~\eqref{nhg} and~\eqref{nldef} are not needed. It is enough to know their general properties. 

These properties can be exploited as follows.
From the background Einstein equations and Bianchi-identity, we always
have
\ben\label{bianchi}
0=\nabla_{[a} C_{bc]de}
\een
and taking $\nabla^a$ of this equation and using again the background Einstein
equations, we get the wave equation
\ben\label{Cwave}
0=\nabla^a \nabla_a C_{bcde} + 2 C^a{}_{bc}{}^f C_{fade} + 2 C^a{}_{[b|d|}{}^f C_{c]afe} + 2 C^a{}_{[b|e|}{}^f C_{c]adf}
-\frac{4\Lambda }{d-1} C_{bcde} \ .
\een
First order perturbed versions of these equations are derived by considering 1-parameter families of background
metrics satisfying the Einstein equations.
By dotting equations~\eqref{bianchi}, \eqref{Cwave} and their perturbed counterparts in all possible
ways into $l^a, n^a, q^a{}_b$, using the optical properties~\eqref{optical} and
the algebraically special properties of the Weyl tensor~\eqref{alg},~\cite{Durkee}
were able to find a decoupled tensorial wave equation for the quantity
\ben
\label{omdef}
\begin{split}
& \Omega_{ab} \equiv \delta C_{cdef} q_a{}^{c} l^d q_b{}^e l^f \\
  =& \bigg\{
            \nabla_d \nabla_{[e} \gamma_{f] c} + \nabla_c \nabla_{[f} \gamma_{e] d}
            + R_{cd[e}{}^g \gamma_{f]g}
            - \frac{4\Lambda}{(d-1)(d-2)}
              \left( g_{c[e} \gamma_{ f]d} - g_{d[e} \gamma_{ f]c} \right)
     \bigg\} q_a{}^{c} l^d q_b{}^e l^f \ ,
%
\end{split}
\een
which is a trace-free symmetric tensor field whose indices are projected by
$q^a{}_b$.
This wave equation can be written as
\ben\label{teukolsky}
(\cO \Omega)_{ab}=0 \ ,
\een
where $\cO$ is the differential operator\footnote{
Note that the highest derivative terms in $2  \cP' \cP + \cD^c \cD_c$ coincide
with those of $\nabla^a \nabla_a$.
}
\bena\label{teukolsky1}
(\cO \Omega)_{ab}  &\equiv&
  \bigg\{ 2  \cP' \cP + \cD^c \cD_c
          - 6  \tau^c \cD_c +  4 C_{cdef} l^c n^e q^{df} - \frac{4d\Lambda}{(d-1)(d-2)} 
  \bigg\} \Omega_{ab} +  \\
   &&  \bigg\{
            4 \tau^e
              ( q^f_e \cD_{(a}  - q_{e(a} \cD^f  )
            + 2l^cn^eC_{cdeg} ( 5q^{fg}q^d_{(a}
                          -  3q^{df}q^g_{(a}
                            )
        \bigg\} \Omega_{b)f} \ . \non
\eena
Our notations in this and the following equations follow \cite{Pravda2}: 
The operators $\cD_a, \cP, \cP'$ depend on a real number $b \in \RR$ (``boost weight'') and
act on tensor fields $t_{a_1\dots a_s}$ whose indices are
projected by $q_a{}^b$. They are defined by
\ben\label{opdef}
\begin{split}
\cD_c t_{a_1\dots a_s} &=q_{a_1}{}^{d_1} \cdots q_{a_s}{}^{d_s}
  \left[
    q_c{}^e \nabla_e  - b \cdot q_c{}^e n^d (\nabla_e l_d)
  \right] t_{d_1\dots d_s} \\
\cP \ t_{a_1\dots a_s} &= q_{a_1}{}^{d_1} \cdots q_{a_s}{}^{d_s}
 \left[
       l^c \nabla_c \ - \ b \cdot l^e n^d (\nabla_e l_d)
 \right] t_{d_1\dots d_s} \ .
\end{split}
\een
The boost weight of a quantity is defined to be its scaling power under $l^a \to \lambda l^a, n^a \to \lambda^{-1} n^a$,
so for example $\Omega_{ab}$ has $b=2$, $\cP$ raises the boost weight by one unit, and $\cP'$ decreases the boost weight by one unit.
We also use the `prime convention', which means that a ${}'$ on any object means that $n^a$ and $l^a$
are to be exchanged in its definition. Since $n^a$ and $l^a$ are on the same footing,
the `primed' version of equation~\eqref{teukolsky} also holds for $\Omega'_{ab}$. We also use the shorthands
\ben
\tau_a = q_a^b n^c \nabla_c l_b \ , \quad
\tau_a' = q_a^b l^c \nabla_c n_b \ ,
\een
where the second expression is an example of the priming operation. 

In $d=4$, a trace-free symmetric tensor field $\Omega_{ab}$ that is projected by
$q^a{}_b$ can be identified with a complex scalar via a choice of complex 2-bein
for $q_{ab}$. Namely, if $q_{ab} = m_{(a} \bar m_{b)}$, then we can write $\Omega_{ab} = \Phi_0 m_a m_b + \bar \Phi_0 \bar m_a \bar m_b$
for some complex scalar function $\Phi_0$. The equation $\cO \Omega_{ab} = 0$ is then equivalent to the Teukolsky equation~\cite{Teukolsky} for $\Phi_0$. For this reason, we will refer
to eq.~\eqref{teukolsky} as the ``Teukolsky equation'' also in $d>4$.

From the quantity $\Omega_{ab}$ in~\eqref{omdef} one can in principle reconstruct
the perturbation $\gamma_{ab}$ itself  up to gauge transformations, and up to a finite dimensional space of special
perturbations. (In our case the latter would be perturbations to other NH-geometries.)
But since $\Omega_{ab}$ involves derivatives of $\gamma_{ab}$, this relationship would necessarily
be non-local, depend on awkward choices of boundary conditions, etc.
For our purpose, it is much better to construct perturbations $\gamma_{ab}$ satisfying
the linearized Einstein equations directly. We will do this by introducing a certain ``potential'' for gravitational perturbations, whose existence, like that of the Teukolsky
equation~\eqref{teukolsky}, is closely related to the optical and algebraically special properties
of the background.
The desired potential $U^{ab}$, called ``Hertz-potential'',
satisfies an equation that is closely related to the operator $\cO$ defined
above in eq.~\eqref{teukolsky1}. If this equation holds, then one can define
a corresponding gravitational perturbation $\gamma_{ab}$ by acting on
$U_{ab}$ with a certain second-order partial differential operator. This gravitational
perturbation then satisfies~\eqref{einstein}.

To set things up, we recall the standard notion of ``adjoint'' of a linear partial differential operator ${\mathcal P}$ from  sections of a real vector bundle $\bbE$ to sections of a real vector bundle $\bbF$ over $\M$. If both bundles are equipped with a metric structure, 
and if $\M$ is equipped with an integration measure -- as will be usually the case for us -- 
then the adjoint ${\mathcal P}^*$ is a differential operator from $\bbF$ to $\bbE$. 
For example, in the case of the operator ${\mathcal P}=\cO$, $\bbE=\bbF$ is the bundle
of contravariant, symmetric, trace-free tensors
that are projected by $q^a{}_b$, whereas in the case of the linearized Einstein operator, $\cL$ \eqref{einstein}, $\bbE=\bbF$ is the space of
contravariant, symmetric tensors. In these cases, the metric structure is given by $q_{ab}$ respectively $g_{ab}$, and the integration 
element is that induced by $g_{ab}$. 

After these preliminaries, we can construct the Hertz-potentials first found in 4 dimensions in~\cite{Hertz1,Hertz2}. We will follow the elegant method of Wald~\cite{Hertz2}, which can be generalized straightforwardly to higher dimensions~\footnote{After we completed our calculation, we have learned that an almost identical analysis had been 
carried out previously by~\cite{Madi}.}. Consider an arbitrary smooth symmetric tensor field $\gamma_{ab}$, and let $J_{ab} \equiv (\cO \Omega)_{ab}$, where $\cO$ is the operator
defined in~\eqref{teukolsky} acting on symmetric tensors projected by $q^a_b$, 
and where $\Omega_{ab}$ is the ``Teukolksy tensor'', defined in terms of $\gamma_{ab}$ by~\eqref{omdef}.  If $\gamma_{ab}$ satisfies 
the linearized Einstein equations~\eqref{einstein}, $(\cL \gamma)_{ab} = 0$, then $J_{ab}=0$. If not, then clearly $J_{ab}$ must have the form of a linear partial differential operator $\cS$ 
applied to $T_{ab} \equiv (\cL \gamma)_{ab}$, that is $J_{ab} = (\cS T)_{ab}$. We need the concrete form of this operator $\cS$. It is found after a lengthy calculation that
\bena\label{intertw}
(\cS T)_{ab} &=& +2\{\cP \cD_{(a} - (2 \tau_{(a} + \tau'_{(a}) \cP - (\cP \tau'_{(a})\} (q_{b)}^c l^d T_{cd}) - \cP^2 (q_a^c q_b^d T_{cd}) + \frac{1}{d-2} q_{ab} \cP^2 (g^{cd} T_{cd}) \non\\
&& - \frac{1}{d-2}  q_{(a}^d q_{b)}^f C_{cdef}l^c n^e T_{mn} l^m l^n \\
&& + \frac{1}{d-2}  q_{ab} \bigg\{2 \, \cP' \cP + \cD^c \cD_c - 6\tau^c\cD_c
+ 4 \, C_{cdef} n^c l^d n^e l^f - \frac{2d\Lambda}{d-1} \bigg\} T_{mn} l^m l^n  \ . \non
\eena
Let $\cT$ be the linear second order differential operator which associates with a symmetric tensor field
$\gamma_{ab}$ the trace-free symmetric tensor field $(\cT \gamma)_{ab} = \Omega_{ab}$ projected by $q^a_b$. $\cT$ is given concretely
by the second line of eq.~\eqref{omdef}. In terms of the operators $\cL, \cS, \cT, \cO$, the relation 
$(\cO \Omega)_{ab} = J_{ab}$ may then be written as 
$(\cO \cT \gamma)_{ab} = (\cS \cL \gamma)_{ab}$. Since this must hold for an arbitrary smooth 
symmetric tensor field $\gamma_{ab}$, we have the operator equation $\cO \cT = \cS \cL$.
Taking the adjoint of this operator relation and
applying the result to a symmetric, trace-free tensor field $U_{ab}$ projected by $q^a_b$, we find
\ben
\cT^* \cO^* U_{ab} = \cL^* \cS^* U_{ab} \ .
\een
The key point is now that the linearized Einstein operator is self-adjoint, $\cL = \cL^*$,
which is a direct consequence of the fact that it arises from an action principle.
Hence, if\footnote{Note that a non-trivial kernel of the operator $\cT^*$ can also give rise to 
solutions to the linearized Einstein equations. This is closely related to the well-known fact 
that the correspondence between solutions and Hertz potentials is not bijective.}
 $(\cO^* U)_{ab}=0$, then $\gamma_{ab} := (\cS^* U)_{ab}$ is
a symmetric tensor satisfying the linearized Einstein equations $(\cL \gamma)_{ab}=0$.
Working out explicitly the operator $\cS^*$ from eq.~\eqref{intertw} gives:
\ben
(\cS^* U)_{ab} = - l_a l_b C_{cedf} l^e n^f U^{cd} + 2 l_{(a} \cP \cD^c U_{b)c} + 2 l_{(a}(2\tau^c + [l,n]^c) \cP U_{b)c} - \cP^2 U_{ab} \ ,
\een
and we conclude (see also~\cite{Madi}):

\begin{lemma}\label{thm1} (Hertz potentials for gravitational perturbations)
Consider a background solution to the vacuum Einstein equations
with $\Lambda$ having null vector fields $l^a, n^a$ with the optical properties~\eqref{optical} and an algebraically special property~\eqref{alg}.
Let $U_{ab}$ be a smooth symmetric, trace-free tensor field satisfying $q^a{}_c q^b{}_d U_{ab} = U_{cd}$, together with
\ben\label{Uab}
(\cO^* U)_{ab}=0 \ .
\een
Here, $\cO^*$ is the transpose
of the operator $\cO$ defined above in eq.~\eqref{teukolsky1} in terms of the operators $\cP, \cD_a$ given in eq.~\eqref{opdef} with $b=2$. Then
\ben\label{hertz}
\gamma_{ab} = - l_a l_b (C_{cedf} l^e n^f U^{cd}) + 2 l_{(a} \cP \cD^c U_{b)c} + 2 l_{(a}(2\tau^c + [l,n]^c) \cP U_{b)c} - \cP^2 U_{ab}
\een
is a solution to the linearized Einstein equation~\eqref{einstein}. We call $U_{ab}$ the
{\em Hertz-potential} for $\gamma_{ab}$. Note that by definition $\gamma_{ab} l^b = 0 =
q^{ab} \gamma_{ab}$.
\end{lemma}

\subsection{Electromagnetic perturbations}

Hertz potentials in higher dimensions can also be introduced in the case of electromagnetic perturbations. The Maxwell equations are
\ben
\label{maxwell}
\nabla^a F_{ab} = 0 \ , \quad \nabla_{[a} F_{bc]} = 0 \ .
\een
Taking derivatives of these equations, there follows the equation
\ben\label{maxwell2}
  \nabla^c \nabla_c F_{ab} + R_{abcd} F{}^{cd} + R_{ad} F_b{}^d + R_{bd} F^d{}_a = 0 \  .
\een
We now assume again the background Einstein equation $R_{ab} - \half g_{ab} R = - \Lambda g_{ab}$ and
that the background has the optical and algebraically special properties as in~\eqref{optical},~\eqref{alg}. We  define $\Omega_a = F_{cb} q^c{}_a l^b$. By
contracting eqs.~\eqref{maxwell2},~\eqref{maxwell} in all possible ways into $n^a, l^a, q_{ab}$, one finds
again that $\Omega_a$ satisfies a decoupled equation analogous to~\eqref{teukolsky1}. It is~\cite{Durkee}:
\bena
\label{teukolsky2}
(\cO \Omega)_a &\equiv&
 \bigg\{
       2 \cP' \cP + \cD_c \cD^c - 4 \tau^c \cD_c
         +q^{ce} l^d n^f C_{cdef}
      - \frac{2\Lambda(2d-3)}{(d-1)(d-2)}
     \bigg\} \Omega_a + \non\\
&&
   \bigg\{
           - 4\tau_d q^d_{[a} \cD_{b]}
           + l^cn^eC_{cdef}( 3 q^d_aq^f_b - q^d_bq^f_a)
    \bigg\}\Omega^b =0 .
\eena
The operators $\cP, \cD_a$ are defined as in eq.~\eqref{opdef} with $b=1$ in the present case.  In order to derive
a Hertz potential for $A_a$, we proceed just as in the case of gravitational perturbations. Let $A_a$ be an arbitrary 1-form, not necessarily satisfying the
Maxwell equations. One derives
\ben\label{rhs}
(\cO \Omega)_a = \cP(q_a^b J_b) - (\cD_a - 2\tau_a - \tau'{}_a)(l^b J_b) \ ,
\een
where $\Omega_a = l^b q_{a}{}^c F_{cb}$ is as above, and where $J_a = \nabla^b F_{ba}$.
As an equation for $A_a$, we write this again in the form $(\cO \cT A)_a = (\cS \cL A)_a$, where $\cL$ is now the Maxwell operator defined by
\ben
(\cL A)_a = 2 \, \nabla^c \nabla_{[c} A_{a]},
\een
where $\cT$ is defined by
\ben
(\cT A)_a =2 \,  l^b q_{a}{}^c \nabla_{[c} A_{b]} \ ,
\een
and where $\cS$ is defined by the right hand side of eq.~\eqref{rhs}. The transpose
of that operator is
\ben
(\cS^* U)_a = -\cP U_a + l_a (\cD_b + \tau_b) U^b \ .
\een
Taking the transpose of the operator equation $\cO \cT  = \cS \cL$ and applying both sides of the
resulting equation to $U_a$ now gives the desired Hertz potential (see also~\cite{Madi}):

\begin{lemma}\label{thm1} (Hertz potentials for electromagnetic perturbations)
Consider a background solution to the vacuum Einstein equations
with $\Lambda$ having null vector fields $l^a, n^a$ with the optical properties~\eqref{optical} and an algebraically special property~\eqref{alg}.
Let $U_{a}$ be a smooth  tensor field satisfying $q^c{}_b U_{c} = U_{b}$, together with
\ben\label{Uab}
(\cO^* U)_{a}=0 \ .
\een
Here, $\cO^*$ is the transpose
of the operator $\cO$ defined above in eq.~\eqref{teukolsky2} in terms of the operators $\cP, \cD_a$ given in eq.~\eqref{opdef} with $b=1$. Then the
field strength $F_{ab} = 2\nabla_{[a} A_{b]}$ of
\bena\label{hertz1}
A_a = -\cP U_a + l_a (\cD_b + \tau_b) U^b
\eena
is a solution to the Maxwell equations~\eqref{maxwell}. We call $U_{a}$ the
{\em Hertz-potential} for $A_a$. Note that by definition $0 = A_a l^a$.
\end{lemma}

\section{Construction of a perturbation with $\E<0$ in the NH geometry} \label{sec3}

\subsection{Gravitational sector}

We will now employ the Hertz potentials to  construct a gravitational perturbation with $\E < 0$
in the NH geometry if the operator \eqref{Adef} has a suitable spectrum. 

We begin by defining null vector fields $n^a,l^a$ by
\ben\label{nldef}
\begin{split}
l &= \frac{1}{L\sqrt{2}} \Big( R \frac{\partial}{\partial R} -
\frac{1}{R} \frac{\partial}{\partial T} - k^I \frac{\partial}{\partial \phi^I} \Big) \ , \\
n &= \frac{1}{L\sqrt{2}} \Big( R \frac{\partial}{\partial R} +
\frac{1}{R} \frac{\partial}{\partial T} + k^I \frac{\partial}{\partial \phi^I} \Big) \ ,
\end{split}
\een
where the coordinates $R,T,\phi^I$ refer to the NH geometry~\eqref{nhg}.
These vector fields can {\em both} be shown to satisfy  the optical properties~\eqref{optical}, and the
algebraically special properties~\eqref{alg}~\cite{Durkee,Durkee2}.
In particular, since both $n^a, l^a$ are twist free we get
$0 = n_{[a} l_b \nabla_c l_{d]} = n_{[a} l_b \nabla_c n_{d]}$, the subspaces
perpendicular to $n^a, l^a$ are integrable (by Frobenius' theorem). The corresponding
family of $(d-2)$-dimensional submanifolds (all diffeomorphic to $\eB$) establish
an isomorphism $\M \cong \hat \M \times \eB$, where $\hat \M$ is the base space of
the foliation. Also, since the properties of being geodesic, null, shear, expansion, and twist-free are geometrical features that are invariant under any isometry, it follows at once
that $\pounds_X n^a = 0$ for any Killing field $X^a$ (and similarly for $l^a$). It
then also follows that the foliation is invariant under all the isometries. The leaves of this foliation
in fact correspond precisely to the surfaces of constant $T,R$ in eq.~\eqref{nhg}, whereas the base space $\hat \M$ 
is parameterized by $T,R$ and corresponds to an $AdS_2$-space. This explains the geometrical significance
of these coordinates from the point of view of algebraically special geometry.

Since the Hertz-potential
$U^{ab}$ has only components tangent to the foliation (because it is projected by
$q^a{}_b$), we may write
\ben
\label{GNC}
U^{ab} = U^{AB} \Big( \frac{\partial}{\partial x^A} \Big)^a
         \Big( \frac{\partial}{\partial x^B} \Big)^b \ , \quad
\een
where $x^A$ are coordinates of $\eB$. We make the separation of variables ansatz
\ben\label{Udef}
U^{AB} = \psi \cdot Y^{AB}  \ ,
\een
where $\psi = \psi(R,T)$ is a function on $AdS_2$, and where
$Y = Y^{AB}(x^C) \partial_A \otimes \partial_B$ is a symmetric trace-free tensor intrinsic to $\eB$ that has ``angular dependence $e^{-i \um \cdot \underline{\phi}}$''
for some set of ``quantum numbers'' $\um \in \ZZ^n$. The last condition is stated more precisely as follows. 
Let $\bbE_2$ be the bundle of symmetric trace-free rank 2 tensor fields over $\eB$, let $\chi_{\um}$ be the character of ${\rm U}(1)^n$
given by $\chi_{\um}(e^{i\phi^1}, \dots, e^{i\phi^n}) = e^{-i\um \cdot \underline{\phi}}$, and let 
\ben\label{cmdef}
\begin{split}
C^\infty(\eB, \bbE_2)^{\um} &= \{ Y \in C^\infty(\eB, \bbE_2) \quad \mid \quad (\tau^*Y)(x) = \chi_{\um}(\tau) Y(x) \quad 
\text{for all $x \in \eB, \tau \in {\rm U}(1)^n$} \} \\
&= \{ Y \in C^\infty(\eB, \bbE_2) \quad \mid \quad \pounds_{\partial/\partial \phi^I} Y(x) = -im_I Y(x) \ \ \ \text{for $I=1,\dots,n$} \}.
\end{split} 
\een
Then we require $Y$ to be in this space. Inserting these definitions and 
using also~\eqref{GNC}, one finds, just as in~\cite{Durkee}, that for such $U$, the action of $\cO^*$ 
is written as
\ben\label{master}
(\hat D^2 - q^2 - \cA) U  = \cO^* U
\een
where $\hat D = \hat \nabla - iq \hat A$ [see eq.\eqref{ads2}] and where $\cA$ is the second order elliptic operator on $\eB$ given by
\ben\label{Adef}
\begin{split}
(\cA Y)^C{}_B &= \Big( -L^2 \bD^A \bD_A  - (\bD_A L^2) \bD^A + L^{-2} (\bD_A L^2)(\bD^A L^2) + \bD^A \bD_A L^2 \Big) Y^C{}_B - \\
& 4 \ (k_{[A} - \bD_{[A} L^2) \bD_{B]} Y^{AC} +  \Big( 6 - a^2 - 4 L^{-2} k_A k^A - 2(d-4) \ \Lambda L^2 \Big) Y^C{}_B + \\
& L^2 \Big(\cR_{AB} + \cR \mu_{AB} \Big) Y^{CA} + L^2 \Big( \cR_{A}{}^C + \cR \mu_A{}^C \Big) Y^{A}{}_B - 2 \ \cR^C{}_{ABD} Y^{AD} . 
\end{split}
\een
Capital Roman indices  are always raised and lowered with $\mu^{AB}, \mu_{AB}$, and 
$\cR_{ABCD} \omega^D = 2 \bD_{[A} \bD_{B]} \omega_C$ is the Riemann tensor of the Levi-Civita connection $\bD_A$ associated with $\mu_{AB}$.
Note that $\cA$ depends explicitly on $\um$ through $a = \uk \cdot \um$. The quantities $\mu_{AB}, k^A, L$ are given concretely by equations~\eqref{comparison} for the MP solutions.
The ``charge'' $q \in \CC$ is given by
\ben\label{qdef}
q=a+ib  \, \qquad \text{where} \quad
\begin{cases} 
b=2 \\
a = \uk \cdot \um .
\end{cases}
\een
As noted~\cite{Durkee}, $\cA$ is self-adjoint on the Hilbert space of square integrable trace-free rank-2 tensors with inner product
\ben\label{scal}
(Y,Y)_{\eB} = \int_\eB |Y|^2 L^2  \, \D vol_\mu \ .
\een
By the standard theory of elliptic self-adjoint operators on compact manifolds, the eigentensors of $\cA$ form an orthonormal basis of 
$L^2(\eB, \bbE_2; L^2 \D vol_\mu)$. Because $\cA$ commutes with the action of ${\rm U}(1)^n$, it maps the subspace of tensors $C^\infty(\eB, \bbE_2)^{\um}$ transforming according to the character $\chi_{\um}$ to itself. We will denote the restriction of $\cA$ to this subspace by $\cA_{\um}$ [where we also set $a=\uk \cdot \um$ in~\eqref{Adef}]. 

The separation of variables ansatz is now used to solve the condition $(\cO^* U)_{ab} = 0$ required from a Hertz-potential. Let $\lambda$ be an eigenvalue of $\cA_{\um}$ and let $Y$ be an 
a corresponding eigentensor in $C^\infty(\eB, \bbE_2)^{\um}$. 
Then, if $\psi$ satisfies the ``charged $AdS_2$ Klein-Gordon equation''
\ben\label{master1}
(\hat D^2 - q^2 - \lambda) \psi =-\frac{1}{R^2} \frac{\partial^2 \psi}{\partial T^2} + \frac{\partial}{\partial R} \Big(R^2 \frac{\partial \psi}{\partial R}\Big) + \frac{2(2-ia)}{R} \frac{\partial \psi}{\partial T} - \lambda \psi = 0 \ ,
\een
it follows that $U^{ab} = U^{AB} (\partial_A)^a (\partial_B)^b$ as in eq.~\eqref{Udef} satisfies $(\cO^* U)_{ab} = 0$. Consequently, for such a $U_{ab}$,
the perturbation $\gamma_{ab}$ given by eq.~\eqref{hertz} satisfies the linearized Einstein equations on the NH geometry. By first solving\footnote{It is straightforward to obtain solutions to~\eqref{master1} as we recall in appendix~\ref{appB}. Solutions
to the eigenvalue problem $\cA Y = \lambda Y$ must in general be found numerically in
concrete examples.} the eigenvalue equation $\cA Y = \lambda Y$ to
get $\lambda$, and then the Klein-Gordon equation~\eqref{master1}, one can thus get
solutions to the linearized Einstein equations. \\

\medskip

We wish to evaluate the canonical energy of such a perturbation in terms of $\psi$ on a Cauchy surface $\Sigma$ of the form 
$\eB \times \hat \Sigma$, where $\hat \Sigma$ is a slice in $AdS_2$ as drawn in 
the Penrose diagram~\ref{MP2}. This is a somewhat lengthy calculation, which we therefore break up into several steps. 
First, we evaluate
symplectic form $W(\gamma_1, \gamma_2)$ of two perturbations, each given by eq.~\eqref{hertz} in terms of two Hertz potentials as
in \eqref{Udef}. We record the lengthy expression in the next lemma:

\begin{lemma}\label{lemsymp}
Let $Y \in C^\infty(\eB, \bbE_2)^{\um}$ such that $\cA Y = \lambda Y$, $\| Y\|_{\eB} =1,$ 
let $\psi_1, \psi_2$ be two (complex) solutions to the equation~\eqref{master1}. Let $U_1, U_2$
be the corresponding (complex) Hertz-potential as in eq.~\eqref{Udef}, and let $\gamma_1, \gamma_2$
be the corresponding (complex) perturbations as in eq.~\eqref{hertz}. Then the symplectic form 
on $\Sigma = \eB \times \hat \Sigma$ is\footnote{For complex perturbations, we continue $W$ {\em anti-} linearly in the
first entry.}
\ben
W(\Sigma, \gamma_1, \gamma_2) =  \int_{\hat \Sigma} \hat \star \hat w (\psi_1, \psi_2)\ ,
\een
where $\hat \star$ is the Hodge operator of $AdS_2$, and where the conserved current $\hat w$ on $AdS_2$ is given up 
to a total divergence (i.e. up to changing $\hat \star \hat w$ by an exact 1-form)
by
\bena
128 \pi \ \hat w &=&
(- R^{-1} \partial_T + R \partial_R + ia)^2 \bar \psi_1
(\D + ia R \D T ) ( - R^{-1} \partial_T + R \partial_R - ia )^2 \psi_2 + \non\\
&& 5 (- R^{-1} \partial_T + R \partial_R + ia ) \bar \psi_1 (\D + ia R \D T) (- R^{-1} \partial_T + R \partial_R-ia) \psi_2 + \non\\
&&4 \, \bar \psi_1  (\D + ia R \D T) \psi_2 + \non\\
&&
8 (- R^{-1} \partial_T + R \partial_R + ia) \bar \psi_1
(  R^{-1} \partial_T + R \partial_R + ia) \psi_2 \, (R \D T + R^{-1} \D R) - \non\\
&&\{3(\lambda + a^2) + 13ia
\} (- R^{-1} \partial_T + R \partial_R + ia) \bar \psi_1  \psi_2 (R \D T + R^{-1} \D R) - \non\\
&&
4(- R^{-1} \partial_T + R \partial_R + ia
) \bar \psi_1  \psi_2 (-R \D T + R^{-1} \D R) + \non\\
&&3ia (- R^{-1} \partial_T + R \partial_R + ia ) \bar \psi_1 (- R^{-1} \partial_T + R \partial_R-ia) \psi_2 (R \D T + R^{-1} \D R)
+ \non\\
&&4ia \bar \psi_1 \psi_2 (R \D T + R^{-1} \D R)
 - (\psi_1 \leftrightarrow \psi_2)^* \ .
\eena
As before, $a = \uk \cdot \um$, and the star $*$ in the last line indicates complex conjugation.
\end{lemma}

{\em Proof}: The formula for $\hat w$ can in principle be obtained by inserting
the ansatz~\eqref{Udef} into \eqref{hertz}, then substituting that into
the symplectic form $\star w$, see eq.~\eqref{omegadef} and~\eqref{symform}, and then carrying out the integration over $\Sigma$, 
taking advantage of $\cA Y = \lambda Y$ and of~\eqref{master1} for $\psi_1, \psi_2$. However, the resulting 
calculations would be extremely tedious, and we therefore present an alternative derivation that seems simpler. 

For this, we split the integration of $\star w$ over $\Sigma$ into successive integrations over $\eB$ and then $\hat \Sigma$. More 
precisely, consider a vector field $\hat X$ tangent to $\hat \M \cong AdS_2$, which we may lift in an obvious way to a vector field $X$ 
on $\M$. We define $\hat w$ by
\ben
(\hat \star \hat w)(\hat X) = \int_{\eB} i_{X}(\star w) \ ,  
\een
where Cartan's operator $i_{X}$ acts by inserting $X^a$ into the first argument of a differential form. Because $\star w$ is a closed
$(d-1)$-form on $\M$ that is locally constructed out of $\gamma_{1 \, ab}, \gamma_{2 \, ab}$ and their derivatives, it easily follows that 
$\hat \star \hat w$ is a closed 1-form on $\hat \M \cong AdS_2$ that is locally constructed out of $\psi_1, \psi_2$ and their derivatives.
In fact, since $w$ is a  bilinear local expression in the perturbations $\gamma_{1 \, ab}, \gamma_{2 \, ab}$  containing precisely one derivative, and since $\gamma_{1 \, ab}, \gamma_{2 \, ab}$ are in turn local expressions in $\psi_1, \psi_2$ containing up to two derivatives
each according to eqs.~\eqref{Udef} and \eqref{hertz}, it follows that $\hat w$ is a local expression in $\psi_1, \psi_2$ containing up to 
five derivatives altogether and at most three derivatives on either $\psi_1$ and $\psi_2$ separately. 
Furthermore, $\hat w(\psi_1, \psi_2)$ must be anti-linear in $\psi_1$ and linear in $\psi_2$, and we must have
\ben \label{symm1}
\hat w(\psi_2, \psi_1) = -[\hat w(\psi_1, \psi_2)]^*
\een
from the anti-symmetry of the symplectic form $W$. 

In order to find out what form $\hat w(\psi_1, \psi_2)$ can take, it is efficient to introduce two operators $\hat l, \hat n$ on $AdS_2$ whose definition dovetails 
that of $l^a, n^a$ above in eqs.~\eqref{nldef}. We set:
\ben\label{NLdef}
\begin{split}
\hat l &= R \frac{\partial}{\partial R} -
\frac{1}{R} \frac{\partial}{\partial T} - ia \ , \\
\hat n &= R \frac{\partial}{\partial R} +
\frac{1}{R} \frac{\partial}{\partial T} + ia \ .
\end{split}
\een
One then easily verifies the identities:
\ben
\begin{split}
\label{manipid}
\hat l \hat n - \hat n \hat l &= \hat l - \hat n + 2ia \ , \\
\hat l \hat n \psi &= (-3 \hat n + 2 \hat l + a^2 + 5ia + \lambda)\psi \ , 
\end{split}
\een
where the second equation holds for any $\psi$ satisfying the equation of motion~\eqref{master1}. The first identity allows one 
to swap $\hat n$ and $\hat l$, and the second one allows one to change $\hat l \hat n \psi$ to an expression involving fewer 
derivatives (a similar expression can easily be derived for $\hat n \hat l \psi$). By going through the definitions, it immediately follows that 
the highest derivative part of $\hat w$ must be proportional to\footnote{Of course, the term involving $\hat A$ could be subsumed into 
the lower order terms, but it is easier for calculations to keep it.}
\ben
\label{leading}
\hat w = 
(\hat l^2 \psi_1)^* (\D - ia\hat A) (\hat l^2 \psi_2) - (\psi_1 \leftrightarrow \psi_2)^* +
(\text{lower order terms}) \ , 
\een
where ``lower order'' refers to the number of derivatives. We now wish to argue that, up to terms representing a total divergence already mentioned in the statement of the theorem, there is a unique set of lower order terms that will turn the right side 
of this expression into a divergence free 1-form on $\hat \M$ satisfying~\eqref{symm1} for any $\psi_1, \psi_2$ that solve~\eqref{master1}. 
To carry through this argument, it is important to first remark that, although $\psi$ is constrained by the Klein-Gordon equation
\eqref{master1}, we may, at each arbitrary but fixed point of $\hat \M$, specify the values of $\hat n^k \psi$ and $\hat l^j \psi$ independently for all $j,k$ (here $\hat n^k$ indicates 
the $k$-th power, not a component). This follows, morally speaking, because $\hat n, \hat l$ represent derivatives in null-directions. We may specify those independently for solutions of~\eqref{master1} as this equation allows for a ``null-initial value formulation''. Our second remark is that any mixed expression  $\cdots \hat l^i \hat n^j \cdots \hat l^k \hat n^l \dots \psi$
can be manipulated using eqs.~\eqref{manipid} into a unique expression involving sums of  
$\hat n^k \psi$ and $\hat l^j \psi$ (for different values of $k,l$), which in turn are freely specifiable at each point by our first remark. 
These remarks give a clear procedure to determine which lower order terms can be added to eq.~\eqref{leading} in order produce 
the most general $\hat w$ such that ${\rm div}_{\hat g} \hat w = 0$ for all $\psi_1, \psi_2$ satisfying~\eqref{master1}. 

Taking a divergence of eq.~\eqref{leading} with respect to the $AdS_2$-metric $\hat g$, we find after a computation using~\eqref{manipid}
that 
\ben
\label{L2}
\begin{split}
&
{\rm div}_{\hat g} 
\left[ (\hat l^2 \psi_1)^* (\D -ia \hat A) (\hat l^2 \psi_2) - (\psi_1 \leftrightarrow \psi_2)^* \right] \\
=&
-10 ia (\hat l \psi_1)^* \hat l \psi_2 + 5 (\hat l^2 \psi_1)^* \hat l \psi_2 - 16 (\hat l \psi_1)^* \hat n \psi_2 + 27ia (\hat l \psi_1)^* \psi_2 \\
& -(\psi_1 \leftrightarrow \psi_2)^* + {\rm div}_{\hat g} [\dots] \ ,  
\end{split}
\een 
where the dots $[\dots]$ in the last line represent terms having lower order than those in $[\dots]$ on the left side. Such lower order terms 
should hence be subtracted from the right side in the expression for $\hat w$ given by~\eqref{leading}. We now seek to compensate the other 
lower order terms on the right side of this equation, in particular the leading term proportional to $(\hat l^2 \psi_1)^* \hat l \psi_2 
- (\psi_1 \leftrightarrow \psi_2)^*$, by adding further terms to the right side of eq.~\eqref{leading}. It is seen that the unique (up to a total 
divergence-) expression which can compensate this leading term must be proportional to $(\hat l \psi_1)^* (\D - ia\hat A) \hat l \psi_2
- (\psi_1 \leftrightarrow \psi_2)^*$. The divergence of that term is in fact found to be 
\ben\label{L1}
\begin{split}
&
{\rm div}_{\hat g} \left[
(\hat l \psi_1)^* (\D -ia \hat A) (\hat l \psi_2) - (\psi_1 \leftrightarrow \psi_2)^* \right] \\
=& -(\hat l^2 \psi_1)^* \hat l \psi_2 + \lambda (\hat l \psi_1)^* \hat l \psi_2 +
4 (\hat l \psi_1)^* \hat n \psi_2 - 7ia (\hat l \psi_1)^* \psi_2 \\
&- (\lambda + a^2) (\hat l \psi_1)^* \psi_2 
-(\psi_1 \leftrightarrow \psi_2)^* 
+ 4ia (\hat l \psi_1)^* \hat l \psi_2  + {\rm div}_{\hat g} [\dots ] \ ,  
\end{split}
\een 
where the dots $[\dots]$ in the last line again denote  terms of lower order than those in $[\dots]$ on the left side. Thus, adding 5 times this term 
to the right side of~\eqref{leading} will therefore get rid of the highest derivative term 
proportional to $(\hat l^2 \psi_1)^* \hat l \psi_2 - (\psi_1 \leftrightarrow \psi_2)^*$ on the right side of eq.~\eqref{L2}. 
In fact, one finds using~\eqref{manipid} that 
\ben
\begin{split}
&{\rm div}_{\hat g} \left[
(\hat l^2 \psi_1)^* (\D -ia \hat A) (\hat l^2 \psi_2) + 5(\hat l \psi_1)^* (d -ia \hat A) (\hat l \psi_2) - (\psi_1 \leftrightarrow \psi_2)^* \right]\\
=& 8 \psi_1^* \hat n \psi_2 - 8\psi_1^* \hat l \psi_2  - 8ia (\hat l \psi_1)^* \psi_2 - 20ia \psi_1^* \psi_2 -(\psi_1 \leftrightarrow \psi_2)^* 
+ {\rm div}_{\hat g} [\dots], 
\end{split}
\een
where the dots $[\dots]$ in the last line again denote terms of lower order than those in $[\dots]$ on the left side. We must next find a
term which, when added to $\hat w$ in eq.~\eqref{leading}, will cancel the terms on the right side that are not already in divergence form. 
This unique term (up to a total divergence) is found to be given by $4 \psi_1^* (\D - ia\hat A) \psi_2 -(\psi_1 \leftrightarrow \psi_2)^*$. 
Our argument shows that the final expression for $\hat w$ is uniquely specified by 
the symmetry condition~\eqref{symm1} and leading term~\eqref{leading}, up to a total divergence. Writing out explicitly all the terms 
and proportionality constants left unspecified in this outline 
leads to the expression of $\hat w$ stated in the lemma. \qed

\medskip
\noindent
We can now evaluate the canonical energy~\eqref{Edef} in terms of $\psi$. As our Cauchy surface,
we take for simplicity $\Sigma = \{T=0\}$, and we take a perturbation $\gamma_{ab}$
constructed from a Hertz-potential $U_{ab}$ as in eq.~\eqref{hertz}. For $U_{ab}$, we make
the separation of variables ansatz~\eqref{Udef} with some $\psi$ of compact support
on the slice $\hat \Sigma = \{T=0,R>0\} = \RR_+$. Recalling that $K=\partial/\partial T$ in Poincar\'e coordinates, we get
from the previous lemma
\ben
\E = W(\Sigma; \gamma, \pounds_K \gamma) =  \int_{\hat \Sigma} \hat \star \hat w(\psi, \tfrac{\partial}{\partial T} \psi)  \ .
\een
We can write $\E$ in terms of $\psi |_{\hat \Sigma}, \partial_T \psi|_{\hat \Sigma}$, because
any $T$-derivative of order $>1$ may be eliminated, using~\eqref{master1}, in favor of terms containing only up to one $T$-derivative. Thus, $\E$ becomes
a quadratic form $\E(f_0, f_1)$ of the initial data
\ben\label{initial}
(f_0, f_1) \equiv \Big( \psi \Big|_{T=0},\tfrac{\partial}{\partial T} \psi \Big|_{T=0} \Big) \in C^\infty_0(\RR_+; \CC) \times C^\infty_0(\RR_+; \CC) \ ,
\een
 at $T=0$.
The resulting formula is rather long and is given in the appendix~\ref{appA1}. It simplifies somewhat for initial data
such that $f_1 = 0$ and $f_0$ is real valued. In this case ($y = \log R$)
\bena
\E &=& \frac{1}{128\pi}\int_{-\infty}^\infty \bigg\{
\bigg( 2\frac{\D^3 f_0}{\D y^3} +
 \frac{\D^2 f_0}{\D y^2} - (2\lambda +1) \frac{\D f_0}{\D y} + \lambda f_0 \bigg)^2 \non\\
&&\hspace{.6cm}+
\bigg( 2 \frac{\D^3 f_0}{\D y^3} + \frac{\D^2 f_0}{\D y^2} - (\lambda + a^2)\frac{\D f_0}{\D y} \bigg)^2 +
5\bigg( \frac{\D^2 f_0}{\D y^2} + \frac{\D f_0}{\D y} - \lambda f_0 \bigg)^2 \non\\
&&\hspace{.6cm}
+(5+4a^2) \bigg( \frac{\D^2 f_0}{\D y^2}\bigg)^2  
+(\lambda-3)\bigg( 2 \frac{\D^2 f_0}{\D y} +  \frac{\D f_0}{\D y} - (\lambda + a^2) f_0 \bigg)^2
\non\\
&&\hspace{.6cm} +(2\lambda -4a^2 + 4a^2\lambda) \bigg( \frac{\D f_0}{\D y} \bigg)^2
 + (2\lambda + 3a^2 )f_0^2  \non\\
&&\hspace{.6cm} -2\bigg( \frac{\D^2 f_0}{\D y^2} + \frac{\D f_0}{\D y} - \lambda f_0\bigg) \frac{\D f_0}{\D y} -
2 (4-3\lambda - 3a^2) \bigg( \frac{\D^2 f_0}{\D y^2} + \frac{\D f_0}{\D y} - \lambda f_0\bigg) f_0
\bigg\} e^y \D y \ . \non
\eena
This expression is not
manifestly positive definite, so there is a possibility of having
$\E < 0$ for suitable $f_0$ and $a,\lambda$.  To this end, we make the  variational ansatz:
\ben
\label{ansatz1}
f_0(R) =
\frac{R^N}{(R + \eps)^{N+1/2}(1+R^N e^{1/(1-R)})} \ , \qquad f_1(R) = 0 \ ,
\een
which depends on $N \in \NN$ and $\epsilon>0$. Here,  $0<R<1$ and the definition is extended smoothly to all $R>0$ by setting $f_i(R)=0$ for $R \ge 1$. 
For gravitational perturbations, we choose $N\ge 3$ because this ensures that derivatives up to order $3$ vanish at $R=0$. 
$f_0(R)$ is a pulse whose maximum moves towards $0$ as $\eps \to 0^+$. Its form is inspired by the mode analysis provided in appendix~\ref{appB}. 
Inserting the ansatz into our expression for $\E$ gives, after a lengthy calculation:
\ben\label{Eapprox}
\E = 
\frac{1}{128\pi}(\lambda + \tfrac{1}{4})(\lambda^2 +2a^2 \lambda + a^4 - 9a^2 + \tfrac{7}{2}) \ \log \eps^{-1} + O(1) ,
\een
where $O(1)$
stands for terms having a finite limit\footnote{These terms depend also on $N$.} as $\epsilon \to 0^+$. With the help of this identity, we can now prove the following theorem:

\begin{thm}\label{el0thm}
Let $\lambda$ be an eigenvalue of the operator $\cA$ acting on tensors in $C^\infty(\eB, \bbE_2)^{\um}$ such 
that 
\ben\label{condition11}
(\lambda + \tfrac{1}{4})(\lambda^2 +2a^2 \lambda + a^4 - 9a^2 + \tfrac{7}{2}) < 0 \ , 
\een 
where $a = \uk \cdot \um$. Then there exists a perturbation  such that $\E < 0$ on the NH geometry whose initial data
are compactly supported on $\Sigma = \{T=0\}$.

This is in particular the case if
\begin{enumerate}
\item[(i)] $\um=0$ and if the lowest eigenvalue $\lambda$ of $\cA$ acting tensors invariant under ${\rm U}(1)^n$ satisfies $\lambda < \lc$, or if 
\item[(ii)] $L^2 g^{IJ} m_I m_J < (k^I m_I)^2$ holds for some $\um \in \ZZ^n$, somewhere on $\eB$. 
\end{enumerate}
\end{thm}

{\em Proof:} If $(\lambda + \tfrac{1}{4})(\lambda^2 +2a^2 \lambda + a^4 - 9a^2 + \tfrac{7}{2}) < 0$, then the 
 right side of eq.~\eqref{Eapprox}  becomes negative for sufficiently small  $\eps > 0$. Hence,
 there are initial data of the form~\eqref{ansatz1} with $\E<0$. Our ansatz~\eqref{ansatz1}  does not have compact support on $R>0$, since
the support clearly includes $R=0$. However,
because $\E$ only depends on up to 3 derivatives with respect to $R$, and
because $f_0(R)$ is a three times differentiable function whose derivatives vanish up to third order at $R=0$, it is possible to slightly translate $f_0(R)$ to the right and
modify so that the new $f_0(R)$ is compactly supported away from $R=0$, smooth, and still has $\E<0$. 

(i) If $\um = 0$, then clearly $a=0$. Then, if $\lambda < \lc$, condition~\eqref{condition11} is obviously satisfied.

(ii) In this case, we must necessarily have $\um \neq 0$. We can label the rotational symmetries so that $m_I \neq 0$ for $0< I \le j$ and $m_I = 0$ for $j< I \le n$. We then view $\eB$ as a compact manifold with 
an action of ${\rm U}(1)^j$ corresponding to the first $j$ rotational symmetries. 
Since $\cA$ is invariant under ${\rm U}(1)^j$, the eigenspaces for each fixed eigenvalue $\lambda$ 
can be decomposed into irreducible representations of this group. Since the latter are labelled by $\um \in \ZZ^j$, we may thus decompose 
\ben
L^2(\eB, \bbE_2; L^2 \D vol_\mu)  \cong \bigoplus_{\um \in \ZZ^j, \lambda \in {\rm spec}(\cA_{\um})} V_{\um, \lambda} \ ,  
\een
where the eigenspace $V_{\lambda, \um}$ is a subspace of the space $C^\infty(\eB,\bbE_2)^{\um}$ of symmetric, trace-free rank 
two tensors with angular dependence $e^{-i\um \cdot \underline{\phi}}$. 
We denote by $\cA_{\um}$ the restriction of $\cA$ to $C^\infty(\eB,\bbE_2)^{\um}$ for our fixed $\um$. 
For any function $f \in C_0^\infty(\RR)$, we can define $f(\cA_{\um})$ as an operator acting on the subspace 
$C^\infty(\eB, \bbE_2)^{\um}$ 
via the spectral theorem. Standard results imply that $f(\cA_{\um})$ has a smooth kernel and is a trace-class operator, i.e. that 
\ben
{\rm tr} f(\cA_{\um}) = \sum_{\lambda \in {\rm spec}(\cA_{\um})} {\rm dim}(V_{\um, \lambda}) \ f(\lambda) < \infty \ .  
\een
We wish to compute this trace for very large values of the ``magnetic quantum numbers'', $\um$. For this, 
we introduce a parameter\footnote{This notation is meant to be suggestive and does not indicate that 
we want to quantize our metric perturbation!} $\hbar > 0$ such that $1/\ \hbar \in \NN$, we rescale $\um \to \um/\ \hbar$ (so that also $a \to a/ \ \hbar$), 
and we consider the operator $\hbar^2 \cA_{\um/\hbar}$. Since the limit $\hbar \to 0$ corresponds to a semi-classical limit,  it is plausible that 
the trace ${\rm tr} f(\ \hbar^2 \cA_{\um/\hbar})$ can be evaluated by semi-classical methods.  Precisely such an analysis has been carried out in \cite{guillemin}. 
To state the relevant result, we first introduce the ``semi-classical principal symbol'' of the operator $\hbar^2 \cA_{\um/\hbar}$, given by 
replacing $i\hbar \partial/\partial x^A$ in the ordinary symbol by $\xi_A$ (where $\xi \in T_x^* \eB$), and then setting $\ \ \hbar = 0$. In the present case, the 
``semi-classical principal symbol'' is ${\frak a}_0(x,\xi) id_{\bbE_2}$, where
\ben
{\frak a}_0(x,\xi) = L(x)^2 \mu^{AB}(x) \xi_A \xi_B - (k^I m_I)^2 \ .
\een
Let us also define, for each $E \in \RR$, the set
\ben
{\mathcal S}(E, \um) = \{ (x, \xi) \in T^* \eB \mid 
{\frak a}_0(x,\xi) \le E \ , \ 
i_{\partial/\partial \phi^I} \xi  = m_I \ , I = 1, \dots, j \}/ {\rm U}(1)^j \ . 
\een
If we view ${\frak a}_0$ as a ``Hamiltonian'', then this set is the part of phase space with energy less than or equal to $E$ and
angular momenta $m_I, I = 1, \dots, j$, divided out by the rotational symmetries. On $T^* \eB$, the symplectic form is defined as usual by $\omega = \D x^A \wedge \D \xi_A$. It can be seen that
$\omega$ induces a 2-form on ${\mathcal S}(E, \um)$, which we denote by the same symbol. We then define
\ben
\nu_{\um}(E) = \int_{{\mathcal S}(E, \um)} \underbrace{\omega \wedge \dots \wedge \omega}_{(d-2-j) \ {\rm times}} \ge 0 \ . 
\een
Using that $m_I \neq 0$ for $0<I \le j$ (which corresponds to their requirement that $\um$ be a ``regular weight'' of ${\rm U}(1)^j$), \cite{guillemin} show\footnote{The paper~\cite{guillemin}
considers ${\rm U}(1)^j$ invariant self-adjoint operators on compact manifolds under certain restrictions on the 
action of  ${\rm U}(1)^j$. The situation considered by~\cite{guillemin} is more general than that encountered here, because these authors allow 
the presence of points with discrete isotropy subgroup, which are absent in our case. Such points give additional terms the asymptotic 
expansion.}
\ben\label{96}
{\rm tr} f(\ \hbar^2 \cA_{\um/\hbar}) = {\rm dim}(\bbE_2) \ (2\pi \ \hbar)^{j-d+2}  \  \int_{-\infty}^{\infty} f(E) \ \D \nu_{\um}(E)  + O(\ \hbar^{j-d+3}) \ ,  
\een
as $\hbar \to 0$. 
(Here ${\rm dim}(\bbE_2)=\frac{1}{2}d(d-3)$ is the dimension of the space of $(d-2)$-dimensional, symmetric, trace-free rank 2 tensors\footnote{This factor is not 
present in~\cite{guillemin}, because these authors deal with scalar operators. The generalization to operators in a vector bundle with 
diagonal leading semiclassical symbol is straightforward.}.)
Suppose now that $L^2 g^{IJ} m_I m_J < (k^I m_I)^2$ somewhere on $\eB$. Within the set $\mathcal{S}(E, \um)$ we have 
by definition $\xi_I = m_I$. It then follows from $\mu^{IJ} = g^{IJ}$ and the definition of the semiclassical principal
symbol that $\partial\nu_{\um}(-E_0)/\partial E$ is non-zero (and positive) for  any sufficiently small $E_0 > 0$. If we now take a $f\ge 0$ which is a peak  
supported in $[-3E_0/2, -E_0/2]$, and if we use that $j < d-2$, then~\eqref{96} shows that $\hbar^2 \cA_{\um/\hbar}$ has 
an eigenvalue in $[-3E_0/2, -E_0/2]$. Hence $\cA_{\um/\hbar}$ has a negative eigenvalue $\lambda(\ \hbar) \in [-3 \ \hbar^{-2}E_0/2, -E_0 \ \hbar^{-2}/2]$. 
Since $a = \um \cdot \uk/ \ \hbar$ holds for the rescaled magnetic quantum numbers, it follows 
 that $(\lambda + \tfrac{1}{4})(\lambda^2 +2a^2 \lambda + a^4 - 9a^2 + \tfrac{7}{2}) \sim \lambda a^4 < 0$
for sufficiently small $\hbar$ and sufficiently small $E_0>0$. Thus, by the first part of the theorem, there exists a perturbation such that $\E < 0$. \qed

\subsection{Electromagnetic sector}\label{elsec}

A similar analysis is possible in the case of electromagnetic perturbations. The Hertz potential is in this case
\ben
U^a = U^A \left(  \frac{\partial}{\partial x^A} \right)^a
\een
and it must satisfy $(\cO^* U)^a = 0$, where $\cO$ was defined in eq.~\eqref{teukolsky2}.
The separation ansatz is now
\ben\label{Udef1}
U^{A} = \psi \cdot Y^{A}  \ ,
\een
where $Y = Y^{A}(x^B) \partial_A \in C^\infty(\eB, \bbE_1), \bbE_1=T\eB$ has angular dependence $e^{-i\um \cdot \underline{\phi}}$ [see eq.~\eqref{cmdef}], and where $\psi=\psi(R,T)$. Inserting these definitions and using also~\eqref{GNC}, one finds, just as in the gravitation case, that for such $U$, the action of $\cO^*$  becomes 
$\cO^* U = (\hat D^2 - q^2 + \cA)U$. In the electromagnetic case $q = a+ib, a = \uk \cdot \um, b=1$, and $\cA$ is 
$\cA$ is now 
\ben\label{Adef1}
\begin{split}
(\cA Y)_A &= -L^{-2} \bD^B (L^4 \bD_B Y^A) + (2 - a^2 - \tfrac{5}{4L^2} k_B k^B - \tfrac{d-6}{2}\Lambda L^2) Y_A \\
& + L^2  (\cR_{AB} - \tfrac{1}{2} \mu_{AB} \cR) Y^B + \Big( -\bD_{[A} k_{B]} + 2(k_{[A} - 2 L \bD_{[A} L) \bD_{B]} - 2L^{-1} \bD_{[A} k_{B]}  \Big) Y^B \ .
\end{split}
\een
$\cA$ is elliptic and self-adjoint with respect to the inner product on $(\eB, \mu)$ given by \eqref{scal}. 
The following lemma parallels lemma~\ref{lemsymp} in the gravitational case:
\begin{lemma}
Let $Y \in C^\infty(\eB, \bbE_1)^{\um}$ such that $\cA Y = \lambda Y$, $\| Y\|_{\eB} =1,$ 
let $\psi_1, \psi_2$ be two (complex) solutions to the equation~\eqref{master1}. Let $U_1, U_2$
be the corresponding (complex) Hertz-potential as in eq.~\eqref{Udef1}, and let $A_1, A_2$
be the corresponding (complex) perturbations as in eq.~\eqref{hertz1}. Then the symplectic form is\footnote{For complex perturbations, we continue $W$ {\em anti-} linearly in the
first entry.}
\ben
W(\Sigma, A_1, A_2) =  \int_{\hat \Sigma} \hat \star \hat w (\psi_1, \psi_2)\ ,
\een
where the conserved current $\hat w$ on $AdS_2$ is given up to a total divergence (i.e. up to changing $\hat \star \hat w$ by an exact 1-form)
by
\bena
8\pi \ \hat w &=& (- R^{-1} \partial_T + R \partial_R+ ia) \bar \psi_1  (\D + ia R \D T) (- R^{-1} \partial_T + R \partial_R - ia) \psi_2 +
\bar \psi_1  (\D + ia R \D T) \psi_2 - \non\\
&& [\bar \psi_1 (R^{-1} \partial_T + R \partial_R + ia) \psi_2 - ia \bar \psi_1 \psi_2 ](R \D T + R^{-1} \D R ) - \overline{(\psi_1 \leftrightarrow \psi_2)} \ ,
\eena
where as before, $a=\um \cdot \uk$.
\end{lemma}

The proof is similar to that given in the gravitational case. \qed

\medskip
\noindent
We can now write the canonical energy~\eqref{Edef1} in terms of $\psi$. We take a perturbation $A_a$
given in terms of a Hertz-potential $U^a$ as in eq.~\eqref{hertz1}. For $U^a$, we make
the separation of variables ansatz~\eqref{Udef1} in terms of some $\psi$ with compact support
on $\hat \Sigma = \{T=0,R>0\} = \RR_+$. Recalling that $K=\partial/\partial T$ in Poincar\'e coordinates, we get
from the previous lemma
\ben
\E = W(\Sigma; A, \pounds_K A) =  \int_{\hat \Sigma} \hat \star \hat w(\psi, \tfrac{\partial}{\partial T} \psi)  \ .
\een
When we evaluate $\E$, we may again use eq.~\eqref{master1} (this time with $q=i+a$) in order to eliminate terms containing more than one $T$-derivative. Thus, $\E$ becomes
a quadratic form $\E(f_0, f_1)$ of the initial data~\eqref{initial} at $T=0$.
The resulting formula is rather long and given in the appendix~\ref{appA2}. It simplifies somewhat for initial data
having $f_1 = 0$ and $f_0$ real valued. A calculation reveals that, in this case
\ben
\E = \frac{1}{8\pi}\int_{-\infty}^\infty \bigg\{
\bigg( \frac{\D^2 f_0}{\D y^2} + \frac{\D f_0}{\D y} - \lambda f_0 \bigg)^2 +
\bigg(  \frac{\D^2 f_0}{\D y^2} \bigg)^2 
+ (\lambda+a^2) \bigg( \frac{\D f_0}{\D y} \bigg)^2
+ \lambda a^2 f_0^2
\bigg\} e^{y} \D y \ . \non
\een
Here, $a = \um \cdot \uk$ as before and $y = \log R$. We substitute the variational ansatz~\eqref{ansatz1} for $f_0, f_1$, taking any $N \ge 2$. A lengthy calculation shows that, for this choice
\ben\label{con1}
\E = \frac{1}{8\pi}(\lambda+\tfrac{1}{4})(\lambda + \tfrac{1}{2} + a^2) \, \log \eps^{-1}  + O(1) \ ,
\een
where $O(1)$ stands for terms that do not diverge as $\epsilon \to 0^+$. A relatively simple form of $\E$ is also found 
for initial data having $f_0=0$ and $f_1$ real valued. In this case:
\ben
\E = \frac{1}{8\pi}\int_{-\infty}^\infty \bigg\{
\bigg( \frac{\D f_1}{\D y} - 2 f_1 \bigg)^2 +
\bigg(  \frac{\D f_1}{\D y} - f_1 \bigg)^2 + (\lambda+4a^2-2) f_1^2
\bigg\} e^{-y} \D y.
\een
We substitute the variational ansatz~\eqref{ansatz1} with $f_1 \leftrightarrow f_0$ and $R \leftrightarrow 1/R$. A calculation shows that, for this choice
\ben\label{con2}
\E = \frac{1}{8\pi}(\lambda + \tfrac{1}{2} + a^2) \, \log \eps^{-1}  + O(1) \ .
\een
The expressions~\eqref{con1},~\eqref{con2} are now used to show the following theorem: \\

\begin{thm}\label{el1thm}
Let $\lambda$ be an eigenvalue of the operator $\cA$ given by~\eqref{Adef1} acting on tensors in $C^\infty(\eB, \bbE_1)^{\um}$ such 
that 
\ben\label{condition11}
(\lambda+\tfrac{1}{4})(\lambda + \tfrac{1}{2} + a^2)<0 \quad \text{or} \quad (\lambda + \tfrac{1}{2} + a^2) < 0
 \ , 
\een 
where $a = \uk \cdot \um$. Then there exists a perturbation  such that $\E < 0$ on the NH geometry whose initial data
are compactly supported on $\Sigma = \{T=0\}$.

This is in particular the case if
\begin{enumerate}
\item[(i)] $\um=0$ and if the lowest eigenvalue $\lambda$ of $\cA$ acting tensors invariant under ${\rm U}(1)^n$ satisfies $\lambda < \lc, \lambda \neq -\frac{1}{2}$, or if 
\item[(ii)] $L^2 g^{IJ} m_I m_J < (k^I m_I)^2$ holds for some $\um \in \ZZ^n$, somewhere on $\eB$. 
\end{enumerate}
\end{thm}

{\em Proof:} The argument is exactly the same as in the case of gravitational perturbations. In particular:

 (i) If $a = 0$ and if $-\half < \lambda < \lc$, then for $\eps > 0$ and sufficiently small, 
we get $\E<0$ from~\eqref{con1}, whereas for $\lambda < - \half$, we get $\E<0$ from~\eqref{con2}. \\

 (ii) Alternatively, suppose $L^2 g^{IJ} m_I m_J < (k^I m_I)^2$ holds for some  $\um \in \ZZ^n$ somewhere on $\eB$.
 We perform the same rescaling trick as in the gravitational case, noting that the leading semiclassical symbol of $\cA$ is again
 ${\frak a}_0(x,\xi) id_{\bbE_1}$. \qed

\section{Construction of a perturbation with $\E<0$ in the extremal BH geometry} \label{sec4}

\subsection{Outline of the construction}\label{outline}

In theorem \ref{el0thm} we have identified cases (depending generically on the properties of the operator $\cA$) in which there is a gravitational 
perturbation $\gamma_{ab}$ of the form \eqref{hertz}, with Hertz potential $U^{ab}$ as in~\eqref{Udef}, which: (i) is of compact support on the Cauchy surface $\Sigma=\{T=0\}$, (ii) satisfies the linearized Einstein equations~\eqref{einstein}, 
and (iii) has $\E<0$ in the {\em NH spacetime}. Starting from such a perturbation, we will construct in this section a perturbation of the corresponding {\em BH spacetime}
which is of compact support on $\Sigma$, which satisfies the perturbed Einstein
equations and still has $\E<0$. This will lead to the main results of this 
paper given in theorem~\ref{thm2} for $\Lambda=0$, assumed from now on. A simple extension to asymptotically $AdS$ solutions ($\Lambda<0$) will give theorem~\ref{thm3}.

We repeat that 
$\gamma_{ab}$ as given by thm.~\ref{el0thm} is, by construction, a solution to the linearized Einstein equations~\eqref{einstein} on the {\em NH background}, but of course {\em not} on
the {\em BH background}, $(\cL \gamma)_{ab}  \neq 0$, where from now on and
in the following $g_{ab}, \nabla_a$ refer to the {\em BH background}, and
$\cL$ in this equation is the linearized Einstein operator of the {\em BH background}, see eq.~\eqref{einstein}.
We will construct the desired perturbation of the BH background in two steps:

\begin{enumerate}

\item We identify tensor fields on the NH geometry (in particular $\gamma_{ab}$) with tensor fields in the BH geometry
by identifying points in both spacetimes (near $\eH^\pm$) if they carry the same
Gaussian null coordinates. Under this identification the slice $\Sigma = \{T=0\}$
in the NH geometry corresponds via eqs.~\eqref{coordinates1} to a slice
$\Sigma$ in the BH spacetime ``running down the throat''.

\item We then apply the scaling isometry $\phi_\epsilon$ [see~\eqref{phieps}]  to $\gamma_{ab}$, and define, for small $\eps>0$
\ben\label{pert1}
\gamma_{ab}(\eps) \equiv \frac{1}{\sqrt{\eps}} \phi^*_{\eps} \gamma_{ab} \ .
\een
(This $\epsilon$ is a new small parameter having nothing to do conceptually with the 
parameter $\epsilon$ appearing in the constructions leading to thm.~\ref{el0thm}!)
Since $\phi_{\eps}$ is an isometry of the NH geometry, $\gamma_{ab}(\eps)$ is 
a new solution to the linearized Einstein equations on the NH geometry, having 
compactly supported initial data on $\Sigma$. The support ``moves down the throat'' as $\eps \to 0$. 
Moreover, since $\phi^*_\eps K = \eps K$, the scaling by $1/\sqrt{\epsilon}$ of our perturbation
\eqref{pert1} ensures that the canonical energy remains unchanged\footnote{It is important to note that the scaling by $1/\sqrt{\epsilon}$ is just a convenient choice in order to simplify 
our discussion. It plays no fundamental role as the equations for the perturbations~\eqref{einstein} are {\em linear}.}, 
\ben
\E(\Sigma,\gamma(\eps)) = \E(\Sigma,\gamma) < 0 \ .
\een

\item Let 
\ben\label{nhpert}
(\delta h_{ij}(\eps), \delta p^{ij}(\eps)) \equiv \text{initial data of $\gamma_{ab}(\eps)$
 on $\Sigma$.}
\een
By construction, these $\eps$-dependent initial data (we omit the reference to $\eps$ in the following) 
satisfy the constraints of the NH-spacetime, but not the BH-spacetime. But we can add 
a small correction (for small $\eps$), such that the modified initial data are
still of compact support, satisfy the constraints of the BH-spacetime [under the identification in 1)], and
still have negative canonical energy in the BH-spacetime. As described in section~\ref{sec1}, 
the time-evolution of these modified initial data in the BH spacetime
cannot settle down to a perturbation that is pure gauge or represents a
perturbation to another stationary black hole in the family. Thus, such a black hole 
is linearly unstable.
\end{enumerate}

\subsection{Correcting the variational ansatz for initial data}\label{correctingID}

We now turn to a more precise explanation of this strategy. Steps 1) and
2) do not require further explanation, but step 3) is of a rather technical nature and
needs to be discussed. Generally speaking, we have the following problem. We have
an ansatz $(\delta h_{ij}, \delta p^{ij})$ -- in our case given by eq.~\eqref{nhpert} -- for the initial data having support in a bounded set $A_0 \Subset \Sigma$
(in the BH spacetime). The linearized constraints are not
satisfied. We would like to modify our ansatz by adding a correction so that the new initial data
are compactly supported in a some (possibly slightly larger) bounded set $A \Supset A_0$, and
solve the linearized constraints. There is a well-known general method for achieving just this, developed
in~\cite{Corvino}, and  also in~\cite{Chrusciel}\footnote{We remark that these references also deal with the 
full non-linear constraint equations.}. We now describe this method. We follow with
minor modifications the original references but pay special attention to the key question by how
much the original ansatz has to be modified depending on how much it violated the linearized constraints.

The constraints are (assuming $\Lambda=0$ from now on):
\ben
{\bf C}
 =   h^\half
  \left(
  \begin{matrix}
   -Scal_h + h^{-1} p_{ij} p^{ij} - \frac{1}{d-2} h^{-1} p^2 \\
    -2D_j(h^{-\half} p^{ij})
    \end{matrix}
\right) = 0  \ .
\label{constraintformula}
\een
The first entry is the Hamiltonian constraint, and the second entry is the vector constraint. We
will generally use boldface letters for a tuple consisting of a scalar (or density) on $\Sigma$, and
a vector (or density) on $\Sigma$.\footnote{In terms of the Einstein tensor $G_{ab}$ and unit normal $\nu^a$,
the Hamiltonian constraint is given by $= G_{ab} \nu^a \nu^b + \Lambda$, whereas the vector
constraint is given as $= G_{cb} \nu^b h_a{}^c$.}
The linearized constraints $\delta {\bf C}$ may be viewed as the
result of acting on $(\delta h_{ij}, \delta p^{ij})$ by a linear operator, $\cC$.
It is explicitly given by
\ben\label{delc}
 \cC \left(
\begin{matrix}
\delta h_{ij}\\
\delta p^{ij}
\end{matrix}
\right) =
\left(
\begin{matrix}
h^\half(D^i D_i \delta h_j{}^j - D^i D^j \delta h_{ij} + Ric(h)^{ij} \delta h_{ij})+\\
h^{-\half}(-\delta h_k{}^k p^{ij} p_{ij}
+ 2p_{ij} \delta p^{ij} + 2p^{ik} p^j{}_i \delta h_{jk}+\\
\frac{1}{d-2} p^k{}_k p^l{}_l \delta h^i{}_i
-\frac{2}{d-2} p^i{}_i \delta p^j{}_j - \frac{2}{d-2}\delta h_{ij} p^{ij} p_k{}^k) \\
\\
-2h^\half D^j(h^{-\half} \delta p_{ij}) +  D_i \delta h_{kj} p^{kj} -
2D_k \delta h_{ij} p^{jk}
\end{matrix}
\right) \ .
\een
Since $\cC$ is a differential operator that
maps the pair $(\delta h_{ij}, \delta p^{ij})$ consisting of a symmetric tensor, $\delta h_{ij}$,
and a symmetric tensor density, $\delta p^{ij}$, on $\Sigma$ into a pair $(u,X_j)$ consisting of a scalar density
and dual vector density on $\Sigma$, its adjoint
differential operator, $\cC^*$, maps a pair $\bX=(u,X_j)$ consisting of
a scalar and vector field on $\Sigma$ into a pair $(\delta h^{ij}, \delta p_{ij})$ consisting of a symmetric tensor density and symmetric tensor on $\Sigma$.
One can straightforwardly calculate that $\cC^*$ is given by
\ben\label{delc*}
 \cC^*  \left(
\begin{matrix}
u\\
X_j
\end{matrix}
\right) = \left(
\begin{matrix}
h^\half(-(D^k D_k u) h^{ij} + D^i D^j u  + Ric(h)_{ij} u)+\\
h^{-\half}( - h^{ij} p^{kl} p_{kl} u + 2 p^{(i}{}_k p^{j)k} u + \frac{1}{d-2} h^{ij} p^k{}_k p^l{}_l u\\
-\frac{2}{d-2}p^{ij}p^k{}_k u - p^{ij} D_k X^k + 2D_k X^{(i} p^{j)k})\\
\\
h^{-\half}(2p_{ij}u
- \frac{2}{d-2} h_{ij} p^k{}_k u)
+\pounds_X h_{ij}
\end{matrix}
\right)
\een
The idea is to make particular ansatz for the correction to $(\delta h_{ij}, \delta p^{ij})$
in order to satisfy the linearized constraints.
Let $s: A \to \RR$ be a function $1 \ge s>0$ such that near the boundary $\partial A$, we have
\ben
s(x) = {\rm dist}_h(x, \partial A) \ ,
\een
where we mean the geodesic distance relative to the metric $h$ on $\Sigma$. We also ask that $s(x) = 1$ in $A_0 \Subset A$.
The ansatz is:
\ben\label{ansatz}
\left(
\begin{matrix}
\delta \tilde h_{ij}\\
\delta \tilde p^{ij}
\end{matrix}
\right) \equiv
\left(
\begin{matrix}
\delta h_{ij}\\
\delta p^{ij}
\end{matrix}
\right)
-
e^{-2/s^\alpha} \left(
\begin{matrix}
s^{4\alpha+4} & 0\\
0 & s^{2\alpha+2}
\end{matrix}
\right)
 \cC^* \left(
\begin{matrix}
u \\
X_j
\end{matrix}
\right)
\een
Cutoff functions involving $s$ have been inserted because we hope to extend
the solution by $0$ across the boundary $\partial A$ in a smooth way. The tensors $\bX \equiv (u, X_j)$ are to be determined.
 The matrix of
 cutoff functions on the right side will appear often,
so we introduce the shorthand:
\ben\label{matrixm}
\Phi \equiv e^{-1/s^\alpha} \left(
\begin{matrix}
s^{2\alpha+2} & 0\\
0 & s^{\alpha+1}
\end{matrix}
\right).
\een
Our ansatz can then be written more compactly as
\ben\label{corrected}
\left(
\begin{matrix}
\delta \tilde h\\
\delta \tilde p
\end{matrix}
\right) =
\left(
\begin{matrix}
\delta h\\
\delta p
\end{matrix}
\right)
-
\Phi^2 \cC^* \bX \ .
\een
We want $(\delta \tilde h_{ij}, \delta \tilde p^{ij})$ to satisfy the
linearized constraints of the BH background. Acting with $\cC$ shows that
$\bX$ must satisfy the equation:
\ben\label{xeps}
\cC \Phi^2 \cC^* \bX  = {\bf f} \ ,
\een
where ${\bf f} \equiv \delta \bC = \cC(\delta h, \delta p)$ is the violation of the linearized constraints of our NH
ansatz $(\delta h_{ij}, \delta p^{ij})$.

The question is of course whether~\eqref{xeps} has a suitable solution at all, which is far from obvious.
In order to construct such a solution, and to control its properties, one uses the technique of
weighted Sobolev spaces. We begin by defining the weighted Sobolev norms
\ben
\| u \|_{W^{p,k,\alpha}}= \left( \sum_{n=0}^k \int_A |D^n u|^p \, s^{pn(\alpha+1)} e^{-2/ s^\alpha} \, \D vol_A \right)^{1/p} \ 
\een
on $C^\infty_0(A)$ tensor fields $u$ on $A \subset \Sigma$. 
We let $W^{p,k,\alpha}_0(A)$ be the completion
of the space of such tensor fields under this norm. Since we will mostly consider $p=2$, and sometimes $\alpha=0$, we introduce the notations
$H^{k,\alpha} = W^{2,k,\alpha}, L^{2,\alpha} = H^{0,\alpha}, L^2 = L^{2,0}$. We also use the notation $H^k$ for the ordinary Sobolev spaces and norms
without any weights.  Our weights differ slightly from those used by~\cite{Corvino}.
The following lemma is the key to prove the existence of a weak solution to
\eqref{xeps}:

\begin{lemma}\label{friedrichs} (Generalized weighted Friedrichs-Poincar\' e inequality)
For sufficiently large $\alpha$, there is a constant $c=c(\alpha,A)$ such that
\ben\label{fp}
c \left\| \Phi \cC^* \bX
\right\|_{L^{2}}
\ge \|\bX - P_A \bX \|_{H^{2,\alpha} \oplus H^{1,\alpha}} \ ,
\een
for any tensor field ${\bf X} \in H^{2,\alpha}_0(A) \oplus H^{1,\alpha}_0(A)$. Here $P_A$ is the orthogonal projector (in $L^{2,\alpha}(A)$) onto the
subspace $\frak k$ spanned by the KVF's, i.e. if ${\bf Y}_i$ is a basis of Killing vector fields on $\M$ that has been orthonormalized (in $L^{2,\alpha}$) via the
Gram-Schmidt process, we have
\ben
P_A \bX = \sum_i {\bf Y}_i  ({\bf Y}_i, \bX)_{L^{2,\alpha}} \  \ .
\een
\end{lemma}
The proof of this lemma is given in appendix~\ref{appD} using a method which is somewhat different from~\cite{Corvino,Chrusciel}.
Using this key lemma, one can show existence:

\begin{lemma}\label{lem4}
Let ${\bf f} \in C^\infty_0(A)$ with support in $A_0 \Subset A$.
Then there exists a solution $\bX$ to~\eqref{xeps} which is in $H_0^{2,\alpha}(A) \oplus H_0^{1,\alpha}(A)$ and which in fact
additionally satisfies for all $k=0,1,2,\dots$
\ben\label{regularity}
\int_A s^{2k\beta} |D^k (\Phi \cC^* \bX)|^2 \D vol_A \le c \| {\bf f} \|_{H^k \oplus H^k}^2
\een
for a sufficiently large $\beta>0$, and a constant $c=c(A,\alpha,\beta, k)$.
\end{lemma}

\medskip
\noindent
{\bf Remarks:} a) Note that in our definition of the corrected initial
data~\eqref{corrected}, we have on the right side the expression $\Phi^2 \cC^* \bX$, i.e. we have the {\em square} of $\Phi$.
Since $\Phi$ is a multiplication operator involving the exponential cutoff factor $e^{-1/s^\alpha}$ [cf.~\eqref{matrixm}], it follows from
the estimate in the previous lemma (because $s^{-N} e^{-1/s^\alpha} \to 0$ for any $N$ when $s\to 0$) that
$s^{-N} \Phi^2 \cC^* \bX$ is in each (unweighted) Sobolev space of arbitrary order $k$ for any $N$. Thus,
by the usual Sobolev embedding theorem, $C^\infty(\bar A) \subset \cap_k H^k(A)$, it follows
that $\Phi^2 \cC^* \bX$ is smooth up to and including the boundary $\partial A$, and that it can in fact be
smoothly extended by $0$ across $\partial A$. Thus, the corrected initial data~\eqref{corrected} are smooth up to and
including the boundary $\partial A$ and can be extended by $0$ across $\partial A$.

b) Below, we will consider applying this result to an annular domain of the form $A = \{ x \in \Sigma \mid y_ 0 - \log \eps < y(x) < y_1 - \log \eps \}$.
We claim that for $\eps \to 0$ (i.e., for the annular domain going down the throat), the constant $c=c(\alpha,\beta, k,A)$ may be chosen to be independent of
$\eps$. This is in essence a direct consequence of the fact that the background $h_{ij}$ and $\chi_{ij}$ (hence also $p^{ij}$)
 are nearly translation invariant under shifts of $y$ in
the throat $(y \to -\infty$), see eq.~\eqref{background}, and follows by inspecting the constants in~\eqref{fp} and~\eqref{regularity}.

\medskip
\noindent
{\em Proof of Lemma~\ref{lem4}}: The lemma is demonstrated using standard tools from PDE-theory for elliptic operators. The
only non-standard feature is the presence of the weight factors, 
and the fact that the operator in question $\cC \Phi^2 \cC^*$ is a matrix of operators of mixed order (up to order 4), see~\cite{Nirenberg,Schechter}
for the corresponding classical results. Existence
is proved with the help of the weighted Poincar\' e-Friedrichs inequality. One considers the weak formulation of the PDE problem~\eqref{xeps}
which consists in finding an element $\bX \in H_0^{2,\alpha}(A) \oplus H_0^{1,\alpha}(A)$ such that
\ben\label{weak}
B[\bX, {\bf Y}] = F[{\bf Y}] \ , \qquad \text{for all ${\bf Y} \in H_0^{2,\alpha}(A) \oplus H_0^{1,\alpha}(A)$,}
\een
where the bilinear form $B$ is $B[\bX, {\bf Y}] = (\Phi \cC^* \bX, \Phi \cC^* {\bf Y} )_{L^2}$ and where the functional $F$ is
$F[\bX] = ({\bf f}, \bX)_{L^2}$. This weak formulation is obtained as usual by formally multipling the PDE with $\bf Y$,
integrating over $A$, and performing (formally) partial integrations to bring the operator $\cC$ to the other factor as $\cC^*$.
The subscript ``0'' in our choice of Sobolev space anticipates/reflects a choice of ``boundary conditions'', 
and the weight $\alpha$ in the Sobolev space corresponds that in $\Phi$, see~\eqref{matrixm}.
Note that if $\bf Y$ corresponds to a KVF, then, since $\cC^* \bf Y = 0$, it follows that $F[{\bf Y}] = 0$ (because $\bf f$ is in the image of 
$\cC$), and it also follows evidently that 
$B[\bX, {\bf Y}]=0$. Thus, it is sufficient to satisfy the above identity for all ${\bf Y} \in H_0^{2,\alpha}(A) \oplus H_0^{1,\alpha}(A)$
that are orthogonal to the span $\frak k$ of KVF's. On that subspace the quadratic form is bounded from below by a
positive multiple of the norm by the Poincar\' e-Friedrichs inequality, whereas $F$ is bounded in the $H_0^{2,\alpha}(A) \oplus H_0^{1,\alpha}(A)$-norm\footnote{Here it is used
that $\bf f$ is supported away from the boundary $\partial A$, so that the weight factors do not play a role.}.
Existence of a weak solution then follows from the standard Lax-Milgram theorem (basically the Riesz-representation theorem, see e.g. \cite{Evans}), and
one has, in fact,
\ben\label{regu}
\| \bX \|_{H^{2,\alpha} \oplus H^{1,\alpha}} \le c_0 \| {\bf f} \|_{L^2 \oplus L^2} \ ,
\een
for some constant $c_0=c_0(\alpha, A)$.

It remains to demonstrate the higher regularity estimates~\eqref{regularity}. Here, we proceed by the standard method of finite difference quotients, combined
with the Poincar\' e-Friedrichs inequality. First, we slice $A$ near the boundary
into onion skins $O_n = A_{2^{-n+1}} \setminus A_{2^{-n-1}}$, where each set $A_\delta$ is characterized by the condition that $s(x)  < \delta$. Next, we choose test functions $\zeta_n \ge 0$ having support in $O_n$, and such that $\sum \zeta_n = 1$. We may assume that $|D^k \zeta_n| \le c_1 \ 2^{nk}$
for some constants $c_1=c_1(k,A)$, and we shall pretend, in order avoid a more cumbersome
notation, that the support of each $\zeta_n$ is contained in a single coordinate chart. This could always be achieved by subdividing $O_n$ further into a fixed (independent of $n$) number of subregions. Points in $O_n$ are then identified with their coordinate
vectors in $ \RR^{d-1}$. The finite difference operator in the $j$-th coordinate direction is defined as
\ben
\Delta^\delta_j u(x) = \frac{u(x) - u(x + \delta e_j)}{\delta} \ .
\een
It satisfies standard properties riminicsient of the `Leibniz rule' and a `partial
integration rule'. One can also establish that $\| \Delta^\delta_j u \|_{L^2} \le c_2 \| \partial_j u\|_{L^2}$ for some constant $c_2$ and sufficiently small $\delta$. Conversely, if we know that $\| \Delta^\delta_j u \|_{L^2}$ is uniformly bounded for sufficiently small $\delta$, then $\| \partial_j u \|_{L^2} \le \limsup_\delta \| \Delta^\delta_j u \|_{L^2}$, i.e. $u$ has a square-integrable weak derivative, see 
sec.~5.8 of~\cite{Evans} for details and proofs. After these preliminaries, we test~\eqref{weak} with the test-function
\ben
{\bf Y} = \Delta^{-\delta}_j (\zeta_n^2 \Delta_j^\delta \bX) \ ,  \quad 0 < \delta \ll 1
\een
so that
\ben\label{test}
F[\Delta^{-\delta}_j (\zeta_n^2 \Delta_j^\delta \bX)] =
B[\Delta^{-\delta}_j (\zeta_n^2 \Delta_j^\delta \bX), \bX] \ .
\een
The right side is now bounded from below as explained e.g. in sec.~6.3.2 of~\cite{Evans},
where the only differences in our case are the presence of weights in $B$, and the fact
that $B$ contains higher derivatives. As in the standard case, the basic idea is
simply to `move $\Delta^\delta_j$ to the other factor' using the `partial integration' and `Leibniz' rules for finite difference operators. One finds, for sufficiently small $\delta>0$:
\ben
\begin{split}
{\rm r.h.s.} \ge & B[\zeta_n \Delta^\delta_j \bX, \zeta_n \Delta^\delta_j \bX] \\
&- c_3 \Big\{
\delta \| \zeta_n \Delta^\delta_j \bX \|^2_{H^{2,\alpha} \oplus H^{1,\alpha}} + 2^{n(1+\alpha)} \| \bX \|_{H^{2,\alpha} \oplus H^{1,\alpha}} \| \zeta_n \Delta^\delta_j \bX \|_{H^{2,\alpha} \oplus H^{1,\alpha}}
\Big\} \ , \non
\end{split}
\een
The factor of $2^{1+\alpha}$ arises from the fact that a `partial integration'
results in a factor $\Delta^\delta_j (s^{p(1+\alpha)} e^{-2/s^\alpha})$, which,
for sufficiently small $\delta>0$ is bounded on $O_n$ by
\ben
|\Delta^\delta_j (s^{p(1+\alpha)} e^{-2/s^\alpha})| \le c_{12} 2^{n(1+\alpha)}s^{p(1+\alpha)} e^{-2/s^\alpha}.
\een
We also have factors of $2^{n}$ arising from the fact that pulling $\zeta_n$ through various derivative operators will result in $D\zeta_n, D^2 \zeta_n$,
which are bounded by a constant times $2^{n}$ respectively $2^{2n} \le 2^{n(1+\alpha)}$, choosing $\alpha>1$. Employing the `Peter-Paul' trick $2|ab| \le
a^2/\eps + \eps b^2$ on the last term (giving small weight
to the norm $b=\| \Delta^\delta_j \bX \|_{H^{2,\alpha} \oplus H^{1,\alpha}}$)
results altogether in
\ben
\begin{split}
{\rm r.h.s.}
\ge&  B[\zeta_n \Delta^\delta_j \bX, \zeta_n \Delta^\delta_j \bX]
-c_3(\delta + \eps) \| \zeta_n \Delta^\delta_j \bX \|_{H^{2,\alpha} \oplus H^{1,\alpha}}^2
-c_3 2^{2n(1+\alpha)} \| \bX \|_{H^{2,\alpha} \oplus H^{1,\alpha}}  \\
\ge& [c_4- c_3(\delta + \eps)] \| \zeta_n \Delta^\delta_j \bX \|_{H^{2,\alpha} \oplus H^{1,\alpha}}^2
-c_3 \eps^{-1} 2^{2n(1+\alpha)} \| \bX \|_{H^{2,\alpha} \oplus H^{1,\alpha}}^2 \\
\ge& [c_4 - c_3(\delta + \eps)] \| \zeta_n \Delta^\delta_j \bX \|_{H^{2,\alpha} \oplus H^{1,\alpha}}^2
-c_5 \eps^{-1} 2^{2n(1+\alpha)} \| {\bf f} \|^2_{L^2 \oplus L^2} \ ,
\end{split}
\een
applying in the second line the Friedrichs-Poincar\' e inequality (giving rise to the constant $c_4$), and in the third line the inequality~\eqref{regu}, 
combining the constants into $c_5$. We choose $\eps, \delta$
so small that $c_4 > c_3(\delta + \eps)$. Then the coefficient in front of the first term on the right side is positive. Using similar arguments, the left side of eq.~\eqref{test} is bounded by
\ben
{\rm l.h.s.} \le  c_6 \Big\{ 2^{2n(1+\alpha)} \|{\bf f}\|^2_{H^1 \oplus H^1} +  \| \bX \|_{H^{2,\alpha} \oplus H^{1,\alpha}}^2 \Big\}
\le  c_7 2^{2n(1+\alpha)} \|{\bf f}\|^2_{H^1 \oplus H^1}, 
\een
using again~\eqref{regu}. 
Combining the bounds for the left and right  sides, we find for some constant $c_8(\alpha,A)$ and sufficiently small $\delta>0$ that
\ben
\|\zeta_n \Delta_j^\delta \bX \|_{H^{2,\alpha} \oplus H^{1,\alpha}}
\le c_8 2^{n(1+\alpha)}\|{\bf f}\|_{H^1 \oplus H^1} \ ,
\een
and the same bound in fact then also holds for $\zeta_n D_j \bX$ by the properties of 
the finite difference quotients and~\eqref{regu}. Therefore
\ben
\begin{split}
\|s^\beta D \bX \|_{H^{2,\alpha} \oplus H^{1,\alpha}} =&
\| s^\beta (\sum_n \zeta_n) D \bX \|_{H^{2,\alpha} \oplus H^{1,\alpha}} \\
\le& \sum_n \| s^\beta \zeta_n D \bX \|_{H^{2,\alpha} \oplus H^{1,\alpha}}\\
\le& c_9 \sum_n 2^{-\beta n} \| \zeta_n D \bX \|_{H^{2,\alpha} \oplus H^{1,\alpha}}\\
\le& c_{10} \sum_n 2^{-\beta n} 2^{n(1+\alpha)}  \| {\bf f} \|_{H^{1} \oplus H^{1}}\\
\le& c_{11}  \| {\bf f} \|_{H^{1} \oplus H^{1}}
\end{split}
\een
assuming $\beta>1+\alpha$ in the last step. This proves the statement of the theorem
for $k=1$, because the $L^2$-norm of $s^\beta D(\Phi \cC^* \bX)$ is bounded by
a constant times the $H^{2,\alpha} \oplus H^{1,\alpha}$-norm of $s^\beta D\bX$. The case of general $k$ is treated with an induction in $k$, considering
in the $k$-th step the test-function
\ben
{\bf Y} = \prod_l^k \Delta^{-\delta}_{j_l} \left( \zeta_n^2 \left[\prod_m^k \Delta_{j_m}^\delta \right] \bX\right) \ .
\een
Since there are no new ideas need in that step, and since the details closely resemble
standard constructions as given e.g. in sec.~6.3 of~\cite{Evans}, we do not elaborate further on these constructions.
\qed


\subsection{Construction of a gravitational perturbation with $\E<0$ in the extremal BH geometry}

After these preliminaries, we turn back to the construction of the modified gravitational
perturbation from step 3) in the outline section~\ref{outline}, using the general construction from the previous subsection.
Let $\delta \bC_\eps$ be the constraints of the perturbation $\gamma_{ab}(\eps)$ [see~\eqref{nhpert}] in the BH background. They are given by
\ben
\delta {\bf C}_\eps =  \cC \left(
\begin{matrix}
\delta h_{ij} (\eps)\\
\delta p^{ij}(\eps)
\end{matrix}
\right)
=
 \left(
\begin{matrix}
\cL \gamma_{ab}(\eps) \nu^a \nu^b \\
\cL \gamma_{ab}(\eps) \nu^b h_c{}^a
\end{matrix}
\right)
\label{L}
\een
where $\cL$ is the linearized Einstein operator~\eqref{einstein} for the BH background, where $\nu^a$ 
is the normal to the Cauchy surface $\Sigma$ in the BH background, 
and where $\cC$ is the linearized constraint operator for the BH background. From now on we 
drop the reference to $\eps$ in the initial data to lighten the notation. The next lemma
tells us that $\delta \bC_\eps$ is small:

\begin{lemma}\label{prevlemma}
We have $|\partial_y^n \delta \bC_\eps | \le c \sqrt{\eps}$ on $\Sigma$, where $c=c(n)$ and 
${\rm supp} \ \delta \bC_\eps \subset A_0$. Here, $A_0 = A_0(\eps)$ is an `annular' domain of the form
\ben\label{annular}
A_0 = \{ x \in \Sigma \mid y_0 - \log \eps < y(x) < y_1 - \log \eps\}
\een
for some $y_0,y_1>0$ independent of $\eps$.
\end{lemma}

{\em Proof:} This lemma relies on the following simple facts. First, by construction the perturbation $\gamma_{ab}$ is of the form
eq.~\eqref{hertz} for a suitable Hertz potential $U^{ab}$. As a consequence,
the perturbation $\gamma_{ab}$ has the schematic
form $\gamma_{ab} = x l_a l_b + y_{(a} l_{b)} + z_{ab}$, where $y_a, z_{ab}$ are projected by $q_{ab}$. Furthermore, it follows from 
eqs.~\eqref{phieps} and~\eqref{coordinates1} that the diffeomorphism
$\phi_\eps$ acts as $(T,R) \mapsto (\eps T, R/\eps)$ or equivalently as $(T,y) \mapsto
(\eps T, y - \log \eps)$. Thus, it is just a shift in $y$, from which the support
property is immediately obvious. Then, e.g. from the explicit
expressions of the dual 1-forms of $n^a, l^a$ [cf. eq.~\eqref{nldef}]
\ben
n = e^y \D T + \D y \ , \qquad l = -e^y \D T + \D y \ ,
\een
it follows that $\phi_\eps^* l = l, \phi^*_\eps n = n$. It is not difficult to see
from these facts that $(\delta h_{ij}, \delta p^{ij})$, defined as in eq.~\eqref{nhpert},
must have coordinate expressions in $(y, x^A)$ that are of order $O(1/\sqrt{\eps})$
together with all their $(y,x^A)$-derivatives.

It also follows by construction
that the background $(h_{ij}, p^{ij})$ of the BH and NH-backgrounds as in eq.~\eqref{background} agree
up to terms of $O(e^{y})$ in the coordinates $(y,x^A)$.
for $y \to -\infty$. Substituting this information into the
definition of the linearized constraint operator on the BH background, and using
that $(\delta h_{ij}, \delta p^{ij})$ is annihilated by the linearized constraint operator
on the NH background, gives the
statement of the lemma.
\qed

With this in mind, we are now ready to state and prove the main two theorems of this paper concerning instability
criteria of extremal BH's in the asymptotically flat- respectively asymptotically $AdS$ case.

\begin{thm}\label{thm2}
Let $(\M,g)$ be an extremal MP black hole  with $\Lambda = 0$. Let $\cA$ be the elliptic operator~\eqref{Adef}
viewed as an operator on tensors $Y \in C^\infty(\eB,\bbE_2)^{\um}$ with angular dependence $e^{-i\um\cdot \underline{\phi}}$. 
 \begin{enumerate}
 \item[(i)] If $\um = \underline{0}$, and if the smallest eigenvalue of $\cA$ on $C^\infty(\eB,\bbE_2)^{\underline{0}}$ satisfies
 $\lambda < -\frac{1}{4}$, there is a perturbation
of compact support on the Cauchy surface $\Sigma$ which cannot settle down to perturbation
to another stationary black hole (or a pure gauge transformation). In other words, the black hole is
linearly unstable. 
\item[(ii)] If there is a $\um \in \ZZ^n$ such that $m_I \Omega^I = 0$ and such that $L^2 g^{IJ} m_I m_J < (k^I m_I)^2$ somewhere on $\eB$,  then the black hole is unstable (in the same sense).
 \end{enumerate}
\end{thm}
\noindent
{\bf Remarks:} 
1) It is very important to remark that this notion of instability does, by itself, not automatically imply the existence of an ``exponentially growing mode'', i.e. a linearized perturbation 
for which a suitably defined gauge invariant norm grows as $e^{|\omega|t}$ in time. In fact, one cannot even exclude a priori that the solutions $\gamma_{ab}$ identified with the 
canonical energy argument will have an oscillating behavior asymptotically of the form $e^{i\omega t}$ (in the terminology of dynamical systems, the background could still be 
``orbitally stable''). Recently, the existence of an exponentially growing mode has been related, under certain conditions, with the existence of modes for which $\E < 0$ by~\cite{Kartik}.  
It is conceivable that these results, when combined with our arguments, 
can be applied to establish that there is an exponentially growing, but this is outside the scope of the present work. \\
2) Case (i) may be called the ``generic case'', because $m_I \Omega^I = 0$ has no solutions except for a 
measure zero set of spin-parameters. When all the spin parameters  $a_I$
are equal (``cohomogeneity-1'' BHs) the lowest eigenvalue $\lambda$ of $\cA$ has been calculated analytically in~\cite{Durkee}. 
These authors also identified the cases where $\lambda < - \frac{1}{4}$ and found agreement with the conclusions of the  
numerical investigations of linear perturbations in cohomogeneity-1 black holes by~\cite{Santos}. 

Case (ii) may be called the ``resonant case''. The condition $m_I \Omega^I = 0$ reads explicitly (with summation sign written out)
\ben\label{omcond}
0 =  \sum_{I=1}^n \frac{m_I a_I}{r_+^2 + a_I^2} 
\een
for the MP solutions. The condition $L^2 g^{IJ} m_I m_J < (k^I m_I)^2$ somewhere on $\eB$ in case (ii) is seen to be satisfied for instance for ``ultra-spinning'' black holes. 
As an illustration, we take $d=6$, and we consider an ultra-spinning extremal MP black hole characterized by $a_1 \to \infty, a_2 \to r_+$. One finds, up to terms of 
order $O(a_2/a_1)$:
\ben
(k^I m_I)^2 = m_1^2 \ ,  \qquad 
L^2 g^{IJ} m_I m_J = \frac{2(1-\mu_1^2-\tfrac{1}{2}\mu_2^2)(1-\frac{1}{2}\mu_2^2)}{\mu_2^2} \ m_1^2 \ , 
\een
where the direction cosines have to satisfy $\mu_1^2 + \mu_2^2 \le 1$ in $d=6$ dimensions. It follows from these expressions that 
$L^2 g^{IJ} m_I m_J < (k^I m_I)^2$ holds e.g. when $\mu_2$ is sufficiently close to $1$, and $\mu_1$ is sufficiently close to $0$ (and 
when $a_1 \gg a_2$).  Hence, we have shown the existence of a ``resonant instability'' of ultra-spinning extremal black holes in $d=6$.\\

\medskip

{\em Proof:} (i)
Since $\lambda < \lc$, (i) of theorem~\ref{el0thm} applies and can construct a smooth  $\eps$-dependent perturbation $\gamma_{ab}(\eps)$ having $\E < 0$ in the NH-geometry,
as described in 1) and 2) in the outline subsection~\ref{outline}. 
Let $(\delta h_{ij}(\eps), \delta p^{ij}(\eps))$ be the  initial data of this
perturbation as in~\eqref{nhpert}, which are compactly supported in an annular domain of the form $A_0$~\eqref{annular} in the slice $\Sigma$ of the NH geometry
(from now on we drop the reference to $\eps$ in the initial data).
By lemma \ref{lem4} and the following remark,
there exists a solution ${\bf X}$ to~\eqref{xeps} with ${\bf f} :\equiv \delta {\bf C}_\eps$ such that
$(\delta \tilde h_{ij}, \delta \tilde p^{ij})$ defined in~\eqref{corrected} are $C^\infty_0$ tensor fields
on $\Sigma$ supported in an annular domain, called $A$, slightly larger than~\eqref{annular}. By the estimate in lemma~\ref{lem4},
the $L^2$-norms of $k$-th derivatives of $\delta \tilde h_{ij} - \delta h_{ij}$ and of $\delta \tilde p^{ij} - \delta p^{ij}$
(equal by definition to those of $\Phi^2 \cC^* \bX$) are bounded by the $H^k \oplus H^k$ Sobolev norms of ${\bf f}$,
which in turn are of order $O(\sqrt{\eps})$ by the previous lemma~\ref{prevlemma}. Thus, in this sense, our correction to the original variational
ansatz~\eqref{L} is small. The background initial data $(h_{ij}, p^{ij})$ of the NH-geometry and
the BH-geometry $(\tilde h_{ij}, \tilde p^{ij})$ are both given by~\eqref{background}
and hence differ by terms of order $O(e^y)$ for $y \to -\infty$, or in other words, by terms of order $O(\eps)$ within $A$.

Now let $\E$ be the canonical energy of $(\delta h_{ij}, \delta p^{ij})$ in the NH-geometry, and let $\tilde \E$ be the canonical energy of
$(\delta \tilde h_{ij}, \delta \tilde p^{ij})$ in the BH-geometry. Using the concrete form of $\E$ respectively $\tilde \E$ in the
NH- respectively BH-geometry given by eq.~\eqref{Eexpr}, using that $N,N^j$ are of order $O(e^y)$ (hence of order $O(\eps)$ within $A$),
using that the $H^1$ norm of $\delta \tilde h_{ij} - \delta h_{ij}$ and the $L^2$-norm  of $\delta \tilde p^{ij} - \delta p^{ij}$ is of order
$O(\sqrt{\eps})$, it
follows that $\tilde \E - \E = O(\eps^2)$. Since it is already known that $\E<0$ (independently of $\eps$), we
conclude that $\tilde \E < 0$ for sufficiently small $\eps$.

We may now appeal to the general arguments of~\cite{HW}. Pick an $\eps$ so that $\tilde \E = \tilde \E(\Sigma) < 0$
for the compactly supported perturbation $(\delta \tilde h_{ij}, \delta \tilde p^{ij})$. Let $\tilde \gamma_{ab}$ be the spacetime perturbation
on the BH-background defined by these initial data obeying the transverse-trace-free gauge condition.
By  eq.~\eqref{omcond}, the Hertz-potential $U^{ab}$  as defined through~\eqref{Udef}
is Lie-derived by $\psi = \Omega^I \partial/\partial \phi^I$ (meaning that $\pounds_\psi U_{ab} = 0$). Therefore, the perturbation $\gamma_{ab}$ 
of the NH-geometry as in~\eqref{hertz} also is Lie-derived by $\psi$. Then, since $\psi$ is tangent to $\Sigma$, also
its initial data $(\delta h_{ij}, \delta p^{ij})$ are Lie-derived by $\psi$. Furthermore, since the construction of the ``corrected''
initial data $(\delta \tilde h_{ij}, \delta \tilde p^{ij})$ is unique and only involves auxiliary data that are Lie-derived by $\psi$, the corrected initial data
as in~\eqref{corrected} are also Lie-derived by $\psi$, and hence also $\tilde \gamma_{ab}$. Thus, the flux lemma~\ref{fluxlemma} applies, and
$\tilde \E(\Sigma') \le   \tilde \E(\Sigma) < 0$ for
any later slice $\Sigma'$ (see fig.~\ref{DOC1} with $\Sigma_1 \to \Sigma, \Sigma_2 \to \Sigma'$). Hence, $\tilde \E$ cannot go to zero on an asymptotically late slice, and, therefore,
as argued in~\cite{HW}, the corresponding perturbation cannot settle down to a perturbation towards another stationary
black hole (modulo gauge). 

(ii) In this case, (ii) of theorem~\ref{el0thm} applies. The rest of the argument is as in (i). \qed

\medskip

We have a similar, but stronger, version of the theorem in the asymptotically $AdS$-case:

\begin{thm}\label{thm3}
Let $(\M,g)$ be an extremal MP black hole  with $\Lambda < 0$. Let $\cA$ be the elliptic operator~\eqref{Adef}
viewed as an operator on tensors $Y \in C^\infty(\eB,\bbE_2)^{\um}$ with angular dependence $e^{-i\um\cdot \underline{\phi}}$. 
 \begin{enumerate}
 \item[(i)] If $\um = \underline{0}$, and if the smallest eigenvalue of $\cA$ on $C^\infty(\eB,\bbE_2)^{\underline{0}}$ satisfies
 $\lambda < -\frac{1}{4}$, there is a perturbation
 of compact support on the Cauchy surface $\Sigma$ which cannot settle down to perturbation
to another stationary black hole (or a pure gauge transformation). In other words, the black hole is
linearly unstable. 
\item[(ii)] If there is a $\um \in \ZZ^n$ 
such that $L^2 g^{IJ} m_I m_J < (k^I m_I)^2$ somewhere on $\eB$,  then the black hole is unstable (in the same sense).
 \end{enumerate}
\end{thm}

\noindent
{\bf Remarks:} 
1) It is possible that the instability identified in part (ii) of this theorem is related to the so called ``superradiant instability'' which has been discussed 
in \cite{HR,KKZ09,Dias:2010gk}. 
At any rate, there should be an independent proof of the superradiant 
instability by the canonical energy method which is not restricted to extremal black holes and does not depend on the use of NH geometries. 
We have learnt from S.~Green that he is working on such a proof~\cite{Greene1}. 

2) At present, it is not known whether extremal  non-rotating (static) $AdS$ 
black holes are linearly unstable. For a review of perturbations of static black holes, see \cite{Ishibashi:2011ws}.

\medskip

{\em Proof:} The statement only differs from that in the asymptotically flat case in that the condition $\um \cdot \underline{\Omega} = 0$ is not needed. 
In the asymptotically flat case this condition was necessary in order to get a perturbation $\gamma_{ab}$ that is Lie-derived by $\psi^a$ (cf.
previous proof). This was needed in turn to apply the flux lemma~\ref{fluxlemma} which in the asymptotically flat case 
only holds for perturbations that are Lie-derived
by $\psi^a$. By contrast, in the asymptotically AdS-case, the flux formula applies also for
perturbations that are not Lie-derived by $\psi^a$, hence the condition $\um \cdot \underline{\Omega} = 0$ is unnecessary. 
\qed

\subsection{Construction of an electromagnetic perturbation with $\E<0$ in the extremal BH geometry}

The strategy of the previous sections can also be applied to electromagnetic perturbations and directly leads to 
the exact analogs of thms.~\ref{thm2} and~\ref{thm3} for electromagnetic fields (except possibly for the case $\lambda = -\half$): Simply replace `gravitational perturbation'
by `electromagnetic perturbation' in these statements [and the operator $\cA$ now refers to \eqref{Adef1}]. Because the strategy is so similar, we only outline the main changes
required for electromagnetic fields. As in the outline subsec.~\ref{outline} for gravitational perturbations, 
there are three steps. The first step 1) is again to take an electromagnetic perturbation $A_a$ on the NH geometry having $\E<0$
and compact support on $\Sigma$, as guaranteed by thm.~\ref{el1thm}. Step 2) proceeds as in the gravitational case, leading to initial data $(E^i(\eps), A_i(\eps))$ that have been 
scaled ``down the throat''. Step 3), i.e. correcting these initial data to give a perturbation satisfying the constraints in the BH spacetime,
is actually easier in the electromagnetic case. Here, the constraint is simply Gauss' law, $D_i(h^{-\half} E^i) = 0$, which does not involve $A_i$ and is linear. 
The ansatz for the corrected initial data on the BH spacetime is now
\ben
\tilde E^i = E^i - s^{2\alpha+2} e^{-2/s^\alpha} h^{\half} D^i u \ ,  
\een
instead of~\eqref{ansatz}, where $u$ is to be determined. Thus, letting $f = D_i(h^{-\half} E^i)$ be the violation  
of the Gauss law constraint of the original initial data $E^i = E^i(\eps)$, in order for $\tilde E^i$ to satisfy Gauss' law, we now need to solve
\ben\label{correctioneq}
D_i( s^{2\alpha+2} e^{-2/s^\alpha} D^i u )= f
\een 
instead of eq.~\eqref{xeps}. The main tool is again a weighted Friedrichs-Poincar\' e inequality, which in this case is:

\begin{lemma}\label{friedrichs} (Weighted Friedrichs-Poincar\' e inequality)
For each $\alpha>0$, there is a constant $c=c(\alpha,A)$ such that
\ben\label{fp}
c \int_A |Du|^2 s^{2\alpha+2} e^{-2/s^\alpha} \D vol_A 
\ge \| u-\langle u \rangle_A \|_{H^{1,\alpha}}^2 \ ,
\een
for any $u \in H^{1,\alpha}_0(A)$. Here the weighted mean value is defined as
\ben
\langle u \rangle_A \equiv 
\int_A u s^{2\alpha+2} e^{-2/\alpha} \D vol_A
\bigg/
\int_A s^{2\alpha+2} e^{-2/\alpha} \D vol_A       . 
\een
\end{lemma}
The proof of this lemma is analogous, but simpler than, that given for the generalized weighted Poincar\' e-Friedrichs inequality~\eqref{fp}. 
(Note that $\langle u \rangle_A$ plays a similar role as $P_A {\bf X}$ in~\eqref{fp}, because it may be viewed as the projection of $u$
onto the `constant mode'.)
With this inequality at hand, one again proves the existence of a suitably regular solution to~\eqref{correctioneq}, with bounds on the norms of $u$
in terms of those of $f$ of the type 
\ben
\int_A s^{2\beta k} |D^k( s^{\alpha+1} e^{-1/s^\alpha} u)|^2 \D vol_A \le c \| f \|_{H^k}^2 \ , 
\een
$c=c(A,\alpha,\beta, k)$, for sufficiently 
large $\beta$ and all $k$. In a similar way as 
in the gravitational case, it follows that $\tilde E_i$ is smooth and of
compact support on $A$. The rest of the proof is also similar to the gravitational case, so 
we omit the details to avoid repetition.  

\section{Conclusions and outlook}\label{conclusions}

\subsection{Generalization to near extremal black holes}

In this paper, we have proved an extension of conjecture~1 for all known extremal stationary asymptotically flat vacuum black holes. A similar, but 
stronger, result was also obtained for asymptotically anti-deSitter black holes.  Due to its conceptual
nature, it should be possible to apply our strategy, in suitably modified form, to a variety of other interesting situations in which some sufficiently simple limiting spacetime 
(analogous to the NH-geometry) can be identified. What we mean more precisely is this. Suppose we would like to study the stability of 
a spacetime with a metric $g_{ab}$ which is a member of a family of metrics labelled by various parameters such as mass, spins, charges etc. 
Suppose we can form out of these parameters a small parameter $\epsilon$ such that the family has a limiting spacetime as 
$\epsilon \to 0$ for which the existence of a suitable perturbation with $\E < 0$ can be shown. (In practice, the limiting spacetime should be simpler in the sense that 
the corresponding perturbation can be constructed by analytic methods.) Then, the original metric should be unstable for 
sufficiently small $\epsilon$. Note that the limiting spacetime does not need to represent a black hole itself, as exemplified e.g. by the near horizon limit.

This kind of reasoning suggests for example that conjecture~1 should also be true for {\em near extremal} black holes, 
i.e. one is tempted to conjecture\footnote{This conjecture already appears in the introduction of \cite{Durkee}.}:

\medskip
\noindent
{\bf Conjecture~2}: {\em Consider an extremal, stationary, asymptotically flat black hole spacetime, and assume that the angular velocities are generic~\footnote{See thm.~\ref{thm2}.}. 
If the lowest eigenvalue  $\lambda$ of the operator
$\cA$  (see eq.~\eqref{Adef}) acting on ${\rm U}(1)^n$-invariant tensors is below the critical value of $\lc$, then there exists a neighborhood of the black hole parameters near 
extremality for which the corresponding regular, non-extremal BHs are linearly unstable.
} 

\medskip

Let us comment on the evidence for this conjecture. We first note that the numerical investigations~\cite{Santos} reported in table~2 of~\cite{Durkee}
in support of the Durkee-Reall conjecture (covering extremal black holes)  
were, actually, carried out for a sequence of non-extremal spacetimes converging to an extremal one. Thus~\cite{Santos,Durkee} should be 
viewed as support for Conjecture~2.  

\begin{figure}
\begin{center}
\begin{tikzpicture}[scale=1.1, transform shape]
\filldraw[fill=gray,opacity=.2,draw=black] (-1,0) -- (-1.5,0.5) -- (0,2) -- (.75,1.25) -- (-.5,0) -- (.75,-1.25)--(0,-2)--(-1.5,-.5) --(-1,0);
\draw[black, very thick] (-1,0) -- (-.5,0);
\draw[black, very thick] (-1.26,.256) -- (-0.3,0.177);
\draw[double] (0,2) -- (2,0) -- (0,-2);
\draw (0,2) -- (-2,0) -- (0,-2);
\draw[gray, opacity=.8,very thick] (-2,0) -- (2,0);
\node at (0,-.4) {$\Sigma$};
\draw[gray, opacity=.8,very thick] (-1.7,.3) -- (2,0);
\node at (0,.5) {$\Sigma'$};
\draw (-2,0) node[draw,shape=circle,scale=0.3,fill=black]{};
\draw (2,0) node[draw,shape=circle,scale=0.3,fill=black]{};
\draw (0,2) node[draw,shape=circle,scale=0.3,fill=black]{};
\draw (0,-2) node[draw,shape=circle,scale=0.3,fill=black]{};
\node at (-1.8,.6) {$\mathscr{B}'$};
\node at (-2.45,0) {$\mathscr{B}$};
\draw (-1.7,.3) node[draw,shape=circle,scale=0.3,fill=black]{};
\end{tikzpicture}
\end{center}
\caption{The slice $\Sigma'$ and the support of the perturbation $\gamma_{ab}$ in the domain of outer communication. 
\label{DD} 
}
\end{figure}
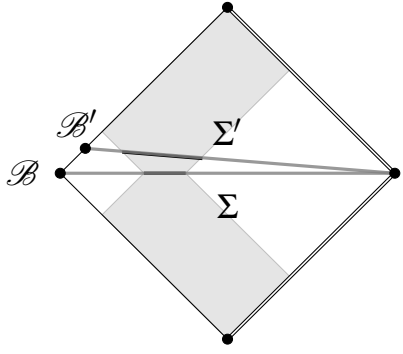

We believe that, in fact, an analytic proof of Conjecture~2 can be given using the general strategy developed in this paper. 
The argument would proceed along the following lines. Consider a 1-parameter family of  asymptotically flat Myers-Perry spacetime metrics $g_{ab}(\eps)$, 
such that $g_{ab}(\eps)$ represents a regular, non-extremal BH for $\eps>0$, and such that
$g_{ab}(\eps=0)$ represents an extremal limit. (We emphasize that the parameter $\eps$ here is conceptually totally different from the parameters $\eps$ used in other 
places in this paper.) Let us assume that the lowest eigenvalue  $\lambda$ of the operator
$\cA$  [see eq.~\eqref{Adef}] is below the critical value of $\lc$ on this extremal limit. We have shown in this paper that 
there exist perturbations $\gamma_{ab}$ having compact support on $\Sigma$ such that $\E(\gamma)<0$ on the extremal limit.
Let $\Sigma'$ be another slice intersecting the future event horizon such that $\gamma_{ab}$ still has compact support on $\Sigma'$ (in particular, it has its support 
bounded away from $\eB'$), see figure~\ref{DD}. 

Near $\eB'$, we introduce Gaussian null coordinates, and we identify points in a neighborhood of $\eB'$ in the extremal 
spacetime with points in corresponding neighborhoods of the non-extremal spacetimes labeled by $\eps$, by declaring that points with the same values of 
$(\rho, u, x^A)$ should be equal. Under this identification the line element of our 1-parameter family takes the form [compare \eqref{gnc}]
\ben
\D s^2(\eps) = 2 \D u[ \D \rho - \tfrac{1}{2} \rho \{ \kappa(\eps)  + \rho \alpha(\eps)\} \, \D u - \rho \beta_A(\eps) \, \D x^A] + \mu_{AB}(\eps) \, \D x^A \D x^B \ 
\een
near $\eB'$.  We have made explicit the dependence upon $\eps$, where $\kappa(\eps)$ is the surface gravity, and where $u$ is the flow parameter of the Killing 
field $K$ on $\eH^+$ (it will not coincide with affine time when $\eps >0$). 
Then, as $\eps \to 0$, all quantities converge smoothly to their extremal limits, e.g. $\kappa(\eps) \to 0$. Under our identification, the slice 
$\Sigma'$ defines a slice in each spacetime of the family at a ``fixed coordinate location'' in Gaussian null coordinates, and 
the perturbation $\gamma_{ab}$ can likewise be viewed as a perturbation on each member of the family\footnote{Here it is tacitly 
assumed that the Gaussian null coordinates cover the part of $\Sigma'$ where $\gamma_{ab}$ is different from zero. This can be justified.}. 
Its initial data on $\Sigma'$ will of course not satisfy the linearized constraints for the background $g_{ab}(\eps)$ when $\eps > 0$. 
But, for $\eps \to 0$, the failure of the constraints must go to $0$, because 
all metric components $\alpha(\eps), \beta_A(\eps), \gamma_{AB}(\eps), \kappa(\eps)$ and their derivatives converge to their extremal limits near $\eB'$. Thus, it is plausible 
that one should be able to correct the initial data of $\gamma_{ab}$ by the Corvino-Schoen method (now on $\Sigma'$ rather than $\Sigma$) in a similar way as described earlier on in subsection~\ref{correctingID}. Furthermore, 
one ought to be able to show using lemma~\ref{lem4} that the correction will become small for small $\epsilon$ in a sufficiently strong sense. This would imply that the canonical energy $\E(\eps)$ of the corrected initial data on the background $g_{ab}(\eps)$ converges to $\E$ on the extremal limit. Since $\E < 0$, we have thereby constructed for sufficiently small $\eps$
a  perturbation on a non-extremal black hole with compact support on $\Sigma'$ and with a negative canonical energy. 
As we have argued before, this black hole must therefore be unstable. 

With regard to asymptotically $AdS$ black holes, there is a corresponding conjecture. The results of~\cite{KLR06}
support this conjecture in the special case of equal angular velocities, and the analytic proof which we have just sketched should go through in the same manner, because it is basically 
local to the horizon and insensitive to the asymptotic region.

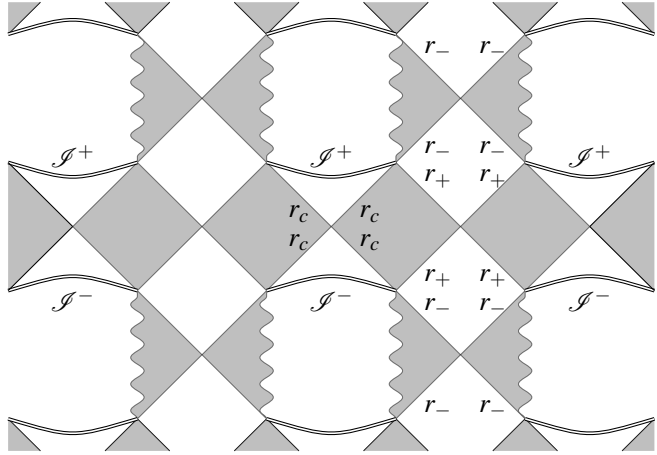
\begin{figure}
\begin{center}
\begin{tikzpicture}[scale=.85, transform shape]
\filldraw[fill=gray,opacity=.5,draw=black] (0,0) -- (1,1) -- (2,0) -- (1,-1) -- (0,0);
\filldraw[fill=gray,opacity=.5,draw=black] (2,0) -- (3,1) -- (4,0) -- (3,-1) -- (2,0);
\filldraw[fill=gray,opacity=.5,draw=black] (0,0) -- (-1,1) -- (-2,0) -- (-1,-1) -- (0,0);
\filldraw[fill=gray,opacity=.5,draw=black] (-2,0) -- (-3,1) -- (-4,0) -- (-3,-1) -- (-2,0);
\filldraw[fill=gray,opacity=.5,draw=black] (1,1) -- (2,2) -- (1,3) decorate[decoration=snake] {-- (1,1)};
\filldraw[fill=gray,opacity=.5,draw=black] (3,1) -- (2,2) -- (3,3) decorate[decoration=snake] {-- (3,1)};
\filldraw[fill=gray,opacity=.5,draw=black] (-1,1) -- (-2,2) -- (-1,3) decorate[decoration=snake] {-- (-1,1)};
\filldraw[fill=gray,opacity=.5,draw=black] (-3,1) -- (-2,2) -- (-3,3) decorate[decoration=snake] {-- (-3,1)};
\filldraw[fill=gray,opacity=.5,draw=black] (1,-1) -- (2,-2) -- (1,-3) decorate[decoration=snake] {-- (1,-1)};
\filldraw[fill=gray,opacity=.5,draw=black] (3,-1) -- (2,-2) -- (3,-3) decorate[decoration=snake] {-- (3,-1)};
\filldraw[fill=gray,opacity=.5,draw=black] (-1,-1) -- (-2,-2) -- (-1,-3) decorate[decoration=snake] {-- (-1,-1)};
\filldraw[fill=gray,opacity=.5,draw=black] (-3,-1) -- (-2,-2) -- (-3,-3) decorate[decoration=snake] {-- (-3,-1)};
\draw (1,3) -- (1.5,3.5);
\draw (1,3) -- (0.5,3.5);
\fill[gray,opacity=.5] (1,3) -- (1.5,3.5) -- (0.5,3.5);
\draw (3,3) -- (3.5,3.5);
\draw (3,3) -- (2.5,3.5);
\fill[gray,opacity=.5] (3,3) -- (3.5,3.5) -- (2.5,3.5);
\draw (-1,3) -- (-1.5,3.5);
\draw (-1,3) -- (-0.5,3.5);
\fill[gray,opacity=.5] (-1,3) -- (-1.5,3.5) -- (-0.5,3.5);
\draw (-3,3) -- (-3.5,3.5);
\draw (-3,3) -- (-2.5,3.5);
\fill[gray,opacity=.5] (-3,3) -- (-3.5,3.5) -- (-2.5,3.5);
\draw (1,-3) -- (1.5,-3.5);
\draw (1,-3) -- (0.5,-3.5);
\fill[gray,opacity=.5] (1,-3) -- (1.5,-3.5) -- (0.5,-3.5);
\draw (3,-3) -- (3.5,-3.5);
\draw (3,-3) -- (2.5,-3.5);
\fill[gray,opacity=.5] (3,-3) -- (3.5,-3.5) -- (2.5,-3.5);
\draw (-1,-3) -- (-1.5,-3.5);
\draw (-1,-3) -- (-0.5,-3.5);
\fill[gray,opacity=.5] (-1,-3) -- (-1.5,-3.5) -- (-0.5,-3.5);
\draw (-3,-3) -- (-3.5,-3.5);
\draw (-3,-3) -- (-2.5,-3.5);
\fill[gray,opacity=.5] (-3,-3) -- (-3.5,-3.5) -- (-2.5,-3.5);
\fill[gray,opacity=.5] (4,0) -- (5,-1) -- (5,1);
\fill[gray,opacity=.5] (-4,0) -- (-5,-1) -- (-5,1);
\draw (-5,3) -- (-4.5,3.5);
\draw (5,3) -- (4.5,3.5);
\draw (-5,-3) -- (-4.5,-3.5);
\draw (5,-3) -- (4.5,-3.5);
\fill[gray,opacity=.5] (-5,3) -- (-4.5,3.5) -- (-5,3.5);
\fill[gray,opacity=.5] (-5,-3) -- (-4.5,-3.5) -- (-5,-3.5);
\fill[gray,opacity=.5] (5,3) -- (4.5,3.5) -- (5,3.5);
\fill[gray,opacity=.5] (5,-3) -- (4.5,-3.5) -- (5,-3.5);
\draw[double] (-1,1) .. controls (0,0.7) and (0,0.7) .. (1,1);
\draw[double] (3,1) .. controls (4,0.7) and (4,0.7) .. (5,1);
\draw[double] (-3,1) .. controls (-4,0.7) and (-4,0.7) .. (-5,1);
\draw[double] (-1,3) .. controls (0,3.3) and (0,3.3) .. (1,3);
\draw[double] (3,3) .. controls (4,3.3) and (4,3.3) .. (5,3);
\draw[double] (-3,3) .. controls (-4,3.3) and (-4,3.3) .. (-5,3);
\draw[double] (-1,-1) .. controls (0,-0.7) and (0,-0.7) .. (1,-1);
\draw[double] (3,-1) .. controls (4,-0.7) and (4,-0.7) .. (5,-1);
\draw[double] (-3,-1) .. controls (-4,-0.7) and (-4,-0.7) .. (-5,-1);
\draw[double] (-1,-3) .. controls (0,-3.3) and (0,-3.3) .. (1,-3);
\draw[double] (3,-3) .. controls (4,-3.3) and (4,-3.3) .. (5,-3);
\draw[double] (-3,-3) .. controls (-4,-3.3) and (-4,-3.3) .. (-5,-3);
\draw (-4,0) -- (-5,1);
\draw (-4,0) -- (-5,-1);
\draw (4,0) -- (5,1);
\draw (4,0) -- (5,-1);
\node at (0,1.1) {${\mathscr I}^+$};
\node at (0,-1.15) {${\mathscr I}^-$};
\node at (4,1.1) {${\mathscr I}^+$};
\node at (4,-1.15) {${\mathscr I}^-$};
\node at (-4,1.1) {${\mathscr I}^+$};
\node at (-4,-1.15) {${\mathscr I}^-$};
\node at (1.65,0.75) {$r_+$};
\node at (2.5,0.75) {$r_+$};
\node at (1.65,1.2) {$r_-$};
\node at (2.5,1.2) {$r_-$};
\node at (1.65,2.75) {$r_-$};
\node at (2.5,2.75) {$r_-$};
\node at (1.65,-0.8) {$r_+$};
\node at (2.5,-0.8) {$r_+$};
\node at (1.65,-1.25) {$r_-$};
\node at (2.5,-1.25) {$r_-$};
\node at (1.65,-2.85) {$r_-$};
\node at (2.5,-2.85) {$r_-$};
\node at (-0.5,0.2) {$r_c$};
\node at (0.6,0.2) {$r_c$};
\node at (-0.5,-0.25) {$r_c$};
\node at (0.6,-0.25) {$r_c$};
\end{tikzpicture}
\end{center}
\caption{
\label{MPdS1}
Conformal diagram of the non-extremal deSitter-MP spacetime. $r_c$ indicates a cosmological horizon, $r_\pm$ an event horizon.  The stationary 
regions are shaded. 
}
\end{figure}

\begin{figure}
\begin{center}
\begin{tikzpicture}[scale=.8, transform shape]
\filldraw[fill=gray,opacity=.5,draw=black] (-4,-1) -- (-3,0) -- (-4,1) decorate[decoration=snake] {--(-4,-1)};
\filldraw[fill=gray,opacity=.5,draw=black] (-2,1) -- (-3,0) -- (-2,-1) decorate[decoration=snake] {--(-2,1)};
\filldraw[fill=gray,opacity=.5,draw=black] (4,-1) -- (3,0) -- (4,1) decorate[decoration=snake] {--(4,-1)};
\filldraw[fill=gray,opacity=.5,draw=black] (2,1) -- (3,0) -- (2,-1) decorate[decoration=snake] {--(2,1)};
\filldraw[fill=gray,opacity=.5,draw=black] (-1,2) -- (0,3) -- (-1,4) decorate[decoration=snake] {--(-1,2)};
\filldraw[fill=gray,opacity=.5,draw=black] (1,2) -- (0,3) -- (1,4) decorate[decoration=snake] {--(1,2)};
\filldraw[fill=gray,opacity=.5,draw=black] (-7,2) -- (-6,3) -- (-7,4) decorate[decoration=snake] {--(-7,2)};
\filldraw[fill=gray,opacity=.5,draw=black] (-5,2) -- (-6,3) -- (-5,4) decorate[decoration=snake] {--(-5,2)};
\filldraw[fill=gray,opacity=.5,draw=black] (7,2) -- (6,3) -- (7,4) decorate[decoration=snake] {--(7,2)};
\filldraw[fill=gray,opacity=.5,draw=black] (5,2) -- (6,3) -- (5,4) decorate[decoration=snake] {--(5,2)};
\filldraw[fill=gray,opacity=.5,draw=black] (1,-2) -- (0,-3) -- (1,-4) decorate[decoration=snake] {--(1,-2)};
\filldraw[fill=gray,opacity=.5,draw=black] (-1,-2) -- (0,-3) -- (-1,-4) decorate[decoration=snake] {--(-1,-2)};
\filldraw[fill=gray,opacity=.5,draw=black] (-7,-2) -- (-6,-3) -- (-7,-4) decorate[decoration=snake] {--(-7,-2)};
\filldraw[fill=gray,opacity=.5,draw=black] (-5,-2) -- (-6,-3) -- (-5,-4) decorate[decoration=snake] {--(-5,-2)};
\filldraw[fill=gray,opacity=.5,draw=black] (7,-2) -- (6,-3) -- (7,-4) decorate[decoration=snake] {--(7,-2)};
\filldraw[fill=gray,opacity=.5,draw=black] (5,-2) -- (6,-3) -- (5,-4) decorate[decoration=snake] {--(5,-2)};
\draw (-8,1) -- (-7,2) -- (-6,1) -- (-5,2) -- (-4,1) -- (-3,2) -- (-2,1) -- (-1,2) -- (0,1) -- (1,2) -- (2,1) -- (3,2) -- (4,1) -- (5,2) -- (6,1) -- (7,2) -- (8,1);
\draw (-8,-1) -- (-7,-2) -- (-6,-1) -- (-5,-2) -- (-4,-1) -- (-3,-2) -- (-2,-1) -- (-1,-2) -- (0,-1) -- (1,-2) -- (2,-1) -- (3,-2) -- (4,-1) -- (5,-2) -- (6,-1) -- (7,-2) -- (8,-1);
\draw[double] (-8,1) .. controls (-7,1.3) and (-7,1.3) .. (-6,1) .. controls (-5,1.3) and (-5,1.3) .. (-4,1);
\draw[double] (-2,1) .. controls (-1,1.3) and (-1,1.3) .. (0,1) .. controls (1,1.3) and (1,1.3) .. (2,1);
\draw[double] (8,1) .. controls (7,1.3) and (7,1.3) .. (6,1) .. controls (5,1.3) and (5,1.3) .. (4,1);
\draw[double] (-8,-1) .. controls (-7,-1.3) and (-7,-1.3) .. (-6,-1) .. controls (-5,-1.3) and (-5,-1.3) .. (-4,-1);
\draw[double] (-2,-1) .. controls (-1,-1.3) and (-1,-1.3) .. (0,-1) .. controls (1,-1.3) and (1,-1.3) .. (2,-1);
\draw[double] (8,-1) .. controls (7,-1.3) and (7,-1.3) .. (6,-1) .. controls (5,-1.3) and (5,-1.3) .. (4,-1);
\draw[double] (-8,1.7) .. controls (-7.8,1.7) and (-7.8,1.7) .. (-7,2);
\draw[double] (-5,2) .. controls (-4,1.7) and (-4,1.7) .. (-3,2) .. controls (-2,1.7) and (-2,1.7) .. (-1,2);
\draw[double] (8,1.7) .. controls (7.8,1.7) and (7.8,1.7) .. (7,2);
\draw[double] (5,2) .. controls (4,1.7) and (4,1.7) .. (3,2) .. controls (2,1.7) and (2,1.7) .. (1,2);
\draw[double] (-8,-1.7) .. controls (-7.8,-1.7) and (-7.8,-1.7) .. (-7,-2);
\draw[double] (-5,-2) .. controls (-4,-1.7) and (-4,-1.7) .. (-3,-2) .. controls (-2,-1.7) and (-2,-1.7) .. (-1,-2);
\draw[double] (8,-1.7) .. controls (7.8,-1.7) and (7.8,-1.7) .. (7,-2);
\draw[double] (5,-2) .. controls (4,-1.7) and (4,-1.7) .. (3,-2) .. controls (2,-1.7) and (2,-1.7) .. (1,-2);
\draw (0,1) node[draw,shape=circle,scale=0.3,fill=black]{};
\draw (0,-1) node[draw,shape=circle,scale=0.3,fill=black]{};
\draw (-6,1) node[draw,shape=circle,scale=0.3,fill=black]{};
\draw (6,1) node[draw,shape=circle,scale=0.3,fill=black]{};
\draw (-6,-1) node[draw,shape=circle,scale=0.3,fill=black]{};
\draw (6,-1) node[draw,shape=circle,scale=0.3,fill=black]{};
\draw (-3,2) node[draw,shape=circle,scale=0.3,fill=black]{};
\draw (3,2) node[draw,shape=circle,scale=0.3,fill=black]{};
\draw (-3,-2) node[draw,shape=circle,scale=0.3,fill=black]{};
\draw (3,-2) node[draw,shape=circle,scale=0.3,fill=black]{};
\node at (-4,2.1) {${\mathscr I}^+$};
\node at (-2,2.1) {${\mathscr I}^+$};
\node at (2,2.1) {${\mathscr I}^+$};
\node at (4,2.1) {${\mathscr I}^+$};
\node at (-7,1) {${\mathscr I}^-$};
\node at (-5,1) {${\mathscr I}^-$};
\node at (-1,1) {${\mathscr I}^-$};
\node at (1,1) {${\mathscr I}^-$};
\node at (7,1) {${\mathscr I}^-$};
\node at (5,1) {${\mathscr I}^-$};
\end{tikzpicture}
\end{center}
\caption{
\label{MPdS3}
Conformal diagram of another possible extremal limit of the deSitter-MP spacetime corresponding to $r_c \to r_+$.  The stationary regions are shaded.
}
\end{figure}
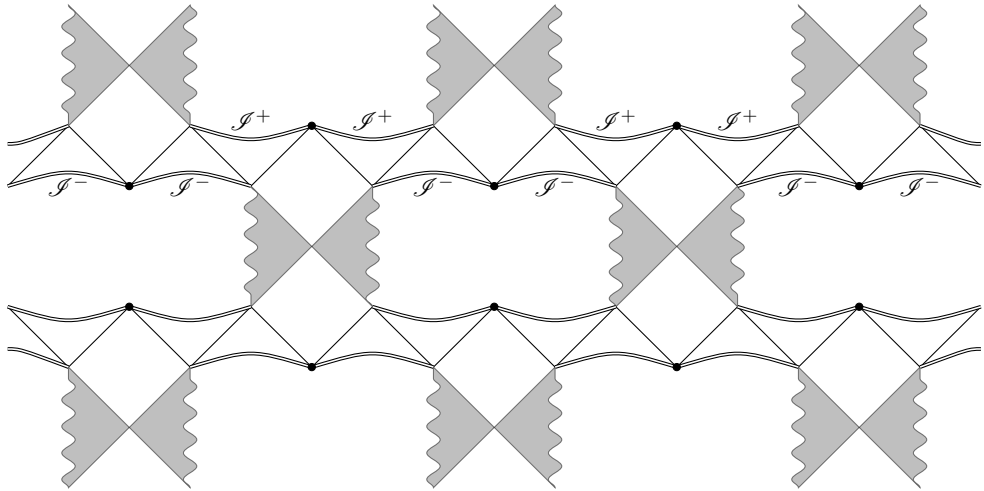

Another interesting question is whether conjecture~1 (and 2) is true also in the deSitter case. A conformal diagram for a deSitter-MP black hole is given in fig.~\ref{MPdS1}. 
There are actually two ways in which to take the extremal limit, depicted in figs.~\ref{MPdS2} and~\ref{MPdS3}. To have a reasonable notion of stability, one should look at the ``stationary regions'' in these extremal limits. It is clear that this makes sense only in the case described in 
the case depicted in fig.~\ref{MPdS2}, wherein one would look at the region whose boundaries are the event- and cosmological horizons. Since the cosmological horizon has 
geometrical properties that are very similar to those of an event horizon, it seems plausible that an analogue of the monotonicity result for $\E$, expressed in lemma~\ref{fluxlemma},
should also hold in the asymptotically deSitter case (the region ${\mathscr I}^+$ in the 
asymptotically flat case would now effectively be replaced by a portion of the cosmological horizon.). 
Hence, it seems likely that conjecture~1 (and 2) continue to be true also in the deSitter case.

\begin{figure}
\begin{center}
\begin{tikzpicture}[scale=.8, transform shape]
\draw (0.5,3.5) -- (2,2);
\draw (-1.5,3.5) -- (2,0);
\draw (-2,2) -- (2,-2);
\draw (-2,0) -- (1.5,-3.5);
\draw (2,-2) -- (0.5,-3.5);
\draw (-0.5,3.5) -- (-2,2);
\draw (1.5,3.5) -- (-2,0);
\draw (2,2) -- (-2,-2);
\draw (2,0) -- (-1.5,-3.5);
\draw (-2,-2) -- (-0.5,-3.5);
\filldraw[fill=gray,opacity=.5,draw=black] (-1.5,-3.5) -- (-1,-3) -- (-2,-2) decorate[decoration=snake] {--(-2,-3.5)};
\filldraw[fill=gray,opacity=.5,draw=black] (-2,0) -- (-1,-1) --  (-2,-2) decorate[decoration=snake] {--(-2,0)};
\filldraw[fill=gray,opacity=.5,draw=black] (-2,2) -- (-1,1) -- (-2,0) decorate[decoration=snake] {--(-2,2)};
\filldraw[fill=gray,opacity=.5,draw=black] (-1.5,3.5) -- (-1,3) -- (-2,2) decorate[decoration=snake] {-- (-2,3.5)};
\filldraw[fill=gray,opacity=.5,draw=black] (1.5,-3.5) -- (1,-3) -- (2,-2) decorate[decoration=snake] {--(2,-3.5)};
\filldraw[fill=gray,opacity=.5,draw=black] (2,0) -- (1,-1) --  (2,-2) decorate[decoration=snake] {--(2,0)};
\filldraw[fill=gray,opacity=.5,draw=black] (2,2) -- (1,1) -- (2,0) decorate[decoration=snake] {--(2,2)};
\filldraw[fill=gray,opacity=.5,draw=black] (1.5,3.5) -- (1,3) -- (2,2) decorate[decoration=snake] {-- (2,3.5)};
\fill[fill=gray,opacity=.5] (-2,2) -- (-1,3) -- (0,2) -- (-1,1);
\fill[fill=gray,opacity=.5] (-2,0) -- (-1,1) -- (0,0) -- (-1,-1);
\fill[fill=gray,opacity=.5] (-2,-2) -- (-1,-3) -- (0,-2) -- (-1,-1);
\fill[fill=gray,opacity=.5] (-1.5,3.5) -- (-0.5,3.5) -- (-1,3);
\fill[fill=gray,opacity=.5] (-1.5,-3.5) -- (-0.5,-3.5) -- (-1,-3);
\fill[fill=gray,opacity=.5] (2,2) -- (1,3) -- (0,2) -- (1,1);
\fill[fill=gray,opacity=.5] (2,0) -- (1,1) -- (0,0) -- (1,-1);
\fill[fill=gray,opacity=.5] (2,-2) -- (1,-3) -- (0,-2) -- (1,-1);
\fill[fill=gray,opacity=.5] (1.5,3.5) -- (0.5,3.5) -- (1,3);
\fill[fill=gray,opacity=.5] (1.5,-3.5) -- (0.5,-3.5) -- (1,-3);
\draw[double] (-1,1) ..controls (0,0.7) and (0,0.7) .. (1,1);
\draw[double] (-1,1) ..controls (0,1.3) and (0,1.3) .. (1,1);
\draw[double] (-1,3) ..controls (0,2.7) and (0,2.7) .. (1,3);
\draw[double] (-1,3) ..controls (0,3.3) and (0,3.3) .. (1,3);
\draw[double] (-1,-1) ..controls (0,-0.7) and (0,-0.7) .. (1,-1);
\draw[double] (-1,-1) ..controls (0,-1.3) and (0,-1.3) .. (1,-1);
\draw[double] (-1,-3) ..controls (0,-2.7) and (0,-2.7) .. (1,-3);
\draw[double] (-1,-3) ..controls (0,-3.3) and (0,-3.3) .. (1,-3);
\draw (-2,2) node[draw,shape=circle,scale=0.3,fill=black]{};
\draw (-2,-2) node[draw,shape=circle,scale=0.3,fill=black]{};
\draw (2,0) node[draw,shape=circle,scale=0.3,fill=black]{};
\draw (2,2) node[draw,shape=circle,scale=0.3,fill=black]{};
\draw (2,-2) node[draw,shape=circle,scale=0.3,fill=black]{};
\draw (-2,0) node[draw,shape=circle,scale=0.3,fill=black]{};
\node at (0,1.5) {${\mathscr I}^+$};
\node at (0,0.6) {${\mathscr I}^-$};
\node at (0,-1.5) {${\mathscr I}^+$};
\node at (0,-0.5) {${\mathscr I}^-$};
\node at (0,2.6) {${\mathscr I}^+$};
\node at (0,-2.5) {${\mathscr I}^-$};
\end{tikzpicture}
\end{center}
\caption{
\label{MPdS2}
Conformal diagram of one possible extremal limit of the deSitter-MP spacetime corresponding to $r_- \to r_+$.   The stationary regions are shaded. 
}
\end{figure}
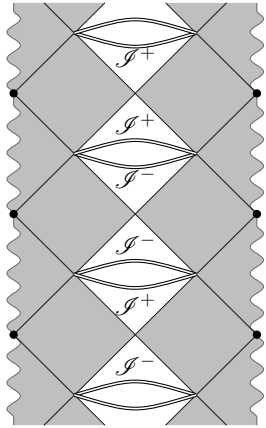

 \subsection{Outlook}
 
We would finally like to comment on the relationship between our results and those of Aretakis\footnote{
Aretakis~\cite{Aretakis} considered a test scalar field in extreme Kerr. The generalization to linear gravitational perturbations 
was given in~\cite{Aretakis1}, and to non-linear graviational perturbations in~\cite{Aretakis2}.
}
~\cite{Aretakis}, and those of Dain et al.~\cite{Dain}.
Dain et al. consider, for the extremal 4-dimensional Kerr spacetime, an ``energy'' for axi-symmetric linear perturbations, which they show to be 
positive definite. We have every reason to believe that their quantity is actually identical to our canonical energy $\E$, when expressed in terms of the 
variables and gauges chosen by Dain et al. If so, this would preclude the possibility of finding compactly supported, axi-symmetric initial data having 
$\E<0$ for extreme Kerr. (Note that this would be consistent with the present paper, because for extreme Kerr, the smallest eigenvalue $\lambda$ happens to be above the effective BF bound, 
and hence our arguments showing $\E<0$ do not apply.)  
As argued by Dain et al., a positive definite $\E$ should translate into pointwise bounds on the perturbation {\em outside} the black hole, indicating that in this sense, 
the extreme Kerr black hole should be regarded as stable. 

On the other hand, Aretakis has argued that sufficiently high transverse derivatives (in fact, second derivatives) of 
linear perturbation with smooth initial data blows up {\em on} the horizon, so that, in this sense, an extremal Kerr black hole\footnote{The analysis by Aretakis has been extended to higher dimensions in~\cite{Aretakis1} under certain reasonable conditions on the background. A similar analysis under somewhat different assumptions on the background [existence of 
a zero eigenvalue of a certain operator related to our operator $\cA$~\eqref{Adef}] has been given by~\cite{murata2}.} is in fact unstable. Aretakis' result
is not in contradiction with that of Dain et al., because their canonical energy (likely to be equal to our $\E$) contains first derivatives of the perturbation only, and hence 
should be insensitive to the phenomenon discovered by Aretakis. Furthermore, the Aretakis-type instability should also be {\em very different} in nature from  
the instabilities identified by conjecture 1 of this paper. One way to see this is that the initial data for the unstable modes identified in our proof of conjecture~1 should, as explained above in the context of conjecture~2, continue to give rise to instabilities even for near extremal 
black holes. By contrast, the phenomenon covered by Aretakis' analysis is restricted strictly to extremal black holes, as exemplified e.g. by the 4-dimensional Kerr metric, which is expected to be stable even at the non-linear level in the non-extremal case. Moreover, unlike our instability, the Aretakis' type instability seems to be present basically for {\em  any} extremal black hole, regardless whether $\lambda$ is or is not below the effective BF-bound entering conjectures 1 and~2.

\medskip

\noindent
{\bf Acknowledgements:} The work of SH is supported in part by ERC starting grant~259562. He has benefitted from 
scientific interactions during YITP workshop YITP-T-14-1 on 
``Holographic vistas on gravity and strings'' in July 2014
 and would like to thank YITP for its hospitality. 
He also thanks Kinki U. for hospitality. The work of AI was supported in part 
by JSPS KAKENHI Grants No. 22540299 and No. 26400280. 
We are greatful to F. Otto from MPI MIS Leipzig for discussions 
on weighted Sobolev spaces and inequalities and to R. Emparan and J. Lucietti for sharing insights about BH stability issues. We also thank H.S. Reall for pointing out 
to us Ref.~\cite{Madi}, and we thank the referees for their constructive criticisms. 

\appendix

\section{Formula for $\E$ in gravitational case}\label{appA1}

Here we give the formula for $\E$ for a perturbation $\gamma_{ab}$ of the form~\eqref{hertz}, where
$U^{ab}$ is given by eq.~\eqref{Udef}, and where $Y^{AB}$ is an eigenfunction of the operator $\cA$ (cf. eq.~\eqref{Adef})
with eigenvalue $\lambda$. As above, we let $a = \uk \cdot \um$. Let $f_0, f_1$ be the initial data of $\psi$ as in
eq.~\eqref{initial}. Both of them are compactly supported, complex valued, smooth functions of the variable $R \in \RR_+$,
and $\E$ is a quadratic functional of these under those conditions.
As above, we find it convenient to express it in terms of the variable $y = \log R \in \RR$.  To write down $\E$, define
\begin{eqnarray*}
X_4 &=& 2e^{-y} \frac{\D^2 f_1}{\D y^2} + (1-2ia) e^{-y} \frac{\D f_1}{\D y} + (4-2ia - \lambda - \alpha^2) e^{-y} f_1 - \lambda f_0 + \non\\
&& \frac{\D}{\D y} \Big[ (1-2\lambda) f_0 - (8-4ia) e^{-y} f_1 - \frac{\D f_0}{\D y} -2 \frac{\D^2 f_0}{\D y^2} \Big] \\
X_5 &=& - \frac{\D^2 f_0}{\D y^2} - \frac{\D f_0}{\D y} +(ia -4) e^{-y} f_1 + \lambda f_0 + e^{-y} \frac{\D f_1}{\D y}  \ , \quad X_6 = e^{-y} f_1 \\
X_3 &=& f_0 \ , \quad  X_1 = 2 \frac{\D^2 f_0}{\D y^2} + (1-2ia) \frac{\D f_0}{\D y} + 5 e^{-y} f_1 - (\lambda + a^2) f_0 - 2\frac{\D f_1}{\D y} e^{-y}  \\
X_2 &=& -e^{-y} f_1 -ia f_0 + \frac{\D f_0}{\D y} \\
X_7 &=& +e^{-y} f_1 +ia f_0 + \frac{\D f_0}{\D y}
\end{eqnarray*}
Then $\E$ is given by $\E= (1/128\pi) \int_{-\infty}^\infty \rho(y) e^{y} \, \D y$, where
\bena
\rho(f_0,f_1) &=&   | \tfrac{d}{d y} X_1|^2 + 5 | \tfrac{d}{d y} X_2|^2 + 4 | \tfrac{d}{d y} X_3|^2 +\non\\
&&  (2\lambda + 3a^2-4) |X_2|^2 + (\lambda -3) |X_1|^2 + 4a^2 |X_3|^2 + |X_4|^2 + 5|X_5|^2 + 4|X_6|^2+\non\\
&&2 \, {\rm Re} \{
-7ia \bar X_5 X_3 + 8ia \bar X_6 X_3
\} + \non\\
&&2 \, {\rm Re} \{
8 \bar X_5 X_7 - 3(\lambda + a^2) \bar X_5 X_3 + 4 \bar X_5 X_3
\} + \non\\
&&2 \, {\rm Re} \{ 3ia \bar X_2 \tfrac{d}{dy} X_2 + 4ia \bar X_3 \tfrac{d}{dy} X_3 \} \ .
\eena

\section{Formula for $\E$ in electromagnetic case}\label{appA2}

Here we give the formula for $\E$ for a perturbation $A_a$ of the form~\eqref{hertz1}, where
$U^{a}$ is given by eq.~\eqref{Udef1}, and where $Y^{A}$ is an eigenfunction of the operator $\cA$ (cf. eq.~\eqref{Adef1})
with eigenvalue $\lambda$. As above, we let $a = \uk \cdot \um$. Let $f_0, f_1$ be the initial data of $\psi$ as in
eq.~\eqref{initial}. Both of them are compactly supported, complex valued, smooth functions of the variable $R \in \RR_+$,
and $\E$ is a quadratic functional of these under those conditions.  To write down $\E$, define
\begin{eqnarray*}
X_1 &=& -\frac{\D^2 f_0}{\D y^2} -\frac{\D f_0}{\D y} + \lambda f_0 - e^{-y} (2-ia) f_1 + e^{-y} \frac{\D f_1}{\D y}\ ,\quad  X_2 = e^{-y} f_1 \\
X_3 &=& -e^{-y} f_1 -ia f_0 + \frac{\D f_0}{\D y} \ , \quad X_4 = f_0 \\
X_5 &=& +e^{-y} f_1 +ia f_0 + \frac{\D f_0}{\D y}
\end{eqnarray*}
Then $\E$ is given by $\E= (1/8\pi)\int_{-\infty}^\infty \rho(y) e^{y} \, \D y$, where
\bena
\rho(f_0,f_1) &=&  |X_1|^2 +  |X_2|^2 +  |\tfrac{d}{dy} X_3|^2 +  |\tfrac{d}{dy} X_4|^2 + (\lambda -1) |X_3|^2 +\non\\
&& a^2 |X_4|^2 - 2 \, {\rm Re} \{ 2ia \bar X_4 X_2 - ia \bar X_4 \tfrac{d}{dy} X_4 + \bar X_5 X_2 \}  \ \ .
\eena

\section{Mode-type solutions to eq.~\eqref{master1}} \label{appB}

Here we study the solutions of eq.~\eqref{master1} of the form $\psi(T,R) = f(R) e^{i\omega T}$ with $\omega \in \CC$. The resulting equation for $f(R)$ is
\ben\label{wit}
0= \frac{\D}{\D R} \Big( R^2 \frac{\D f}{\D R} \Big) - (\lambda + q^2)f + \Big(
q + \frac{\omega}{R} \Big)^2 f \ .
\een
Solutions of the equation can be given in terms of the hypergeometric function 
${}_1 F_1(a,b;z)$ or Whittaker functions~\cite{Lang,Horowitz-Marolf,Dias:2009ex}. 
Except for degenerate cases which are not relevant for this paper, the general solution is a linear
combination of
\ben
f_\pm(R) = e^{i\omega/R} \Big( \frac{-i\omega}{R} \Big)^{1/2 \pm i\delta} \ {}_1 F_1
\left(
\pm \delta + \half -iq, 1 \pm 2\delta; \frac{-2i\omega}{R} \right) \ ,
\een
where
\ben
\delta = \sqrt{\frac{1}{4} + \lambda} \ .
\een
In the case of most interest for this paper, $\lambda < \lc$, so $\delta$ is imaginary.
The solutions to~\eqref{wit} behave generically as linear combinations of $e^{\pm i\omega/R}$ for $R \to 0$ and as
$R^{\half \pm \delta}$ for $R \to \infty$. In order to get a solution whose derivatives vanish at $R=0$ (i.e. the horizon), 
we need to take a particular linear combination of $f_\pm$, and we need to take $\omega$ to have a non-vanishing imaginary part. If, for
example, we let ${\rm Im}(\omega) > 0$, then the desired linear combination having
vanishing derivatives at $R=0$ to all orders is
\ben
f(R) = A_+ f_+(R) + A_- f_-(R) \ ,
\een
where $A_\pm = 2^\delta \Gamma(-2\delta)/\Gamma(\half \pm \delta -iq)$. With this choice, $f(R)$
behaves as $\sim e^{i\omega/R}$ near $R=0$, and as $\sim A_+ R^{-\half + \delta} + A_- R^{-\half - \delta}$ as $R \to \infty$. For example, take $\omega = i$. The corresponding solution $\psi(T,R)$ is
\ben
\begin{split}
\psi(T,R) &= e^{T-1/R} \Bigg[
A_+ R^{-1/2 - i\delta} \ {}_1 F_1\left(+\delta + \half -iq, 1+2\delta; \frac{2}{R} \right) \\
&\hspace{1.6cm} +
A_- R^{-1/2 + i\delta} \ {}_1 F_1\left(-\delta + \half -iq, 1-2\delta; \frac{2}{R} \right)
\Bigg] .
\end{split}
\een
This solution is exponentially growing in $T$, and regular on the future horizon
(as is seen e.g. by the fact that $R, u=T-1/R$ provide regular coordinates on the
future horizon), but not $L^2$-normalizable near infinity.

\section{Proof of generalized Friedrichs-Poincar\' e inequality, lemma~\ref{friedrichs}}\label{appD}

We repeat the statement of lemma~\ref{friedrichs}:

\begin{lemma} (Generalized Friedrichs-Poincar\' e inequality)
For sufficiently large $\alpha$, there is a constant $c=c(\alpha,A)$ such that
\ben\label{poincare}
c \left\| \Phi \cC^* \bX
\right\|_{L^{2}}
\ge \|\bX - P_A \bX \|_{H^{2,\alpha} \oplus H^{1,\alpha}} \ ,
\een
for any tensor field ${\bf X} \in H^{2,\alpha}_0(A) \oplus H^{1,\alpha}_0(A)$. Here $P_A$ is the orthogonal projector (in $L^{2,\alpha}(A)$) onto the
subspace $\frak k$ spanned by the KVF's, i.e. if ${\bf Y}_i$ is a basis of Killing vector fields on $\M$ that has been orthonormalized with the
Gram-Schmidt process, we have
\ben
P_A \bX = \sum_i {\bf Y}_i  ({\bf Y}_i, \bX)_{L^{2,\alpha}} \  \ .
\een
\end{lemma}

The proof is divided into several steps. 

\medskip
\noindent
Step 1) We recall the 1-dimensional weighted Hardy inequality by Kufner, see sec.~5 of~\cite{Kufner}: Let $\sigma$ be a non-negative, smooth function on the interval $[0,1]$ such that
$\sigma(0) = 0$ whereas $\sigma(s) >0$ for $s>0$. For $1<p<\infty$, define
\ben
\phi(s) = (p-1) \frac{\int_0^s \sigma(t) dt}{\sigma(s)} \ .
\een
Then there holds the inequality
\ben
\int_0^1 |u(s)|^p \sigma(s) \D s \le \left( \frac{p}{p-1} \right)^p \int_0^1 |\phi u'(s)|^p \sigma(s) \D s \ ,
\een
for all $u \in C_0^1(0,1)$ i.e. with vanishing boundary values at the ends of the interval. We apply this inequality to the case when $\sigma(s) = e^{-2/s^\alpha}$ for some $\alpha>0$. One finds
that $|\phi(s)| \le (p-1) s^{\alpha+1}/\alpha$. This gives rise to the inequality
\ben
\int_0^1 |u(s)|^p e^{-2/s^\alpha} \D s \le \left( \frac{p}{\alpha} \right)^p \int_0^1 |u'(s)|^p s^{p(\alpha+1)} e^{-2/s^\alpha} \D s \ ,
\een
for the same class of functions $u$.

\medskip
\noindent
Step 2) We now generalize this inequality to tensors $u \in C^1_0(A)$ on some open domain $A$ with smooth boundary in a Riemannian manifold $(\Sigma, h)$.
We let $0 \le s \le 1$ be a function which is equal to $s(x) = {\rm dist}_h(x, \partial A)$ in a neighborhood of the boundary, and which is positive and smooth in
the interior. We denote $A_\eps = \{ x \in A \mid s(x) < \eps \}$ and
first consider only tensors $u$ compactly supported in $A_\eps$. Using parallel transport, we may identify such a tensor field with an $s$-dependent tensor field that is defined on $\partial A$. Then for each fixed $y \in \partial A$, we can apply the 1-dimensional Hardy inequality from step 1) to the function $u(y,s)$ of $s$, and afterwards
integrate with respect to $y$ using the integration element on $\partial A$. This gives
\bena
\int_A |u(x)|^p e^{-2/s^\alpha} \D vol_A &=& \int_{\partial A} \left( \int_0^\eps |u(s,y)|^p J(s) e^{-2/s^\alpha} \D s \right) \D vol_{\partial A} \non\\
&\le & \left( \frac{p}{\alpha} \right)^p \int_{\partial A} \left( \int_0^\eps \bigg|\frac{\partial}{\partial s} [uJ^{1/p}(s,y)] \bigg|^p s^{p(\alpha+1)} e^{-2/s^\alpha}  \D s \right) \D vol_{\partial A} \non
\eena
where
\ben
J(s,y) = \frac{\D vol_{\partial A(s)}}{\D vol_{\partial A}} \ ,
\een
and with $\partial A(s) = \{ x \in A \mid {\rm dist}(x,\partial A) = s\}$. We distribute the $s$-derivatives using the Leibniz rule, we use $\partial \log J/\partial s = \vartheta$ (i.e. equal to the expansion
of the generators of geodesics orthogonal to $\partial A$), and we use the Minkowski inequality for $L^p$-norms. We find:
\ben
\begin{split}
\int_A |u|^p e^{-2/s^\alpha} \D vol_A  &\le \frac{1}{\alpha^p} \bigg\{ \eps^{p(1+\alpha)} (\sup_{s \le \epsilon} |\vartheta|)^p \int_A |u|^p e^{-2/s^\alpha} \D vol_A + \\
& \hspace{1.5cm} \int_A |D_{\partial/\partial s} u|^p s^{p(\alpha+1)} e^{-2/s^\alpha} \D vol_A
\bigg\} \ .
\end{split}
\een
We clearly have $|D_{\partial/\partial s} u| \le |D u|$, and if we furthermore choose $\epsilon$ so small that
\ben
(\eps^{\alpha+1} \sup |\vartheta|/\alpha)^p < 1/2
\een
 then we get
\ben\label{scalar}
\int_A |u|^p e^{-2/s^\alpha} \D vol_A  \le \frac{2}{\alpha^p} \int_A |D u|^p s^{p(\alpha+1)} e^{-2/s^\alpha} \D vol_A \ ,
\een
holding for all $u \in C^\infty_0(A)$ whose support is contained in $A_\eps$. One can apply the same kind of estimate again
to the right side (noting that it holds for tensors). For example, for $p=2$, we get in this way
\ben
\begin{split}
&\int_A |Du|^2 s^{2(\alpha+1} e^{-2/s^\alpha} \D vol_A  \\
\le& \frac{2}{\alpha^2} \int_A |D(s^{1+\alpha} Du)|^2 s^{2(\alpha+1)} e^{-2/s^\alpha} \D vol_A\\
\le& \frac{2}{\alpha^2} \int_A |D^2 u|^2 s^{4(\alpha+1)} e^{-2/s^\alpha} \D vol_A + \frac{2(\alpha+1)^2 \eps^{2\alpha}}{\alpha^2} \int_A
|Du|^2 s^{2(\alpha+1)} e^{-2/s^\alpha} \D vol_A \ .
\end{split}
\een
We now let $\eps$ be so small that $\frac{2(\alpha+1)^2 \eps^{2\alpha}}{\alpha^2}<1/2$. Then the second term on the right side can be absorbed by the left side, resulting in
\ben
\int_A |Du|^2 s^{2(\alpha+1)} e^{-2/s^\alpha} \D vol_A
\le \frac{4}{\alpha^2} \int_A |D^2 u|^2 s^{4(\alpha+1)} e^{-2/s^\alpha} \D vol_A ,
\een
for smooth tensor fields $u$ supported in $A_\eps$. Combining this with eq.~\eqref{scalar}, we may write, for some $c_1$,
\ben\label{result1}
\|u\|_{H^{2,\alpha}}^2 \le c_1 \int_A |D^2 u|^2 s^{4(\alpha+1)} e^{-2/s^\alpha} \D vol_A
\een
for smooth $u$ compactly supported in $A_\eps$. (Here  we recall the notations $H^{k,\alpha} = W^{2,k,\alpha}$ and $L^{2,\alpha} = H^{0,\alpha}$.)

\medskip
\noindent
Step 3) We now wish to obtain, in the case $p=2$, an inequality similar to step 2) for vector fields $u=X$ but with $DX$
replaced by $\pounds_X h$. Note that such an inequality does not follow directly from  \eqref{scalar}, because
$(\pounds_X h)_{ij} = D_i X_j + D_j X_i$, whereas $Du=DX$ corresponds to $D_i X_j$ with tensor indices written out, i.e.
we have an additional symmetrization of the tensor indices. In fact, we have instead
\ben
|D X|^2 = \half |\pounds_X h|^2 - (div_h X)^2 + Ric_h(X,X) - div_h(D_X X  - X div_h X) \ ,
\een
by an elementary computation. Using this identity on the r.h.s. of eq.~\eqref{scalar} gives us
\ben\label{first}
\begin{split}
& \int_A |X|^2 e^{-2/s^\alpha} \D vol_A  \\
\le &\frac{2}{\alpha^2}\int_A |DX|^2 s^{2(\alpha+1)} e^{-2/s^\alpha} \D vol_A \\
\le & \frac{2}{\alpha^2} \int_A \Big\{ \half |\pounds_X h|^2 + Ric_h(X,X) - div_h(D_X X  - X div_h X) \Big\} s^{2(\alpha+1)} e^{-2/s^\alpha} \D vol_A
\end{split}
\een
The total divergence terms are treated with a partial integration, which yields
\ben
\begin{split}
&- \int_A div_h(D_X X  - X div_h X) s^{2(\alpha+1)} e^{-2/s^\alpha} \D vol_A \\
&= \int_A \langle \D s, D_X X  - X div_h X \rangle \frac{d}{ds}  \{ s^{2(\alpha+1)} e^{-2/s^\alpha} \} \D vol_A\\
& = \int_A \Big\{ \langle \D s, D_X X \rangle - \langle \D s, X \rangle div_h X  \Big\}(2\alpha + (2\alpha+2) s^\alpha )  s^{\alpha+1} e^{-2/s^\alpha}  \D vol_A\\
\end{split}
\een
To estimate  the second term under the integral, we now use the Cauchy-Schwarz-inequality, together with the elementary inequality $( div_h X)^2 \le \frac{d-1}{4} |\pounds_X h|^2$.
To estimate the first term, we use that
\ben
\begin{split}
\langle \D s, D_X X \rangle =& \langle \D s, X \rangle \{ (\pounds_X h)(\D s, \D s) - div_h X \} - \\
& H_{\partial A}(X,X) - \vartheta \langle \D s, X \rangle^2 + div_{\partial A} (X_{\partial A} \langle \D s, X \rangle )
\end{split}
\een
where in the last term, $X_{\partial A}$ denotes the projection of $X$ along the surfaces of constant $s$, and $div_{\partial A}$ is the intrinsic divergence on these surfaces (so that this term
does not contribute under an integral). $H_{\partial A}$ is the extrinsic curvature of these surfaces, and $\vartheta$ half its trace (expansion). Using also the Cauchy-Schwarz inequality gives the bound
\ben
\begin{split}
&- \int_A div_h(D_X X  - X div_h X) s^{2(\alpha+1)} e^{-2/s^\alpha} \D vol_A\\
\le &\frac{(4\alpha + 2)(d+2)}{4} \int_A |\pounds_X h|^2 s^{2(\alpha+1)} e^{-2/s^\alpha} \D vol_A + \\
& + (4\alpha+2)\Big\{ 2+ \eps^{1+\alpha} \sup_{s<\eps} (|H|+|\vartheta|) \Big\} \int_A |X|^2  e^{-2/s^\alpha} \D vol_A \ .
\end{split}
\een
We choose $\eps$ so small that $\eps^{1+\alpha} \sup (|H|+|\vartheta|) < 1$. We also have
\ben
\int_A Ric(X,X) s^{2(\alpha+1)} e^{-2/s^\alpha} \D vol_A \le \eps^{2\alpha + 2} \sup_{s<\eps} |Ric_h|  \int_A |X|^2 e^{-2/s^\alpha} \D vol_A ,
\een
and we additionally choose $\eps$ so small that $\eps^{2+2\alpha} \sup |Ric_h| < 1$. Combining these inequalities with eq.~\eqref{first}, we get
\ben\label{second}
\begin{split}
\int_A |X|^2 e^{-2/s^\alpha} \D vol_A  \le &  \frac{1+(2\alpha + 1)(d+2)}{\alpha^2}  \int_A |\pounds_X h|^2  s^{2(\alpha+1)} e^{-2/s^\alpha} \D vol_A +\\
&+ \frac{24\alpha+14}{\alpha^2} \int_A |X|^2 e^{-2/s^\alpha} \D vol_A \ .
\end{split}
\een
We now choose $\alpha$ so large that $(24\alpha+14)/\alpha^2<1/2$. In that case, we get from~\eqref{second} the relation
\ben
\int_A |X|^2 e^{-2/s^\alpha} \D vol_A  \le  c_0  \int_A |\pounds_X h|^2  s^{2(\alpha+1)} e^{-2/s^\alpha} \D vol_A \ .
\een
In fact, because we have really estimated $\int_A |DX|^2 s^{2(\alpha+1)} e^{-2/s^\alpha} \D vol_A$ by the above argument as well, we get the same type of upper bound
for that quantity, too. Thus, we can write
\ben\label{result2}
\|X\|_{H^{1,\alpha}}^2 \le c_2 \int_A |\pounds_X h|^2  s^{2(\alpha+1)} e^{-2/s^\alpha} \D vol_A \ ,
\een
which holds for all $X \in C^\infty_0(A)$ with support in $A_\eps$, and a sufficiently large $\alpha>0$.

\medskip
\noindent
Step 4) Our next aim is to combine eqs.~\eqref{result1} and~\eqref{result2} to get, for $\bX = (u,X)$ supported in $A_\eps$, the inequality
\ben\label{result3}
\|u\|_{H^{2,\alpha}}^2 + \| X\|_{H^{1,\alpha}}^2 \equiv
\|\bX\|_{H^{2,\alpha} \oplus H^{1,\alpha}}^2 \le c_3 \|
\Phi \cC^* \bX \|_{L^{2} \oplus L^{2}}^2
\een
where $\cC^*$ is the adjoint of the constraint operator defined above in eq.~\eqref{delc*}, and where we recall that $\Phi$ is
the matrix multiplication operator~\eqref{matrixm}.
 Since the highest derivative parts of $\cC^*$ on $u$ respectively $X$ are
$D^2 u$ respectively $\pounds_X h$ it follows immediately from the definition of $\cC^*$ and the definition of the norms $\| \ . \ \|_{H^{k,\alpha}}$ that inequality~\eqref{result3}
holds true in the special cases when $\bX$ is either $(u,0)$ or $(0,X)$. To deal with the general case, one only has to take care of the `cross terms' between $X$ and $u$ in
a fairly straightforward way. For this, we first use the Cauchy-Schwarz inequality and that $\bX$ has its support for $s<\eps$, to get:
\ben\label{intermediate}
\begin{split}
\|
\Phi
\cC^* \bX
\|_{L^{2} \oplus L^{2}}^2 \ge & \int_A |\pounds_X h|^2  s^{2(\alpha+1)} e^{-2/s^\alpha} \D vol_A + \int_A |D^2 u|^2 s^{4(\alpha+1)} e^{-2/s^\alpha} \D vol_A \\
& - c_4 \eps^{1+\alpha} \Big\{
   \sup_{s<\eps} (|Ric_h| + |p|^2) \cdot \| u \|_{H^{2,\alpha}} \| u \|_{L^{2,\alpha}} + \\
& \hspace{1.7cm} \sup_{s<\eps} |p| \cdot \| u \|_{H^{2,\alpha}} \| X \|_{H^{1,\alpha}} + \\
& \hspace{1.7cm} \sup_{s<\eps} (|Ric_h|^2 + |p|^2 + |Ric_h| |p|^2) \cdot \| u \|_{L^{2,\alpha}}^2 + \\
& \hspace{1.7cm} \sup_{s<\eps} |p| |Ric_h| \cdot \| X \|_{H^{1,\alpha}} \| u \|_{L^{2,\alpha}} \\
& \hspace{1.7cm} \sup_{s<\eps} |p|^2 \cdot \| X \|_{H^{1,\alpha}}^2 \Big\} \ .
\end{split}
\een
The first two terms on the r.h.s. represent the `diagonal terms' and are bounded from below respectively by~\eqref{result1} and~\eqref{result2}. The terms in curly brackets represent the `cross terms' and are bounded from
below by the trivial inequality $-2ab\ge -a^2-b^2$. Then the desired in equality~\eqref{result3} immediately follows for sufficiently small $\eps > 0$.

\medskip
\noindent
Step 5) The next step is to establish an inequality of the form~\eqref{result3} on the `complement' of the set $A_\eps$ (inside $A$). More precisely, let
$O_\eps = A \setminus A_{\eps/2}$, so that $A=A_\eps \cup O_\eps$, and so that $O_\eps$ and $A_\eps$ have an open overlap.
On $O_\eps$, we clearly have $s>\eps/2>0$, so the weight functions involving $s$ appearing in the various integrals are bounded away from zero and
basically have no influence. Alternatively speaking, for tensors supported in $O_\eps$, the $H^{k,\alpha}$ norms are all equivalent to the norms with $\alpha = 0$,
i.e. ordinary Sobolev space $H^k$-norms with no weight factors. In this setting, inequalities~\eqref{result1} and~\eqref{result2} are standard consequences of the
ellipticity of the operators $D^j D^i D_j D_i u$ and $D^i D_{(i} X_{j)}$ together with the fact that none of these operators has a kernel in $H_0^2(O_\eps)$ respectively
$H^1_0(O_\eps)$, as such objects would correspond to KVF's $D^j u$ respectively $X^i$ with vanishing boundary values on $\partial O_\eps$, which do not exist. To
derive from this a result of the type~\eqref{result3} for $\bX$ supported in $O_\eps$, one can proceed in a similar way as in step 4) and estimate `diagonal terms' and `cross terms',
the details of which are given in lemma~2.8\footnote{Our situation corresponds to $\psi=e^{-1/s^\alpha}, \phi=s^{1+\alpha}$ in the notation of that lemma,
which satisfies assumption A.2 of~\cite{Chrusciel}.} of~\cite{Chrusciel}. One obtains:
\ben\label{result4}
\|\bX\|_{H^{2,\alpha} \oplus H^{1,\alpha}}^2 \le c_5 \left\{ \|
\Phi
\cC^* \bX
\|_{L^{2} \oplus L^{2}}^2 + \|\bX\|^2_{H^{1,\alpha} \oplus H^{0,\alpha}} \right \} \ ,
\een
holding for $\bX \in C^\infty(A)$ supported in $O_\eps$.

\medskip
\noindent
Step 6) One next combines \eqref{result3} (for $\bX$ supported in $A_\eps$) and
\eqref{result4} (for $\bX$ supported in $O_\eps$). Suppose that the Poincar\' e-Friedrichs
inequality does {\em not} hold. Then there is a sequence
$\bX_n \in H^{2,\alpha}_0(A) \oplus H^{1,\alpha}_0(A)$, such that $\|\bX_n - P_A \bX_n\|_{H^{2,\alpha} \oplus H^{1,\alpha}}=1$, but $\| \Phi \cC^* \bX_n\|_{L^2 \oplus L^2} \to 0$. Let $\chi_1 + \chi_2=1$ be a partition of unity such
that ${\rm supp} \chi_1 \subset O_\eps, {\rm supp} \chi_2 \subset  A_\eps$. Then we estimate
\bena\label{kondra}
\|\bX_n\|_{H^{2,\alpha} \oplus H^{1,\alpha}} &\le& \|\chi_1 \bX_n\|_{H^{2,\alpha} \oplus H^{1,\alpha}} + \|\chi_2 \bX_n\|_{H^{2,\alpha} \oplus H^{1,\alpha}} \non\\
&\le& c_6 \|\cC^* (\chi_1 \bX_n) \|_{L^2 \oplus L^2} + c_6 \| \chi_1 \bX_n \|_{H^{1,\alpha} \oplus H^{0,\alpha}}
+ c_6\|\Phi \cC^* (\chi_2 \bX_n) \|_{L^2 \oplus L^2} \non\\
&\le&  c_7 \|\chi_1 \cC^* \bX_n \|_{L^2 \oplus L^2} + c_7\|[\cC^*, \chi_1] \bX_n\|_{L^2 \oplus L^2} +
c_7\| \chi_1 \bX_n \|_{H^1 \oplus H^0} + \non\\
&&
c_8\|\chi_2 \Phi \cC^* \bX_n \|_{L^2 \oplus L^2} + c_8\|\Phi [\cC^*, \chi_2] \bX_n\|_{L^2 \oplus L^2}\non\\
&\le&  c_9 \|\Phi \cC^* \bX_n \|_{L^2(A) \oplus L^2(A)} + c_9\| X_n \|_{H^1(O) \oplus H^0(O)}
\eena
with possibly new constants in each line. In the last step we used that the commutator $[\cC^*, \chi]$
with a smooth compactly supported function $\chi$ decreases the order of each entry of the matrix
operator $\cC^*$ by one unit (unless the order of the entry is already $=0$), so that
$[\cC^*, \chi]: H^2 \oplus H^1 \to H^1 \oplus H^0$ is bounded. Now, by assumption,
$\| \Phi \cC^* \bX_n\|_{L^2 \oplus L^2} \to 0$ for the first term on the right side.
On the other hand, since $\{\bX_n\}$ is by assumption bounded in $H^2(O_\eps) \oplus H^1(O_\eps)$, it follows
from the Rellich-Kondrachov compactness theorem~(see e.g.~\cite{Evans}) that $\{\bX_n\}$ (or a subsequence thereof) is
Cauchy in $H^1(O_\eps) \oplus H^0(O_\eps)$. Hence, the above inequality shows that a subsequence of $\{\bX_n\}$
is Cauchy in $H^{2,\alpha}(A) \oplus H^{1,\alpha}(A)$, hence convergent with limit $\bX$ in this space.
By the continuity of $\Phi \cC^*: H^{2,\alpha}(A) \oplus H^{1,\alpha}(A) \to L^2(A) \oplus L^2(A)$,
we learn that $\Phi\cC^*\bX=0$. So $\bX$ must be equal, almost everywhere, to a non-trivial Killing
vector field. But then, clearly $\|\bX - P_A \bX\|_{H^{2,\alpha} \oplus H^{1,\alpha}}=0$ which is in contradiction with
our assumption $\|\bX_n - P_A \bX_n\|_{H^{2,\alpha} \oplus H^{1,\alpha}}=1$ and the convergence of the sequence.

\section{Behavior of electromagnetic perturbations near $\mathscr I$ in higher dimensions} \label{app:E}

Here we study the behavior of electromagnetic perturbations near null infinity in asymptotically flat backgrounds solving  the vacuum 
Einstein equations following the method described in~\cite{Hollands-Ishibashi} for gravitational perturbations. 
We impose the Lorentz gauge $\nabla^a A_a = 0$, and define the unphysical metric as $\tilde g_{ab} = f^2 g_{ab}$, where 
$f$ is a conformal factor such that, at $\mathscr I$, $f=0$ and $\tilde \nabla_a f \neq 0$. 
We also set 
\ben
\tilde A_a = f^{-(d-4)/2} A_a, \quad \tilde \phi = f^{-1} \tilde n^a \tilde A_a \ , \quad \tilde n^a = \tilde g^{ab} \tilde \nabla_b f  
 \ ,
\een
and we adopt the convention that indices on tensors with a tilde are raised and lowered using $\tilde g^{ab}$. The smoothness of $\tilde g_{ab}$ at 
$\mathscr I$ evidently implies the smoothness of the corresponding Ricci tensor, $\tilde R_{ab}$, at $\mathscr I$. It follows from the background 
Einstein equations that $f^{-1}{\tilde n}^a{\tilde n}_a$ 
is smooth at $\mathscr I$, and hence that $\tilde n^a$ is null there. After a lengthy calculation using the background Einstein equations, 
one finds that the Maxwell equation $\nabla^a \nabla_{[a} A_{b]} = 0$ in Lorentz gauge is equivalent to the coupled system of equations
\bena
\tilde \nabla^b \tilde \nabla_b \tilde A_a &=& 2 \tilde \nabla_a \tilde \phi + 
\frac{2}{d-2} \tilde R_a{}^b \tilde A_b + \frac{d-4}{2(d-1)(d-2)} \tilde R \tilde A_a 
\\
\tilde \nabla^b \tilde \nabla_b \tilde \phi &=& 
-\frac{2}{d-2} \tilde R^{ab} \tilde \nabla_a \tilde A_b - \frac{1}{2(d-1)} \tilde A^a \tilde \nabla_a \tilde R + \frac{d^2}{2(d-1)(d-2)} \tilde R \tilde \phi 
\eena
for $(\tilde A_a, \tilde \phi)$ in the unphysical spacetime metric $\tilde g_{ab}$. The first key point is that these equations have the character of wave equations, i.e. the highest derivative 
part is $\tilde \nabla^b \tilde \nabla_b$. Therefore, the initial value problem is well posed in the unphysical spacetime $\tilde \M$. The second key point is that 
all inverse powers of $f$ have cancelled out (!), meaning that, on the right side, all coefficient tensors are manifestly smooth at $\mathscr I$. Hence, 
if $A_a$, and hence $(\tilde A_a, \tilde \phi)$,
have initial data of compact support on some Cauchy surface as drawn in fig.~\ref{DOC1}, then the solution $(\tilde A_a, \tilde \phi)$ 
to the above system of equations will be smooth at $\mathscr I$. 
In particular, it follows that $A_a$ decays as $f^{(d-4)/2}$ near $\mathscr I$, and it follows 
from the definition of $\tilde \phi$ that $\tilde n^a \tilde A_a$ has to vanish on $\mathscr I$. 

Consider now a quadrangle shaped domain as in fig.~\ref{DOC1}. The fall-off behavior at $\mathscr I$ of $A_a$ implied 
by the smoothness of $\tilde A_a$ and the vanishing of $\tilde n^a \tilde A_a$ on $\mathscr I$ allow one to write
\ben
\int_{{\mathscr I}_{12}} \star w =  \frac{1}{2\pi} \int_{\eI_{12}} (\pounds_{\tilde n} \tilde A^a) \pounds_{\tilde n} \tilde A_a- C(\eC_1, A) + C(\eC_2, A) \ , 
\een
where the boundary terms are as in eq.~\eqref{Cdef1}, and where the symplectic current $w_a$ is as in eq.~\eqref{omegadef1}. Similarly, using the 
gauge condition $A_a K^a |_{\eH} = 0$, one can write, with $n^a = K^a$, 
\ben
\int_{{\mathscr H}_{12}} \star w =  \frac{1}{2\pi} \int_{\eH_{12}} (\pounds_{n} A^a)\pounds_n A_a - B(\eB_1, A) + B(\eB_2, A) \  ,  
\een
where the boundary terms are as in \eqref{Bdef1}. 
Now integrate $\D \star w=0$ over the quadrangle, and use Gauss' theorem to write the integral as a sum of boundary integrals over $\Sigma_1, \Sigma_2, \eH_{12}, {\mathscr I}_{12}$. 
The last  two integrals were just evaluated, whereas the first two give the canonical energy associated with $\Sigma_1, \Sigma_2$, respectively. Equation~\eqref{flux1} follows.

\end{document}